\documentstyle[aps,pre,floats,epsfig]{revtex}

\newcommand{\be}{\begin{equation}}
\newcommand{\ee}{\end{equation}}
\newcommand{\bea}{\begin{eqnarray}}
\newcommand{\eea}{\end{eqnarray}}
\def\(#1){(\ref{#1})}

\newcommand{\eg}{{\em e.g.}}
\newcommand{\ie}{{\em i.e.}}

\newcommand{\bi}{\begin{itemize}}
\newcommand{\ei}{\end{itemize}}

\newcommand{\rh}{\rho}
\newcommand{\sig}{\sigma}
\newcommand{\rhosig}{\rh(\sig)}
\newcommand{\rhonsig}{\rh\pn(\sig)}
\newcommand{\rhoalsig}{\rh\pa(\sig)}
\newcommand{\prior}{R(\sig)}
\newcommand{\effprior}{\tilde{R}(\sig)}
\newcommand{\musig}{\mu(\sig)}
\newcommand{\intsig}{\int\!d\sig\,}
\newcommand{\intL}{\int\!dL\,}
\newcommand{\w}{w}
\newcommand{\wi}{w_i(\sig)}

\newcommand{\m}{\rho}
\newcommand{\mi}{\m_i}
\newcommand{\mn}{\m_0}
\newcommand{\mo}{\m_1}
\newcommand{\mt}{\m_2}

\newcommand{\nm}{m}
\newcommand{\nmi}{\nm_i}
\newcommand{\nmn}{\nm_0}
\newcommand{\nmo}{\nm_1}

\newcommand{\mhat}{\lambda}
\newcommand{\mhatn}{\mhat_0}
\newcommand{\mhato}{\mhat_1}
\newcommand{\mhati}{\mhat_i}

\newcommand{\fexc}{\tilde{f}}
\newcommand{\gexc}{\tilde{g}}
\newcommand{\fideal}{f_{\rm id}}
\newcommand{\sideal}{s_{\rm id}}
\newcommand{\muexci}{\tilde{\mu}_i}

\newcommand{\fproj}{f_{\rm pr}}
\newcommand{\sproj}{s_{\rm pr}}
\newcommand{\fcomb}{f_{\rm comb}}
\newcommand{\scomb}{s_{\rm comb}}

\newcommand{\fmom}{f_{\rm m}}
\newcommand{\gmom}{g_{\rm m}}
\newcommand{\smom}{s_{\rm m}}
\newcommand{\pimom}{\Pi_{\rm m}}

\newcommand{\shat}{\m_0}
\newcommand{\al}{\alpha}
\newcommand{\ph}{v}
\newcommand{\phal}{\ph\pa}
\newcommand{\pn}{^{(0)}}
\newcommand{\po}{^{(1)}}
\newcommand{\pa}{^{(\al)}}
\newcommand{\pb}{^{(\beta)}}
\newcommand{\pt}{^{(2)}}

\newcommand{\prob}{n}
\newcommand{\probsig}{\prob(\sig)}
\newcommand{\probnsig}{\prob\pn(\sig)}
\newcommand{\nsig}{\probsig}
\newcommand{\nnsig}{\probnsig}
\newcommand{\nosig}{\prob\po(\sig)}
\newcommand{\ntsig}{\prob\pt(\sig)}

\newcommand{\dens}{r}

\newcommand{\mmmat}{\mbox{\boldmath $M$}}
\newcommand{\mident}{\mbox{\boldmath $1$}}
\newcommand{\hess}{\mbox{\boldmath $\tilde{F}$}}
\newcommand{\zerovect}{{\bf 0}}
\newcommand{\delm}{\delta\m}
\newcommand{\delmvect}{\mbox{\boldmath $\delta\m$}}
\newcommand{\mom}{moment density}
\newcommand{\moms}{moment densities}

\newcommand{\patrick}{combinatorial}
\newcommand{\Patrick}{Combinatorial}
\newcommand{\tpd}{t}

\newcommand{\rhovect}{\mbox{\boldmath $\rh$}}
\newcommand{\nvect}{\mbox{\boldmath $\prob$}}
\newcommand{\delrhovect}{\delta\mbox{\boldmath$\rh$}}
\newcommand{\muvect}{\mbox{\boldmath $\mu$}}
\newcommand{\order}{{\mathcal O}}

\newcommand{\eqref}[1]{\(#1)}
\newcommand{\kB}{k_{\text{B}}}
\newcommand{\kT}{T}
\newcommand{\sN}{s}
\newcommand{\smix}{s_{\rm mix}}
\newcommand{\myP}{{\cal P}}
\newcommand{\parent}{\probnsig}
\newcommand{\meanm}{m\pn}
\newcommand{\varm}{v\pn}
\newcommand{\loge}{\ln}
\newcommand{\prtns}{_{\mathrm{prtns}}}
\newcommand{\textint}{\int}

\newcommand{\parentL}{\prob\pn(L)}
\newcommand{\gener}{\psi}

\newcommand{\chiab}{\chi'}
\newcommand{\ls}{L_{\rm s}}
\newcommand{\tot}{_{\rm tot}}
\newcommand{\fd}{F_{\rm in}}

\newcommand{\fhalpha}{a}
\newcommand{\fhbeta}{b}
\newcommand{\fhrho}{\mn}
\newcommand{\fhphi}{\mo}
\newcommand{\fhnmo}{\nmo}

\newcommand{\logerr}{\delta}

\renewcommand{\L}{n}

\title{Moment free energies for polydisperse systems}

\author{Peter Sollich$^*$}
\address{Department of Mathematics, King's College London, Strand,
London WC2R 2LS, U.K.}

\author{Patrick B. Warren}
\address{Unilever Research Port Sunlight,
Bebington, Wirral, CH63 3JW, U.K.}

\author{Michael E. Cates}
\address{Department of Physics and Astronomy, University of
Edinburgh, Edinburgh EH9 3JZ, U.K.}

\draft 

\begin{document}

\maketitle

\begin{abstract}

A polydisperse system contains particles with at least one
attribute $\sigma$ (such as particle size in colloids or chain length
in polymers) which takes values in a continuous range. It therefore
has an infinite number of conserved densities, described by a density
{\em distribution} $\rhosig$. The free energy depends on all details
of $\rhosig$, making the analysis of phase equilibria in such systems
intractable. However, in many (especially mean-field) models the {\em
excess} free energy only depends on a finite number of (generalized)
moments of $\rhosig$; we call these models truncatable. 
We show, for these models, how to derive approximate
expressions for the {\em total} free energy which only depend on such
moment densities. Our treatment unifies and explores in detail two
recent separate proposals by the authors for the construction of such
moment free energies. We show that even though the moment free energy
only depends on a finite number of density variables, it gives the
same spinodals and critical points as the original free energy and
also correctly locates the onset of phase coexistence. Results from
the moment free energy for the coexistence of two or more phases
occupying comparable volumes are only approximate, but can be refined
arbitrarily by retaining additional moment densities. Applications to
Flory-Huggins theory for length-polydisperse homopolymers, and for
chemically polydisperse copolymers, show that the moment free energy
approach is computationally robust and gives new geometrical insights
into the thermodynamics of polydispersity.
\end{abstract}


\section{Introduction}
\label{sec:intro}

The thermodynamics of mixtures of several chemical species is, since
Gibbs, a well established subject (see \eg~\cite{DeHoff92}). But many
systems arising in nature and in industry contain, for practical
purposes, an infinite number of distinct, though similar, chemical
species. Often these can be classified by a parameter, $\sigma$, say,
which could be the chain length in a polymeric system, or the particle
size in a colloid; both are routinely treated as continuous
variables. In other cases (see
\eg~\cite{Bauer85,RatWoh91,NesOlvCri93,Solc93}) $\sigma$ is instead a
parameter distinguishing species of continuously varying chemical
properties. In the most general case, several attributes may be
required to distinguish the various particle species in the system
(such as length and chemical composition in length-polydisperse random
copolymers) and $\sig$ is then a collection of
parameters~\cite{discrete_sigma_footnote}. The thermodynamics of
polydisperse systems (as defined above) is of crucial interest to wide
areas of science and technology; it is sometimes also referred to as
``continuous thermodynamics'' (see \eg~\cite{RatWoh91}).

Standard thermodynamic procedures~\cite{DeHoff92} for constructing
phase equilibria in a system of volume $V$ containing $M$ different
species can be understood geometrically in terms of a free energy
surface $f(\rho_j)$ (with $f=F/V$, and $F$ the Helmoltz free energy) 
in the $M$-dimensional space of number densities
$\rho_j$. Tangent planes to $f$ define regions of
coexistence, within which the free energy of the system is lowered by
phase separation. The volumes of coexisting phases follow from the
well-known ``lever rule''\cite{DeHoff92}. Here ``surface'' and ``plane'' are
used loosely, to denote manifolds of appropriate dimension. This
procedure becomes unmanageable, both conceptually and numerically, in
the limit $M\to\infty$ which formally defines a polydisperse system.
There is now a separate conserved density $\rhosig$ for each value of
$\sigma$; $\rhosig$ is in fact a {\em density distribution} and the
overall number density of particles is written as $\rho=\intsig \rhosig$.
The free energy surface is $f = f[\rhosig]$ (a functional) which
resides in an infinite-dimensional space. Gibbs' rule allows the
coexistence of arbitrarily many thermodynamic phases.

To make phase equilibria in polydisperse systems more accessible to
both computation and physical intuition, it is clearly helpful to
reduce the dimensionality of the problem.  Theoretical work in this
area has made significant progress in finding simplified forms of the
conditions for phase coexistence, thus making these more numerically
tractable~\cite{RatWoh91,NesOlvCri93,Solc93,IrvKen82,%
SalSte82,SolKon95,GuaKinMor82,IrvGor81,BeeBerKehRat86,%
Hendriks88,HenVan92,CotPra85,ShiSanBeh87,Bartlett97}.  Our aim is to
achieve a similar reduction in dimensionality on a higher level, that
of the free energy itself. We show that, for a large class of models,
it is possible to construct a reduced free energy depending on only a
small number of variables. From this, meaningful information on phase
equilibria can be extracted by the usual tangent plane procedure, with
obvious benefits for a more intuitive understanding of the phase
behavior of polydisperse systems. As an important side effect, this
procedure also leads to robust algorithms for calculating polydisperse
phase equilibria numerically. In particular, such algorithms can
handle coexistence of more than two phases with relative ease compared
to those used
previously\cite{IrvKen82,Hendriks88,CotPra85,ShiSanBeh87,KinShoFes89,%
Michelsen87,Michelsen82,CotBenPra85,Michelsen86,Michelsen86b,%
Michelsen94,Michelsen94b}.

A clue to the choice of independent variables for such a reduced free
energy comes from the representation of experimental data. We recall
first the definition of a ``cloud point'' (see
\eg~\cite{RatWoh91,IrvKen82,ClaMcLeiJen95}): This is the point at
which, for a system with a given density distribution $\rhosig$, phase
separation first occurs as the temperature $T$ or another external
control parameter is varied. The corresponding incipient phase is
called the ``shadow''. Now consider diluting or concentrating the
system, \ie, varying its overall density $\rho$ while maintaining a
fixed ``shape'' of polydispersity $\probsig = \rhosig/\rho$. We will
find it useful later to refer to the collection of all systems
$\rhosig=\rho\probsig$ which can be obtained by this process as a
``dilution line'' (in the space of all density distributions
$\rhosig$).  Plotting the cloud point temperature $T$ versus density
$\rho$ then defines the cloud point curve (CPC), while plotting $T$
versus the density of the shadow gives the shadow
curve~\cite{dew_bubble}.
The differences in the shapes $\probsig$ of the density distributions
in the different phases are hidden in this representation; only the
overall densities $\rho$ in each phase are tracked. The density $\rho$
is a particular {\em moment} of $\rhosig$ (the zeroth one); higher
order moments would be given by $\intsig \sig^i \rhosig$. Generalizing
slightly, this suggests that our reduced free energy should depend on
several (generalized) {\em \moms} $\mi=\intsig \wi \rhosig$, defined
by appropriate (linearly independent) weight functions $\wi$. Ordinary
power-law \moms\ are included as the special case $\wi=\sig^i$.
Whatever the choice of moments, we insist on $w_0 = 1$ so that in all
cases the zeroth moment density $\rho_0$ coincides with the overall
number density $\rho$.

As an illustration, consider the simplest imaginable case where the
true free energy $f$ already has the required form, \ie, it depends only
on a finite set of $K$ \moms\ of the density
distribution:
%
\be
f = f(\mi), \qquad i=1\ldots K \quad {\mathrm or} \quad 
i = 0 \ldots K-1
\label{puremoment}
\ee
where the range of $i$ depends on whether $\rho_0\equiv \rho$ is among
the $K$ moment densities on which $f$ depends. (From now on, we use
the notation $i=1\ldots K$ inclusively, to cover both possibilities.)
In coexisting phases one demands equality of particle chemical
potentials, defined as $\musig={\delta f/\delta\rhosig}$, for all
$\sig$. Because
\[
\musig\equiv\frac{\delta f}{\delta\rhosig}=\sum_i
\frac{\partial f}{\partial \mi} \wi=\sum_i \mu_i \wi
\]
this implies that all ``moment'' chemical potentials,
$\mu_i\equiv{\partial f}/{\partial \mi}$, are likewise equal among
phases. The second requirement for phase coexistence is that the
pressures or osmotic pressures~\cite{osmotic_footnote}, $\Pi$, of all
phases must also be equal. But from
\[
-\Pi = f-\intsig \musig\rhosig=f - \sum_i\mu_i\mi
\]
one sees that this again involves only the \moms\ $\mi$ and their
chemical potentials $\mu_i$.  Finally, the \moms\ also obey the
``lever rule'', as follows. Let the overall density distribution of a
system of volume $V$ be $\rhonsig$; we call this the ``parent''
distribution. If (after a lowering of temperature, for example) this
parent has split into $p$ coexisting ``daughter'' phases with
$\sigma$-distributions $\rhoalsig$, each occupying a fraction $\phal$
of the total volume ($\al=1\ldots p$), then particle conservation
implies the usual lever rule (or material balance) among species:
$\sum_\al \phal \rhoalsig = \rhonsig, \forall \sigma$.  Multiplying
this by a weight function $\wi$ and integrating over $\sig$ shows that
the lever rule also holds for the \moms:
\be 
\sum_{\al=1}^p \phal \mi\pa = \mi\pn
\label{lever_moments}
\ee
These results express the fact that any linear combination of
conserved densities (a generalized \mom) is itself a conserved density
in thermodynamics. We have shown, therefore, that if the free energy
of the system depends only on $K$ \moms\ $\mo\ldots \m_K$ we can view
these as the densities of $K$ ``quasi-species'' of particles, and
construct the phase diagram via the usual construction of tangencies
and the lever rule.  Formally this has reduced the problem to finite
dimensionality, although this is trivial here because
$f$, by assumption, has no dependence on any variables other than the
$\mi$ ($i=1\ldots K$).

Of course, it is uncommon for the free energy $f$ to
obey~\(puremoment). In particular, the entropy of an ideal mixture
(or, for polymers, the Flory-Huggins entropy term) is definitely not
of this form. On the other hand, in very many thermodynamic
(especially mean field) models the {\em excess} (\ie, non-ideal) part
of the free energy does have the simple form~\(puremoment). In other
words, if we decompose the free energy as (setting $k_{\rm B}=1$)
%
 \be f = -T\sideal + \fexc, \quad
\sideal = -\intsig \rhosig \left[\ln \rhosig-1\right]
\label{free_en_decomp_gen}
 \ee 
then the excess free energy $\fexc$ is a function of some \moms\ only:
\be
\fexc = \fexc(\mi), \qquad i=1\ldots K
 \label{puremoment_excess}
\ee
Examples of models of this kind include polydisperse hard spheres
(within the generalization by Salacuse and Stell~\cite{SalSte82} of
the BMCSL equation of state~\cite{Boublik70,ManCarStaLel71}),
polydisperse homo- and
copolymers~\cite{Bauer85,RatWoh91,NesOlvCri93,Solc93,SolKon95}, and
van der Waals fluids with factorized interaction
parameters~\cite{GuaKinMor82}. With the exception of a brief
discussion in Sec.~\ref{sec:conclusion}, this paper concerns models
with free energies of the
form~(\ref{free_en_decomp_gen},\ref{puremoment_excess}), which we call
%
%
``truncatable''. (This terminology emphasizes that the number $K$ of
\moms\ appearing in the excess free energy of truncatable models is
finite, while for a non-truncatable model the excess free energy
depends on all details of $\rhosig$, corresponding to an {\em
infinite} number of \moms.) In what follows, we regard each model free energy
as given, and do {\em not} discuss the issue of how good a description
of the real system it offers, nor how or whether it can be
derived from an underlying microscopic Hamiltonian. Whenever we refer to
``exact'' results, we mean the exact thermodynamics of such a model as
specified by its free energy.

Different authors refer in different terms to $\sideal$
in~(\ref{free_en_decomp_gen}). Throughout this paper will call the
quantity
\[
\sideal = -\intsig \rhosig \left[\ln \rhosig-1\right]
\]
(which is, up to a factor of $-T$, the ideal part of the free energy density)
the ``entropy of an ideal mixture''. By writing
$\rhosig=\rh\probsig$, where $\probsig$ is the normalized distribution
of $\sig$, this can be decomposed as
\be
\sideal = -\rh(\ln\rh-1) - \rh\intsig\probsig\ln\probsig
\label{sideal_decomp}
\ee
The first term on the right hand side is the ``entropy of an ideal gas'',
while the second term gives the ``entropy of mixing''. The prefactor $\rh$
reflects the fact that both are expressed per unit volume rather than
per particle.

In principle, the entire phase equilibria for any truncatable system
(obeying~(\ref{puremoment_excess})) can be computed exactly by a
finite algorithm. Specifically, the spinodal stability criterion
involves a $K$-dimensional square
matrix~\cite{IrvGor81,BeeBerKehRat86,Hendriks88,HenVan92} whereas
calculation of $p$-phase equilibrium involves solution of $(p-1)(K+1)$
coupled nonlinear equations (see Sec.~\ref{sec:coex_beyond}). This
method has certainly proved
useful~\cite{Bauer85,RatWoh91,NesOlvCri93,SolKon95,GuaKinMor82,HenVan92},
but is cumbersome, particularly if one is interested mainly in cloud-
and shadow-curves, rather than coexisting compositions deep within
multi-phase
regions~\cite{Bauer85,NesOlvCri93,SolKon95,ClaMcLeiJen95}. Various
ways of simplifying the procedure
exist~\cite{IrvKen82,IrvGor81,CotPra85,ShiSanBeh87,Bartlett97}, but
there has previously not been a systematic alternative to the full
computation. Note also that the nonlinear phase equilibrium equations
permit no simple geometrical interpretation or qualitative insight
akin to the familiar rules for constructing phase diagrams from the
free energy surface of a finite mixture. This further motivates the
search for a description of polydisperse phase equilibria in terms of
a reduced free energy which depends on only a small number of density
variables.

In previous work, the authors originally arrived independently at two
definitions of a reduced free energy in terms of \moms\ ({\em moment
free energy} for short)~\cite{SolCat98,Warren98}. Though based on
distinctly different principles, the two approaches led to
very similar results. We explain this somewhat
surprising fact in the present work; at the same time, we
describe the two methods in more detail and explore the relationship
between them. We also discuss issues related to practical applications
and give a number of example results for simple model systems.

The first route to a moment free energy takes as its starting
point the conventional form of the ideal part of the free energy,
$\fideal= T \intsig \rhosig [\ln \rhosig-1]$, as
in~\(free_en_decomp_gen). This can be thought as defining a
(hyper-)surface, with the ``horizontal'' co-ordinate being $\rhosig$
and the ``height'' of the surface $\fideal[\rhosig]$. As explained in
Sec.~\ref{sec:projection}, this surface can then be projected
geometrically onto one with only a finite set of horizontal
co-ordinates: these are chosen to be
the \moms\ appearing in the excess free
energy. Physically, this corresponds to minimizing the free energy
with respect to all degrees of freedom in $\rhosig$ that do not affect
these \moms. We call this first route the ``projection'' method.

The second approach rederives the entropy of mixing in the ideal part
of the free energy in a form that depends explicitly only on the
chosen \moms. As described in Sec.~\ref{sec:combinatorial}, the
expression that results is intractable in general because it still
contains the full complexity of the problem.  However, in situations
where there are only infinitesimal amounts of all but one of the
phases in the systems, the entropy of mixing can be evaluated in a
closed form. Applying this functional form in regimes where it is not
strictly valid (\ie, when phases of comparable volumes coexist)
generates a moment free energy by this ``\patrick'' method.

We show the equivalence of the two approaches in
Sec.~\ref{sec:equivalence}. There, we first demonstrate that the general
form of the entropy of mixing obtained by the \patrick\ method
can be transformed to the standard expression $-\intsig \probsig \ln
\probsig$. Second, we show that the moment free energies arrived at by
the two methods are in fact equal, with the projection method being
slightly more generally applicable.

In Sec.~\ref{sec:properties}, we then discuss the properties of the
moment free energy (as obtained from either the projection or the
\patrick\ method). By construction, it depends only on a (finite)
number of \moms, and so achieves the desired reduction in
dimensionality. The \moms\ can be treated as densities of
``quasi-species'' of particles, and the standard procedures of the
thermodynamics of finite mixtures can be applied to the moment free
energy to calculate phase equilibria. This is only useful, however, if
the results are faithful to the actual phase behavior of the
underlying polydisperse system (as modeled by the given truncatable
free energy). We show that this is so: In fact, the construction of
our moment free energy is such that exact binodals (cloud-point and
shadow curves), critical (and multi-critical) points and spinodals are
obtained. Beyond the onset of phase separation, where coexisting
phases occupy comparable volumes, the results are not exact, but can
be refined arbitrarily by adding extra \moms. This procedure is
necessary also to ensure that, in regions of multi-phase coexistence,
the correct number of phases is found. In
Sec.~\ref{sec:implementation}, we discuss the practical implementation
of our method, followed by a number of examples
(Sec.~\ref{sec:examples}). Our results are summarized in
Sec.~\ref{sec:conclusion}, where we also outline perspectives for
future work.

Note that throughout this work, we focus on the case of phase
coexistence at fixed {\em volume}. However, as described in
App.~\ref{app:const_p}, the formalism can easily be applied to
scenarios where the (mechanical or osmotic) {\em pressure} is fixed
instead.

\section{Derivation of moment free energy}
\label{sec:derivation}

\subsection{Projection method}
\label{sec:projection}

The starting point for this method is the
decomposition~\(free_en_decomp_gen) for the free energy of truncatable
systems~\(puremoment_excess):
\be
f = T \intsig \rhosig \left[\ln \frac{\rhosig}{R(\sigma)}
-1\right] + \fexc(\mi)
\label{free_en_decomp}
\ee
As explained in the previous section, truncatable here means that the
excess free energy $\fexc$ depends only on $K$ \moms\ $\mi$. Note
that, in the first (ideal) term of~(\ref{free_en_decomp}), we have
included a dimensional factor $R(\sigma)$ inside the logarithm. This
is equivalent to subtracting $T\intsig\rhosig\ln \prior$ from the free
energy. Since this term is {\em linear} in densities, it has no effect
on the exact thermodynamics; it contributes harmless additive constants to
the chemical potentials $\musig$. However, 
in the projection route to a moment free energy, it will play a central role.

We now argue that the most important \moms\ to treat correctly are
those that actually appear in the excess free energy $\fexc(\mi)$.
Accordingly we divide the infinite-dimensional space of density
distributions into two complementary subspaces: a ``moment subspace'',
which contains all the degrees of freedom of $\rhosig$ that contribute
to the \moms\ $\mi$, and a ``transverse subspace'' which contains all
remaining degrees of freedom (those that can be varied without
affecting the chosen \moms\
$\mi$)~\cite{moment_equivalence_class_footnote}. Physically, it is
reasonable to expect that these ``leftover'' degrees of freedom play a
subsidiary role in the phase equilibria of the system, a view
justified {\em a posteriori} below. Accordingly, we now allow
violations of the lever rule, so long as these occur {\em solely in
the transverse space}. This means that the phase splits that we
calculate using this approach obey the lever rule for the \moms, but
are allowed to violate it in other details of the density distribution
$\rhosig$ (see Fig.~\ref{fig:lever_rule} in
Sec.~\ref{sec:dense_copolymer} for an example). These ``transverse''
degrees of freedom, instead of obeying the strict particle
conservation laws, are chosen so as to minimize the free energy: they
are treated as ``annealed''.  If, as assumed above, $\fexc=\fexc(\mi)$
only depends on the \moms\ retained, this amounts to {\em maximizing
the ideal mixture entropy} ($\sideal$) in (\ref{free_en_decomp}),
while holding fixed the values of the \moms\ $\mi$.  Note that we are
allowed, if we wish, to include among the retained densities
``redundant'' moments on which $\fexc$ has a null dependence. We will
have occasion to do this later on.

At this point, the factor $\prior$ in (\ref{free_en_decomp}), which is
immaterial if all conservation laws are strictly obeyed, becomes
central. Indeed, maximizing the entropy over all distributions
$\rhosig$ at fixed \moms\ $\mi$ yields
\be
\rhosig=\prior\exp\left(\sum_i \mhati\wi\right)
\label{family}
\ee
where the Lagrange multipliers $\mhati$ are chosen to give the desired
\moms
\be
\mi = \intsig\wi\,\prior\exp\left(\sum_i \mhati\wi\right)
\label{moments_from_lambda}
\ee
The corresponding minimum value of $f$ then defines our projected
free energy as a function of the \moms\ $\mi$:
\be
\fproj(\mi) = - T \sproj(\mi) + \fexc(\mi), \qquad
\sproj= \shat - \sum_i \mhati\mi
\label{annd}
\ee
Here $\sproj$ is the projected entropy of an ideal mixture. The first
term appearing in it, $\mn = \intsig \rhosig$, is the ``zeroth
moment'', which is identical to the overall particle density $\rho$
defined previously. If this is among the \moms\ used for the
projection (or more generally, if it is a linear combination of them),
then the term $-T\mn$ is simply a linear contribution to the projected
free energy $\fproj(\mi)$ and can be dropped because it does not
affect phase equilibrium calculations.  Otherwise, $\mn$ needs to be
expressed -- via the $\mhati$ -- as a function of the $\mi$ and its
contribution cannot be ignored. We will see an example of this in
Sec.~\ref{sec:examples}.
 
The projection
method yields a free energy
$\fproj(\mi)$ which only depends on the chosen set of
\moms .
These can now be viewed as
densities of `quasi-species' of particles, and a finite-dimensional
phase diagram can be constructed from $\fproj$ according to the usual
rules, ignoring the underlying polydisperse nature of the system.
Obviously, though, the results now depend on $\prior$ which is
formally a ``prior distribution'' for the entropy maximization. To
understand its role, we recall that the projected free energy
$\fproj(\mi)$ was constructed as the minimum of $f[\rho(\sigma)]$ at
fixed $\mi$; that is, $\fproj$ is the lower envelope of the projection
of $f$ onto the moment subspace.  Crucially, the shape of this
envelope depends on how, by choosing a particular prior distribution
$\prior$, we ``tilt'' the infinite-dimensional free energy surface {\em
before} projecting it. This {\em geometrical} point of view is illustrated,
for a mixture of only two species, in Fig.~\ref{fig:proj_demo}.

\newlength{\figwidth}
\setlength{\figwidth}{4.3cm}
\setlength{\unitlength}{1pt}

\begin{figure}

\begin{center}
\epsfig{file=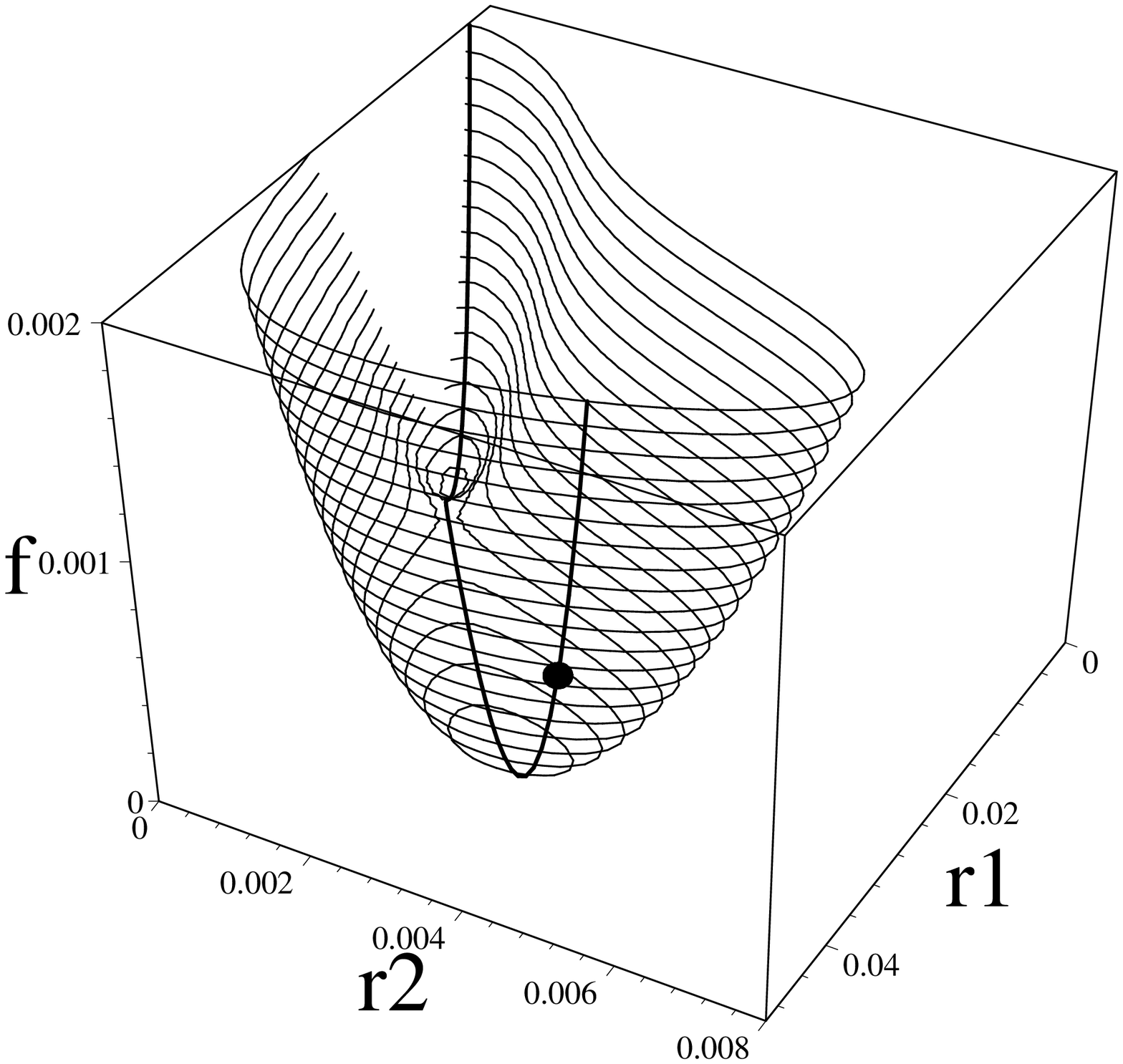,width=\figwidth}%
\begin{picture}(0,0)\put(-67,123){\mbox{(a)}}\end{picture}%
\epsfig{file=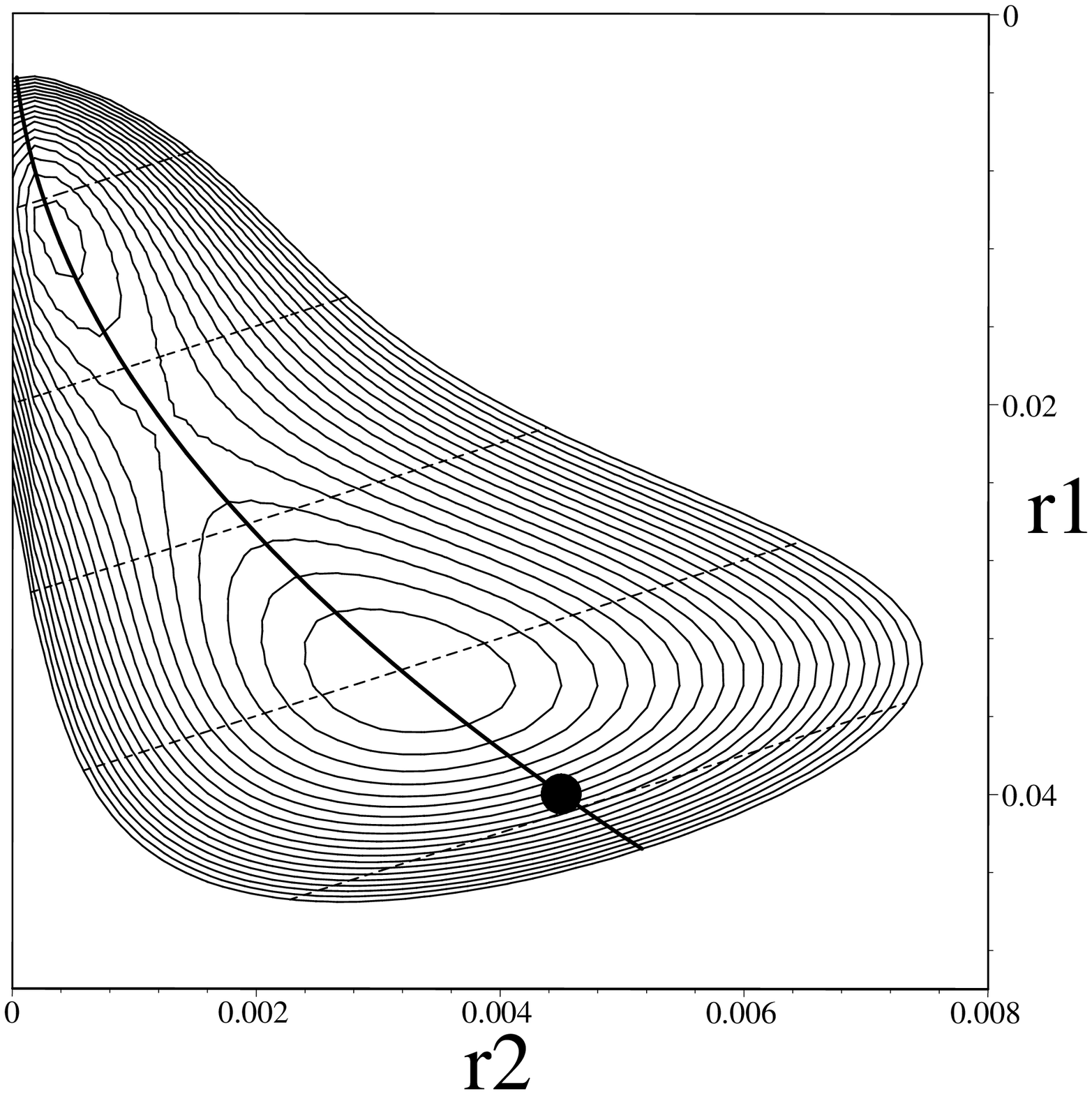,width=\figwidth}%
\begin{picture}(0,0)\put(-67,123){\mbox{(b)}}\end{picture}%
\hspace*{1mm}
\epsfig{file=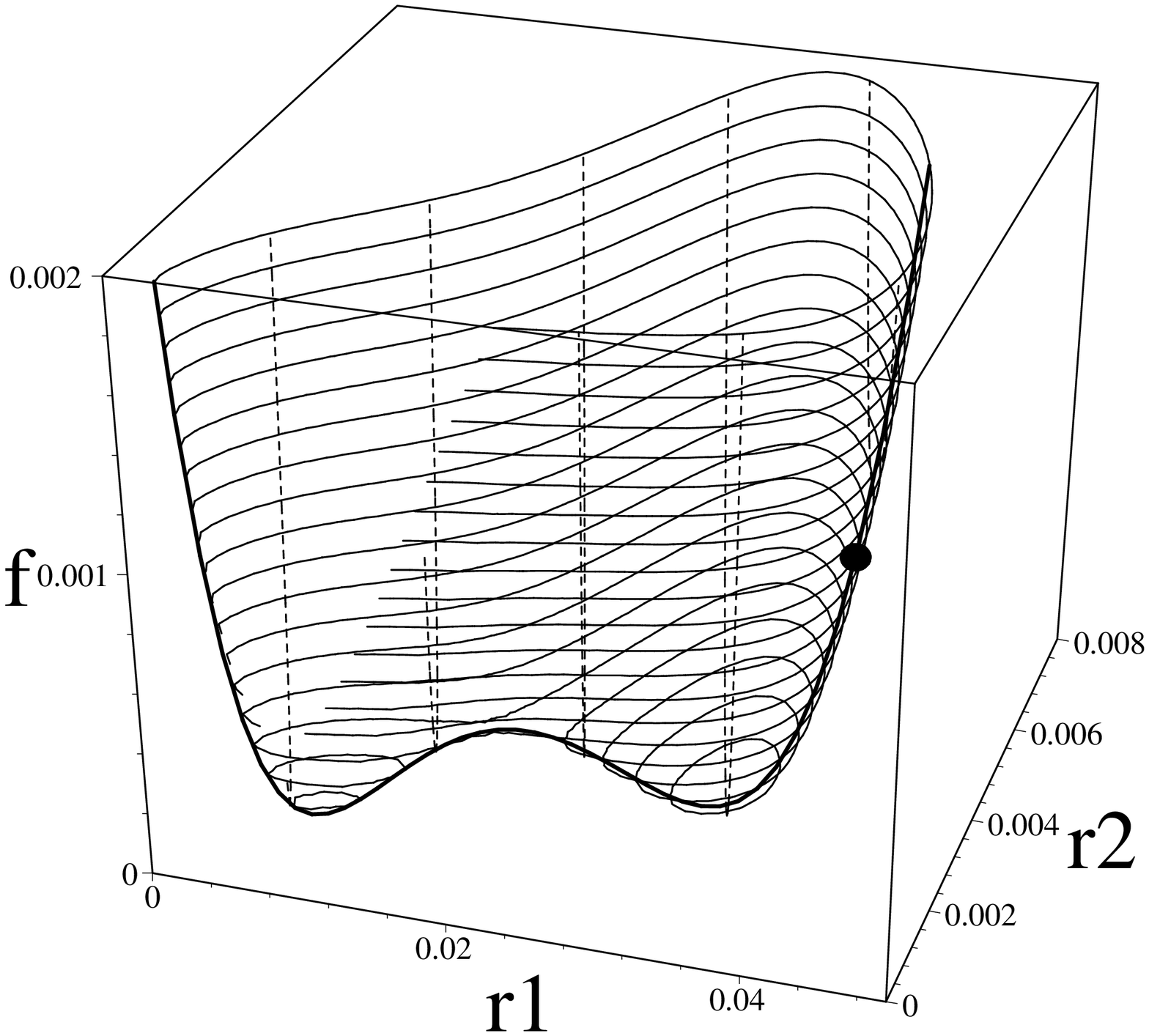,width=\figwidth}%
\begin{picture}(0,0)\put(-67,123){\mbox{(c)}}\end{picture}%
\epsfig{file=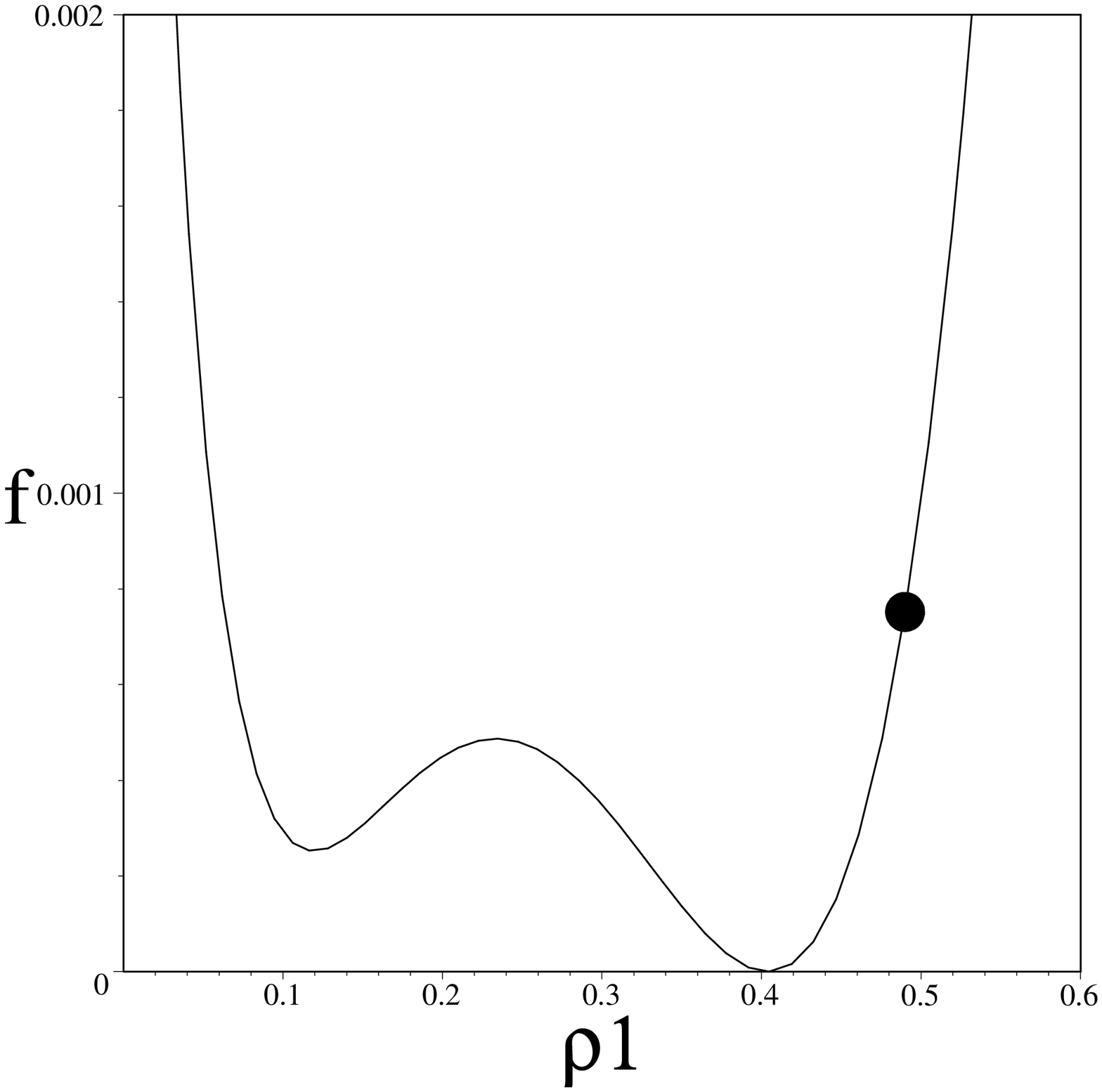,width=\figwidth}%
\begin{picture}(0,0)\put(-67,123){\mbox{(d)}}\end{picture}

\vspace*{8mm}

\epsfig{file=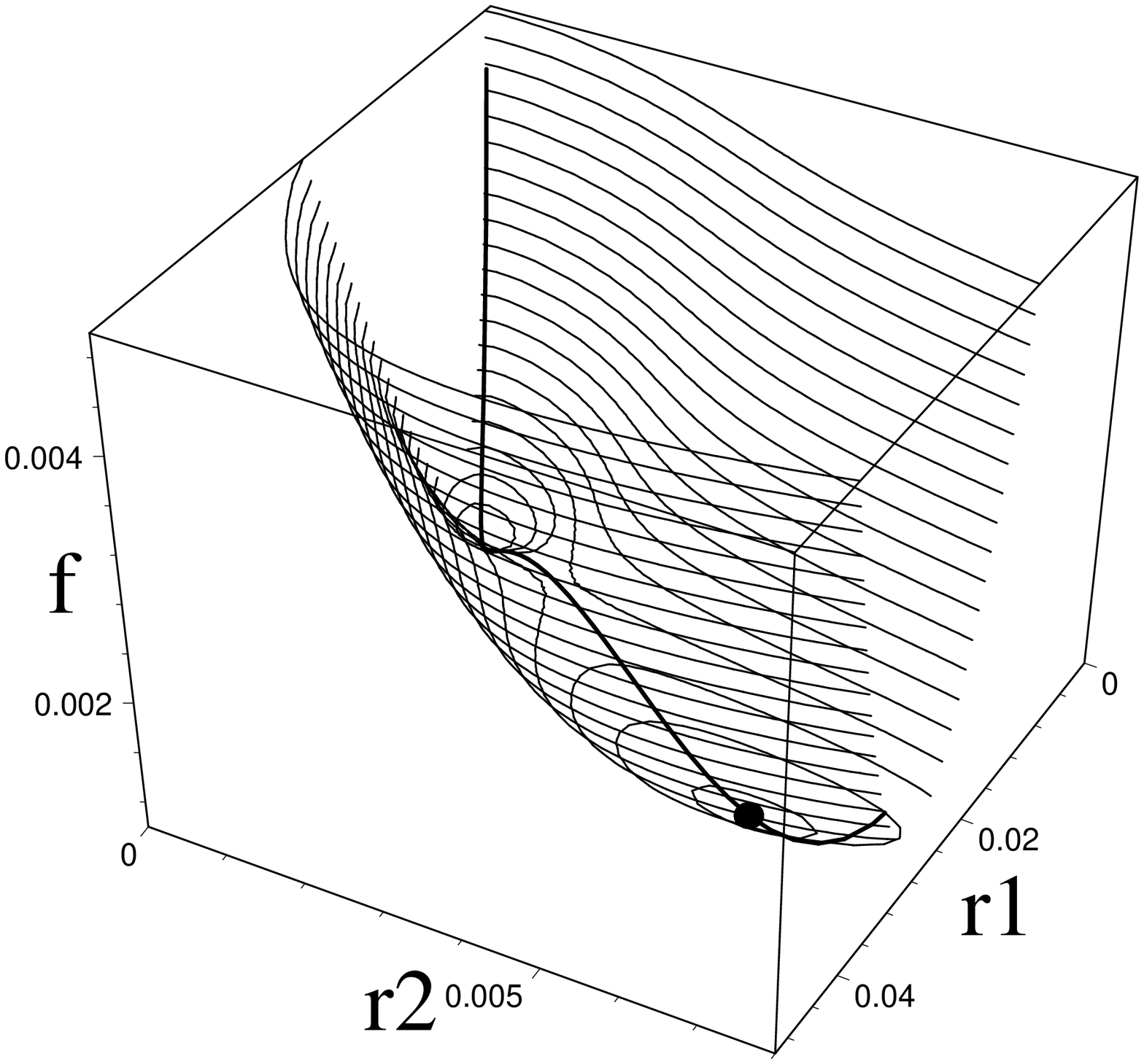,width=\figwidth}%
\begin{picture}(0,0)\put(-67,123){\mbox{(e)}}\end{picture}%
\epsfig{file=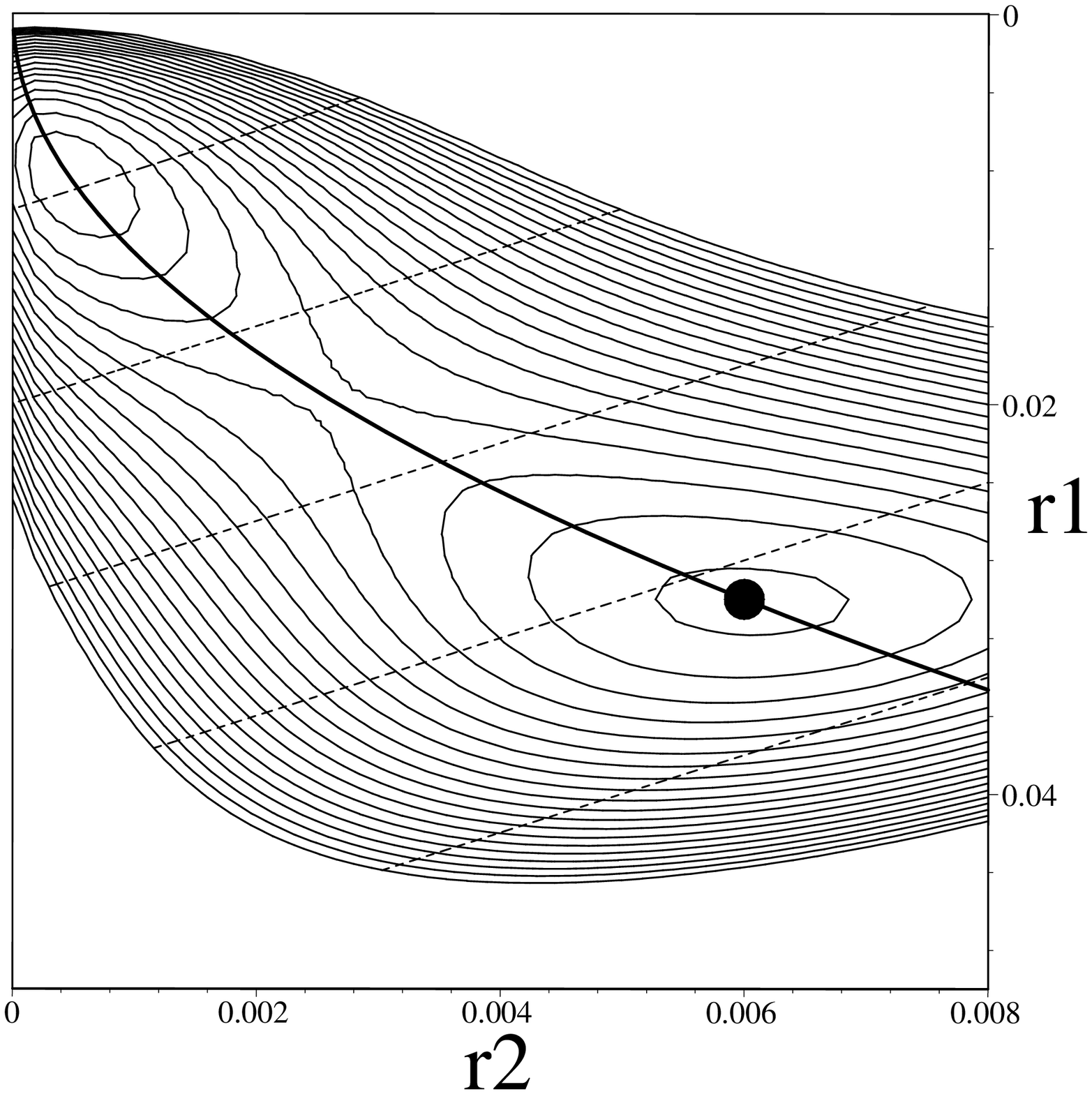,width=\figwidth}%
\begin{picture}(0,0)\put(-67,123){\mbox{(f)}}\end{picture}%
\hspace*{1mm}
\epsfig{file=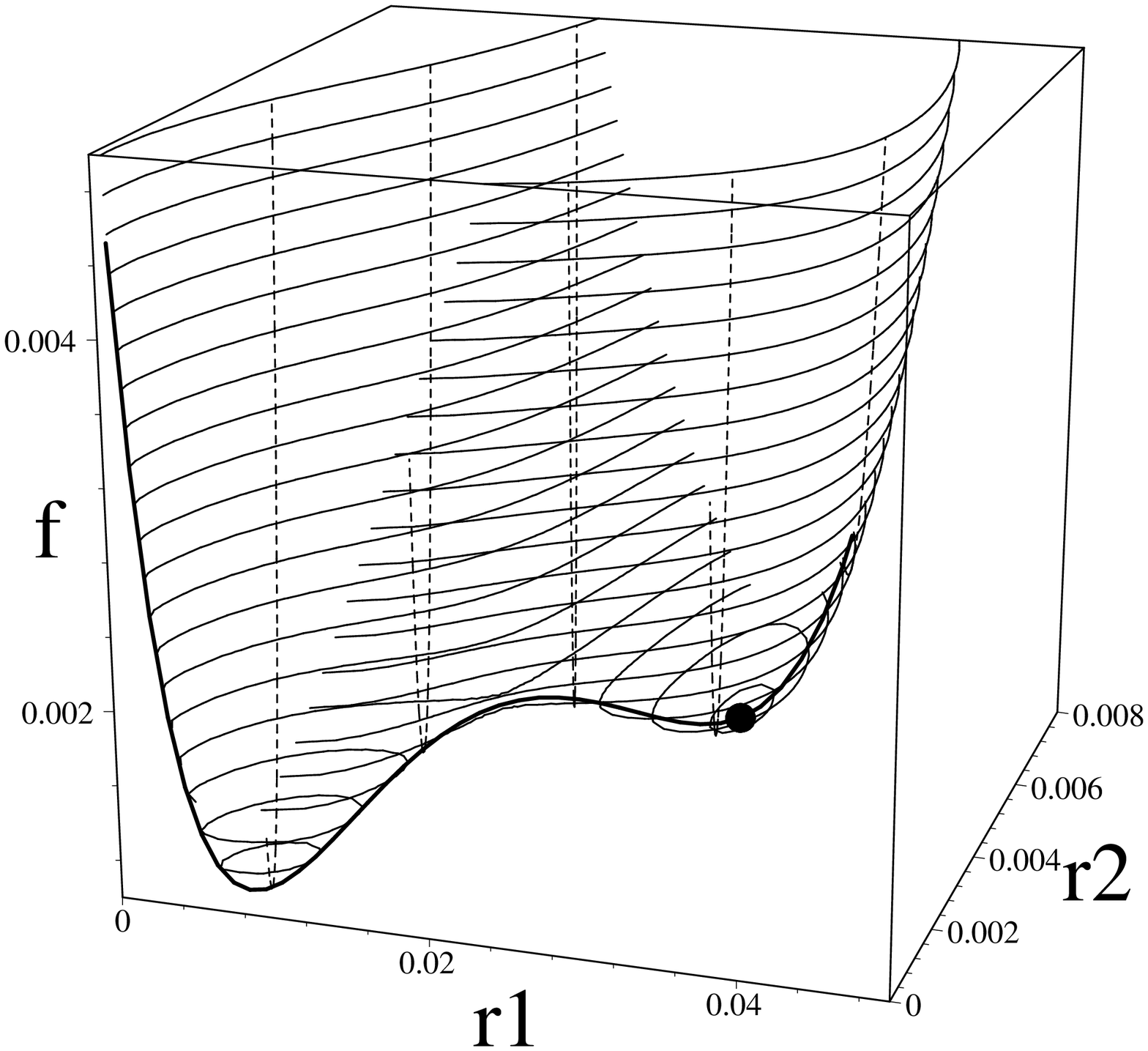,width=\figwidth}%
\begin{picture}(0,0)\put(-67,123){\mbox{(g)}}\end{picture}%
\epsfig{file=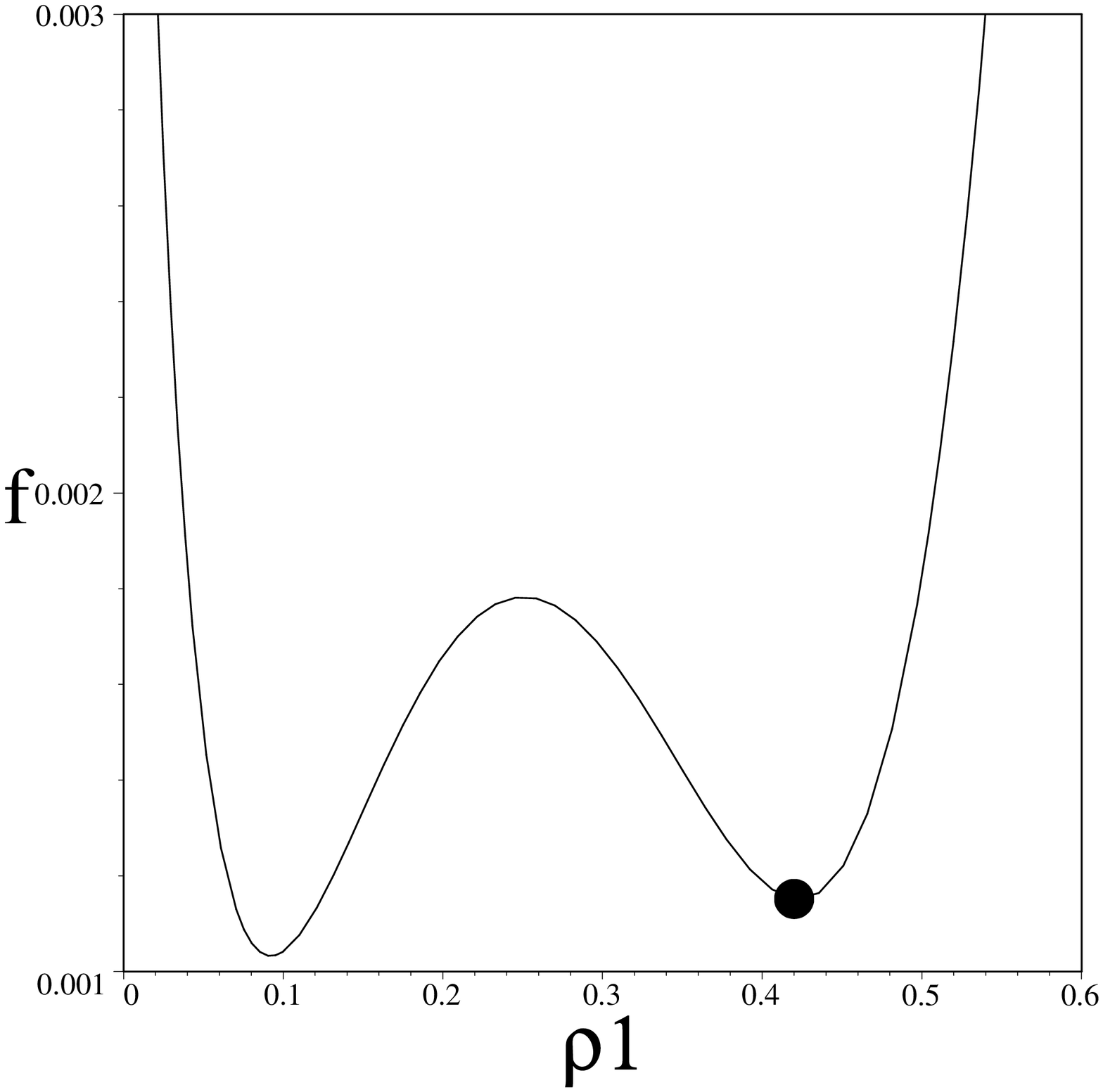,width=\figwidth}%
\begin{picture}(0,0)\put(-67,123){\mbox{(h)}}\end{picture}%
\end{center}

\caption{Illustration of free energy projection. The full
infinite-dimensional free energy surface $f[\rhosig]$ for a truly
polydisperse system cannot be represented graphically, so a system
with only two particle species is used instead. A single phase is then
described by its density pair $(\dens_1,\dens_2)$. (The projection
method remains applicable, although the polydispersity parameter
$\sigma$ is replaced by a discrete species index $a\in\{1,2\}$, and
integrals $\intsig\rhosig \ldots$ by $\sum_a r_a\ldots$) To illustrate
the method we choose a model with an excess free energy depending on a
single \mom\ $\mo= L_1 r_1+L_2 r_2$ (in fact Flory-Huggins theory for
length polydispersity, discussed in more detail in
Sec.~\ref{sec:examples}; see~\(bidisp_FH)). Both the transverse and
the moment subspace are then one-dimensional. (a) 3d-view of free
energy surface. The circle marks the position of the parent
($\dens\pn_1,\dens\pn_2$); the thick line traces out the free energy
of density pairs ($\dens_1,\dens_2$) in the corresponding
family~(\protect\ref{pfamily_precap}). (b) Top view, showing as dashed
lines the transverse directions along which the moment density $\mo$
is constant.  (c) Side view of the free energy surface, ``looking
down'' the transverse direction. (d) The projected free energy is
simply the lower envelope of the free energy surface as seen from this
transverse direction. (e-h) The same free energy surface, now tilted
differently by using a different parent (or prior). While this does
not affect the location of any double tangent planes drawn to the full
surface, it {\em does} produce a different ``lower envelope'' and
hence a different projected free energy. For continuum polydispersity,
the role of the density pair $(r_1,r_2)$ is played by a density
distribution $\rhosig$ of a continuous parameter $\sigma$.
\label{fig:proj_demo}
}
\end{figure}

To understand the effect of the prior $\prior$ {\em physically}, we
note that the projected free energy is simply the free energy of
phases of a system in which the density distributions $\rhosig$ are of
the form~\(family). The prior $\prior$ determines which distributions
lie within this ``maximum entropy family'' (or ``family'' for short), and
it is the properties of phases with these distributions that the
projected free energy represents. Typically, one is interested in a
system where a fixed overall ``parent'' (or ``feed'') distribution
$\rhonsig$ becomes subject to separation into various phases. In such
circumstances, we should generally choose this parent distribution as
our prior, $\prior=\rhonsig$, thereby guaranteeing that it is
contained within the family~\(family). Having done this, we note that
the projection procedure will be {\em exactly valid}, to whatever
extent the density distributions {\em actually arising} in the
various coexisting phases of the system under study {\em are members
of the corresponding family}
\be
\rhosig=\rhonsig\exp\left(\sum_i \mhati\wi\right).
\label{pfamily_precap}
\ee
In fact, the condition just described holds whenever all but one of a
set of coexisting phases are of infinitesimal volume compared to the
majority phase. This is because the density distribution, $\rhonsig$,
of the majority phase is negligibly perturbed, whereas that in each
minority phase differs from this by a Gibbs-Boltzmann factor, of
exactly the form required for~\(pfamily_precap); we show this formally in
Sec.~\ref{sec:properties}. Accordingly, our projection method yields
{\em exact} cloud point and shadow curves.  By the same argument,
critical points (which in fact lie at the intersection of these two
curves) are exactly determined; the same is true for tricritical and
higher order critical points. Finally, spinodals are also found
exactly. We defer explicit proofs of these statements to
Sec.~\ref{sec:properties}.

The projection method does, however, give only approximate results for
coexistences involving finite amounts of different phases. This is
because linear combinations of different density distributions from
the family~\(pfamily_precap), corresponding to two (or more) phases
arising from the same parent $\rhonsig$, do not necessarily add to
recover the parent distribution itself:
\be
\sum_\alpha \phal\rhonsig\exp\left(\sum_i \mhati\pa\wi\right) \neq
\rhonsig
\label{dont_sum_to_parent}
\ee
unless all except one of the $\phal$ are infinitesimal.  Moreover,
according to Gibbs' phase rule, a projected free energy depending on
$\L$ \moms\ will not normally predict more than $\L+1$ coexisting
phases, whereas a polydisperse system can in principle separate into
an arbitrary number of phases. We explain in Sec.~\ref{sec:properties}
how both of these shortcomings can be overcome by systematically
including extra \moms\ within the projection procedure. How quickly
convergence to the exact results occurs depends on the choice of
weight functions for the extra \moms; we discuss this point further in
the context of the examples of Sec.~\ref{sec:examples}.

\subsection{\Patrick\ method}
\label{sec:combinatorial}

Now we turn to the second method of constructing a moment free
energy. As before, we recognize from the outset that the physics of
the problem is contained in the excess free energy and therefore the
most important variables are the \moms\ which feature in this. On the
other hand, the ideal part of the free energy (essentially the entropy
of mixing, up to the trivial ideal gas term) is a book-keeping device
that accounts for the number of ways of partitioning {\em a priori} a
given size distribution between two or more coexisting phases. (To
focus the discussion, we identify $\sig$ as particle size in this
section; the arguments do of course remain valid for a general
polydispersity parameter $\sig$.) The entropy of mixing has its origin
in $1/N!$ factors in the partition function (where $N$ is a particle
number) which are usually derived from the classical limit of quantum
statistics. The origin of the $1/N!$ factors in classical statistical
mechanics has been the subject of debate ever since
Gibbs~\cite{pbw_gibbs}. The problem is also connected with the
question of the extensivity of entropy, and the Gibbs paradox. Below
we present a completely classical derivation of the entropy of mixing
which has the additional benefit of indicating the appropriate
generalisation for moment densities. To start with, follow Gibbs and
define a non-extensive free energy $F'$ as a configuration space
($\Gamma$) integral
\begin{equation}
e^{-F'/\kT}=\int d\Gamma\,e^{-H/\kT}.
\label{pbw_prescreq}
\end{equation}
No $1/N!$ appears in this as all particles are distinguishable in
classical statistical mechanics.

Now consider two phases in coexistence as one joint system. Assume
they occupy volumes $V\po$ and $V\pt$, respectively, and contain
$N\po$ and $N\pt$ particles. Following literally the prescription in
eq.~(\ref{pbw_prescreq}), the free energy of the joint system of
$N\po+N\pt=N$ particles is found from
\begin{equation}
e^{-F'/\kT}=\sum\prtns e^{-({F'}\po+{F'}\pt)/\kT}
\label{pbw_fpeq}
\end{equation}
where the configuration space integral has been done in two parts.
Firstly, for each way of partitioning the particles between the two
phases, the individual configuration space integrals give a product of
the individual partition functions.  Secondly, and crucially, one must
sum over the $N!/N\po!N\pt!$ ways of partitioning the particles
between phases.  Now define an extensive free energy by reinserting
the $1/N!$ {\em as though} the particles were indistinguishable:
\begin{equation}
e^{-\fd/\kT}=\frac{1}{N!}\int d\Gamma\,e^{-H/\kT},
\label{pbw_fconv}
\end{equation}
Eq.~\eqref{pbw_fpeq} can then be written as 
\begin{equation}
e^{-\fd/\kT}=\langle e^{-(\fd\po+\fd\pt)/\kT}\rangle\prtns
\label{pbw_prtns}
\end{equation}
where the average is taken over all partitions, with equal {\em a
priori} probabilities.  This result is the cornerstone of the
\patrick\ method. If we separate $\fd$ into the entropy of an ideal
gas and the remaining non-ideal (excess) parts, it is written as
\be
\fd = NT[\ln(N/V)-1]+\tilde{F}
\label{fd_def}
\ee
with $\tilde{F}$ being the excess free energy. Note that $\fd$ does
not yet contain an entropy of mixing term because it treats the
particles as though they are indistinguishable. 
The entropy of mixing will reappear when we
come to do the average over partitions in~\(pbw_prtns).
Note also that conventional thermodynamics follows from~\(pbw_prtns) if all
particles are identical as far as their mutual interactions are
concerned. There is then no entropy of mixing to consider, so that
$\fd$ is just the conventional free energy $F$; and since the
average over partitions is trivial one has $F=F\po+F\pt$. Similarly,
as we show in Sec.~\ref{sec:equivalence}, one can recover
from~\(pbw_prtns) the conventional form of the entropy of mixing as
given in~\(sideal_decomp).

We now show how the average over partitions in~\eqref{pbw_prtns}
results in an expression for the entropy of mixing which depends
explicitly on moment variables. The key is to note that, by our
assumption that we are dealing with a truncatable model, the excess
free energy $\tilde{F}=V\fexc(\mi)$ depends only on a limited number
of \moms\ $\mi$ (as well as volume $V$, temperature $T$, and other
state variables which we suppress below). If the density $\mn$ is
among the moment densities $\mi$ -- which we assume throughout this
section -- then specifying $V$ and the $\mi$ is equivalent to
specifying $N$, $V$ and the {\em normalized} moments $\nmi = \mi/\mn$
($i>0$). So we can think of $\tilde{F}$ as a function of $N$, $V$ and
the $\nmi$. From~\(fd_def), $\fd$ then depends on the same set of
variables. The choice of the $\nmi$, which are moments of the {\em
normalized} particle size distribution $\probsig$, as independent
variables is natural in the present context because we are considering
the particle number $N$ and volume $V$ of the coexisting phases to be
fixed.

To avoid notational complexity, and without loss of generality, we now
specialize to the case where there is only one normalized moment
$\nmi$, a generalized mean size 
\[
m=\intsig w(\sig)\nsig
\]
We shall also write $\rho_m=\rho m=\intsig w(\sig)\rhosig$ for the
corresponding moment density, and $m\po$ and $m\pt$ for the values of
the generalized mean size in the first and second phase. The overall
(``parent'') size distribution is $\nnsig$ with mean $m\pn$. In each
of the two phases, in which $N$ and $V$ are held fixed, $\fd$ only
depends on $m$, and so the average over partitions becomes
\begin{equation}
\langle e^{-(\fd\po+\fd\pt)/\kT}\rangle\prtns
\to \int dm\po\,\myP(m\po)\,e^{-(\fd\po+\fd\pt)/\kT}
\label{m_one_int}
\end{equation}
Here $\myP(m\po)$ is the probability distribution for the generalized
mean size in the first phase, taken over partitions with fixed $N\po$
and $N\pt$, with equal {\em a priori} probabilities.  Note that given
$m\po$, $m\pt$ is fixed in the second phase by the moment equivalent
of particle conservation: $N\po m\po+N\pt m\pt=Nm\pn$. The integral
in~\eqref{m_one_int} can be replaced by the maximum of the integrand
in the thermodynamic limit, because $\loge\myP(m\po)$ is an extensive
quantity. Introducing a Lagrange multiplier $\mu_m$ for the above
moment constraint then shows that the quantity $\rho_m$ has the same
status as the density $\rho\equiv\mn$ itself: both are thermodynamic
density variables. This reinforces the discussion in the
introduction, where we showed that \moms\ can be regarded as densities
of ``quasi-species'' of particles.

In what follows, we only need to refer to $m\po$ (not $m\pt$) and
therefore drop the superscript for brevity. Since $\loge\myP(m)$ is
extensive, we can write $\loge\myP(m)=N\sN(m)$ where $\sN$ is the
entropy of mixing {\em per particle} expressed as a function of the
moment $m$ (or, more generally, of the full set of normalized
moments).  The total free energy of the system then takes the form
\be
\fd = \min_m \fd\po + \fd\pt - NT\sN(m),
\label{total_free_en}
\ee
and we need to calculate the entropy of mixing $\sN(m)$.
We recognize that this quantity depends not only on the generalized
mean size, but also on $x=N\po/N$, the fraction of particles in the
first phase. A formal result for $\sN(m)$ can be obtained
by first calculating the joint probability distribution $\myP(xm,x)$
of $xm$ and $x$. From this, we can find $\myP(m)$ according
to~\cite{conditioning_footnote}
\be
\myP(m)= \frac{\myP(xm,x)}{\int\!dm'\,\myP(xm',x)} =
\frac{\myP(xm,x)}{\myP(x)/x}
\label{Pm_norm}
\ee
Writing the log probabilities in~\(Pm_norm) as
$\loge\myP(xm,x)=N\sN(xm,x)$ and $\loge\myP(x)=N\sN(x)$, we then have
in the thermodynamic limit
\be
\sN(m)=\sN(xm,x)-\sN(x)
\label{sN_norm}
\ee
The method for calculating $\myP(xm,x)$ proceeds by introducing an
indicator function $\epsilon_i$ for each particle, deemed to be $1$ if
the particle is in the first phase, and $0$ if the particle is in the
second~\cite{pbw_schrodinger}. Then $Nx=\sum_{i=1}^N\epsilon_i$ and
$Nxm=\sum_{i=1}^N\epsilon_i w(\sigma_i)$. We now write the {\em moment
generating function} for $\myP(xm,m)$ as follows
\begin{eqnarray}
\int\!d(xm)\,dx\ \myP(xm,x)\,\exp[{N(\theta x+\lambda xm)}]
&&=\Bigl\langle\exp\Bigl[{\textstyle\theta\sum_{i=1}^N\epsilon_i
+\lambda\sum_{i=1}^N\epsilon_iw(\sigma_i)}\Bigr]
\Bigr\rangle\prtns\nonumber\\
&&=\Bigl\langle\prod_{i=1}^N
\exp\{\epsilon_i[\theta+\lambda  w(\sigma_i)]\}
\Bigr\rangle\prtns
=2^{-N}\prod_{i=1}^N\{1+\exp[\theta+\lambda w(\sigma_i)]\}\nonumber\\
&&=2^{-N}\,\exp\left(\sum_{i=1}^N\loge\{1+\exp[\theta+\lambda
w(\sigma_i)]\}\right).
\label{moment_gen_fn}
\end{eqnarray}
In the second equality we have used the fact that the $\epsilon_i$ are
independent discrete random variables taking the values $0$ and $1$
with {\em a priori} equal probabilities. Taking logarithms
of~\eqref{moment_gen_fn} and dividing by $N$, we obtain on the right
hand side an average over $i=1\ldots N$ of a function of
$\sigma_i$. Since the $\sigma_i$ are drawn independently from
$\parent$, the law of large numbers guarantees that this average tends
to the corresponding average over $\parent$ in the limit $N\to\infty$
(compare the discussion in Ref.~\cite{Salacuse84}). We therefore have
the final result
\begin{equation}
\frac{1}{N}\ln\int\!d(xm)\,dx\,\exp\{N[\sN(xm,x)+\theta x+\lambda xm]\}
=\int d\sigma\, \parent\,\loge\{1+\exp[\theta+\lambda w(\sigma)]\}
\end{equation}
where we have dropped an unimportant additive constant ($-\ln 2$) and
have replaced $\myP(xm,x)$ by $\exp[N\sN(xm,x)]$.

Now, the left hand side can be evaluated, by the steepest descent
method, in the limit $N\to\infty$. Defining
\[
\gener(\theta,\lambda) = \intsig\parent\,\loge\{1+\exp[\theta+\lambda
w(\sigma)]\}
\]
we get $\sN(xm ,x)+\theta x+\lambda xm  = \gener(\theta,\lambda)$ at
the point where $\partial\sN/\partial
x+\theta=\partial\sN/\partial(xm )+\lambda =0$. We recognize that the
relationship between $\sN(x,xm )$ and $\gener(\theta,\lambda)$ is a
double Legendre transform. Inverting this shows that
\be
\sN(xm,x) = \gener(\theta,\lambda) - \theta x - \lambda xm
\label{sNxmx_exact}
\ee
where
\be
x = {\partial \gener}/{\partial\theta} \qquad
xm = {\partial \gener}/{\partial\lambda}
\label{saddle_point}
\ee
Eq.~\(sNxmx_exact) is essentially the desired result; to obtain
$\sN(m)$ from~\(sN_norm), only $\sN(x)$ remains to be determined. It
is obtained from $\myP(x)$ via
$\exp[N\sN(x)]=\myP(x)=\int\!d(xm)\,\myP(xm,x)$ which
by inspection of the moment generating function is seen to correspond
to the point $\lambda=0$. At this point, $\gener =
\loge(1+e^{\theta})$ and $x = \partial \gener/\partial\theta =
e^{\theta}/(1+e^{\theta})$. We find that $\theta = \loge[x/(1-x)]$ and
after some manipulation,
\be
\sN(x) = \gener-x\theta = -x\loge x-(1-x)\loge(1-x)
\label{sNx_exact}
\ee
This is recognized as the standard entropy of mixing that is lost when
a total of $N$ particles is partitioned into $N\po=xN$ and
$N\pt=(1-x)N$ particles in two phases. The simple form of the result
shows that the calculation leading to~\(sNxmx_exact) is essentially a
generalization of the Stirling approximation.

Inserting~\(sNxmx_exact) and~\(sNx_exact) into~\(sN_norm), we finally
have the desired result for $\sN(m)$,
\be
\sN(m) = \gener(\theta,\lambda) - \theta\,x - \lambda\,xm + 
x\loge x + (1-x)\loge(1-x)
\label{sN_exact}
\ee 
Unfortunately this is mainly formal because neither the integral
defining $\gener$ nor the Legendre transform are likely to be
tractable. However, we show in Sec.~\ref{sec:equivalence}
that~\eqref{sN_exact} is equivalent to the more conventional form of
the entropy of mixing as given by the second term in in
eq.~\(sideal_decomp).  From a conceptual point of view, it should be
noted that the conventional form is normally derived by ``binning''
the distribution of particle sizes $\sig$ and taking the number of
bins to infinity {\em after} the thermodynamic limit $N\to\infty$ has
been taken (see \eg~\cite{SalSte82}). In our above derivation, on the
other hand, we have assumed that even for finite $N$ all particles
have different sizes $\sig_i$, drawn randomly from $\probnsig$; the
``polydisperse limit'' is thus taken simultaneously with the
thermodynamic limit. The relation between these two approaches --
which lead to the same results -- has been discussed in detail by
Salacuse~\cite{Salacuse84}. Note that the first limit is physically
more plausible for many homopolymer systems (where there may only be
thousands or millions of species, with many particles of each) whereas
the second limit is more natural for colloidal materials (and also
some random copolymers) in which no two particles present are exactly
alike, even in a sample of macroscopic size.


Although the full result~\(sN_exact) is intractable, progress can be
made for $x\to 0$, when the number of particles in one phase is much
smaller than in the other. The limit $x\to 0$ implies
$\theta\to-\infty$ and hence $\gener(\theta,\lambda) \to
\exp[\theta+h(\lambda)]$ where $h = \loge\intsig\parent e^{\lambda
w(\sigma)}$ is a generalized {\em cumulant} generating function for
$\parent$. From this we derive $x = \partial\gener/\partial\theta =
\psi$, giving $\theta=\ln x - h$. The generalized mean size is given
by $xm = \partial\gener/\partial\lambda = \psi\,\partial
h/\partial\lambda$ and hence $m=\partial h/\partial\lambda$. Inserting
these results into~\eqref{sNxmx_exact} then shows that
\[
\sN(xm,x) = x-x\loge x + x(h - \lambda m),
\qquad m = \frac{\partial h}{\partial\lambda}
\]
and hence from~\(sN_exact)
\be
\sN(m)= x(h - \lambda m) + \order(x^2).
\label{sm_one}
\ee
Noting that $xN=N\po$ is the number of particles in the small phase,
the total free energy~\(total_free_en) can then be written as
\be
\fd = \fd\pt + \fd\po - N\po T (h - \lambda m)
\label{total_free_en_small}
\ee

The term $(h-\lambda m)$, being multiplied by $-N\po T$, can now be
interpreted as the entropy of mixing \emph{per particle in
the small phase}. It arises from the deviation of the (generalized)
mean size $m\equiv m\po$ in the small phase from the mean size $m\pn$
of the parent, and is given by the Legendre transform of the
generalized cumulant generating function:
\begin{equation}
\smix(m)=h-\lambda m,\quad\text{where}\quad 
m=\frac{\partial h}{\partial\lambda},\quad
h = \loge\textint d\sigma\,\parent\,e^{\lambda w(\sigma)}.
\label{pbw_legendre}
\end{equation}

Let us now examine the relation between $\smix(m)$, thereby defined,
and $\sN(m)$ introduced previously. By construction, $Ns(m)$ is the
entropy of mixing of the system as a whole; both phases contribute to
this. The result~(\ref{sm_one}), which can also be written as
\be
Ns(m)=N\po\smix(m)
\label{sm_two}
\ee
appears to contradict this, seeing as it does not contain a term
proportional to $N\pt$. The resolution of this paradox comes
from the neglected $\order(x^2)$ terms in~\(sm_one). In fact, using
the single-phase entropy of mixing defined in eq.~\(pbw_legendre) --
and reinstating on the r.h.s.\ the superscript on $m\equiv m\po$ -- we
can write~\(sm_two) in the more symmetric form
\[
Ns(m)=N\po\smix(m\po)+N\pt\smix(m\pt)
\]
This is still correct to leading (linear) order in $x$ because the
added term is $\order(x^2)$. [To see this, use the fact that
$m\pt=m\pn-xm\po/(1-x)$ and hence
$m\pt-m\pn=\order(x)$. From~\(pbw_legendre) it can then be deduced
that $\smix(m\pt) \sim (m\pt-m\pn)^2 = \order(x^2)$.]  Similarly, we
can rewrite the total free energy~\(total_free_en_small) as
\be \fd = [\fd\po - N\po T\smix(m\pt)] + [\fd\pt - N\pt T \smix(m\pt)]
\label{total_free_en_new}
\ee
This shows that in the limit where one of the two phases is much
larger than the other one, the free energy of each of the phases --
now {\em including} the entropy of mixing -- is given by
\be
F = \fd - NT\smix(m) = -NT\left[-\ln({N}/{V})+1+\smix(m)\right] + \tilde{F}
\label{pbw_mom_free_en}
\ee
This expression, which only depends on the particle size distribution
through the generalized mean size $m$, is our desired moment free
energy.

The Legendre transform result~\(pbw_legendre) for the (single-phase)
entropy of mixing is appealingly simple; we will illustrate its
application in Sec.~\ref{sec:length_homopolymer}, using polydisperse
Flory-Huggins theory as an example. As discussed briefly in
App.~\ref{app:LDT}, eq.~\(pbw_legendre) also establishes an
interesting connection to large deviation theory. However, the most
important aspect of the
result~(\ref{pbw_legendre},\ref{pbw_mom_free_en}) is that -- as we
have shown -- it gives exact results for the limiting case $N\po\ll
N\pt$ (that is, $x\to0$). This includes two important classes of
problems, that are also handled exactly by the projection method (for
essentially the same reason). The first is the determination of
spinodal curves and critical points.  Intuitively these are exact
because they are related to the stability of the system with respect
to small variations in its density or composition, which can be probed
by allowing fluctuations to take place in a vanishingly small
subregion. The second concerns the cloud point and shadow curves.
Again intuitively, these are exact because by definition only an
infinitesimal amount of a second phase has appeared.  Formal proofs of
these statements are given in Sec.~\ref{sec:properties}.

If the
result~(\ref{pbw_legendre},\ref{total_free_en_new},\ref{pbw_mom_free_en})
is applied in the regime where it is no longer strictly valid, \ie,
where the two coexisting phases contain comparable number of particles
(or, equivalently, occupy comparable volumes), then approximate
two-phase coexistences can be calculated. It is also straightforward
to show that the analog of
eqs.~(\ref{pbw_legendre},\ref{total_free_en_new},\ref{pbw_mom_free_en})
holds in the case where two or more small phases coexist with a much
larger phase; application in the regime of comparable phase volumes
then provides approximate results for multi-phase coexistence.

To end this section, let us state in full the analogue
of~(\ref{pbw_legendre},\ref{pbw_mom_free_en}) for the case of several
moment densities, restoring the notation used in the previous
sections. The square bracket on the r.h.s.\ of~\(pbw_mom_free_en) is
the moment expression for the entropy of an ideal mixture.  If, as in
Sec.~\ref{sec:projection}, we measure this entropy {\em per unit
volume} (rather than per particle, as previously in the current
Section) and generalize to several \moms, we find by the \patrick\
approach the following moment free energy:
\be
\fcomb = -T\scomb + \fexc, \qquad 
\scomb = -\mn(\ln\mn-1) + \mn 
\left({h-\sum_{i\neq 0}\mhati\nmi}\right),
\label{comb_entropy}
\ee
with
\be
h = \ln {\intsig} \probnsig \exp
\left({\sum_{i\neq 0}\mhati\wi}\right)
\label{h_def}
\ee
Here $\mn=N/V$ is the particle density as before.

\subsection{Relation between the two methods}
\label{sec:equivalence}

Although eq.~\(comb_entropy) looks somewhat different from the
corresponding result~\(annd) for $\fproj$ obtained by the projection
method, the two methods are, mathematically, almost equivalent, as we
now show. (For brevity we return to the case of a single moment $m$
and continue to refer to the polydisperse feature $\sigma$ as `size'.)
First, consider the exact expression for the entropy of
mixing~\eqref{sN_exact} derived within the \patrick\ method. This
should be equivalent to the conventional result used as the starting
point~\(sideal_decomp) in the projection method. To see this, write
the Legendre transform conditions~\(saddle_point) out explicitly:
\be
x=\intsig\parent \frac{\exp[\theta+\lambda w(\sigma)]}
{1+\exp[\theta+\lambda w(\sigma)]}, \qquad
x m\po=\intsig w(\sigma) \parent \frac{\exp[\theta+\lambda w(\sigma)]}
{1+\exp[\theta+\lambda w(\sigma)]}
\label{saddle_explicit}
\ee
Here we have reinstated the superscript to show that $m\po$ is the
value in phase one of the (generalized) moment $m$:
\[
m\po=\intsig  w(\sig) \nosig
\]
Comparing with~\(saddle_explicit), we can identify the
(normalized) particle size distribution in that phase
as~\cite{exact_form_footnote}
\[
\nosig = \frac{\parent}{x}\frac{\exp[\theta+\lambda w(\sigma)]}
{1+\exp[\theta+\lambda w(\sigma)]}
\]
Particle conservation $x\nosig+(1-x)\ntsig=\nnsig$ implies a
similar form for the size distribution in phase two:
\[
\ntsig = \frac{\parent}{1-x}\frac{1}
{1+\exp[\theta+\lambda w(\sigma)]}.
\]
It then takes only a few lines of algebra to show that the entropy of
mixing~\(sN_exact) can be written (for any $x$) as
\begin{eqnarray}
\sN(m) &=& - x\intsig\nosig\ln\frac{\nosig}{\nnsig}
- (1-x)\intsig\ntsig\ln\frac{\ntsig}{\nnsig}\\
&=& - x\intsig\nosig\ln\nosig
- (1-x)\intsig\ntsig\ln\ntsig
+ \intsig\nnsig\ln\nnsig
\label{sN_rhologrho}
\end{eqnarray}
The second equation, with the third (constant) term discarded, is the
standard result (compare
eqs.~(\ref{sideal_decomp},\ref{free_en_decomp},\ref{g_decomp})). However
the first expression appears more natural, and is retained, within the
combinatorial derivation; it embodies the intuitively reasonable
prescription that entropy is best measured relative to the parent
distribution $\nnsig$. This was also an essential ingredient of the
projection method, as described in Sec.~\ref{sec:projection}, where
the ``prior'' in the entropy expression was likewise identified with
the parent. Retaining the parent as prior also avoids subtleties with
the definition of the integrals in~\(sN_rhologrho) in the case where
the phases contain monodisperse components, corresponding to
$\delta$-peaks in the density distributions~\cite{SalSte82}.

Having established that the rigorous starting points of the \patrick\
and projection method are closely related, we now show that the
subsequent approximations also lead to essentially the same results.
The relevant approximations are use of~\(comb_entropy) for $\fcomb$
(which adopts the small $x$ form for $\smix$) in the \patrick\ case
and use of~\(annd) for $\fproj$ (which minimizes over transverse
degrees of freedom) in the projection case.  First note that the
density $\mn$ of phases within the family~\(pfamily_precap) is given
by
\[
\mn = \intsig\rhonsig\exp\left(\sum_i \mhat_i\w_i(\sig)\right)
\]
Comparing this with~\(h_def), and using the fact that
$\probnsig=\rhonsig/\mn\pn$, one sees that
\[
\mn = \intsig \rhonsig
\exp\left(\sum_{i}\mhati\wi\right) = \mn\pn e^{\mhatn+h}
\]
where the sum over $i$ now includes $i=0$.  Solving for $h$, one
has $h = -\mhatn + \ln(\mn/\mn\pn)$. The ideal mixture entropy derived
by the combinatorial route, eq.~\(comb_entropy), can thus be rewritten
as
\[
\scomb = 
-\mn(\ln\mn-1) + \mn\ln(\mn/\mn\pn) - \mhatn\mn 
-\sum_{i\neq 0}\mhati\mn\nmi = 
\mn - \sum_{i}\mhati\mi - \mn\ln\mn\pn.
\]
This is identical to the projected entropy $\sproj$, eq.~\(annd),
except for the last term. But by construction, the \patrick\ entropy
assumes that $\mn$ -- the overall density -- is among the \moms\
retained in the moment free energy. The difference $\scomb-\sproj =
-\mn\ln\mn\pn$ is then linear in this density, and the \patrick\ and
projection methods therefore predict exactly the same phase behavior.

In summary, we have shown that the projection and \patrick\ methods
for obtaining moment free energies give equivalent results. The only
difference between the two approaches is that within the projection
approach, one need not necessarily retain the zeroth moment, which is
the overall density $\mn = \rho$, as one of the moment densities on
which the moment free energy depends. If $\mn$ does not appear in the
excess free energy, this reduces the minimum number of independent
variables of the moment free energy by one (see
Sec.~\ref{sec:examples} for an example).

\section{Properties of the moment free energy}
\label{sec:properties}

In the previous sections, we have derived by two different routes
(namely~\(annd) and~\(comb_entropy)) our moment free energy for
truncatable polydisperse systems. We now investigate the properties of
this moment free energy, which we henceforward denote $\fmom$ and
write in the form~\(annd)
\be
\fmom(\mi) = - T \smom(\mi) + \fexc(\mi), \qquad
\smom= \shat - \sum_i \mhati\mi\ ;
\label{mom_free_en}
\ee
here $\smom$ is the ``moment entropy'' of an ideal mixture.  In
particular, we formally compare the phase behavior predicted from this
free energy (by treating the $\mi$ as densities of ``quasi-species'' of
particles, and applying the usual tangency construction) with that
obtained from the exact free energy of the underlying (truncatable)
model
\be
f[\rhosig] = T \intsig \rhosig \left[\ln \rhosig -1\right] + \fexc(\mi).
\label{exact_free_en}
\ee

Let us first collect a few simple properties of the moment free
energy~\(mom_free_en) which will be useful later. Recall
that~\(mom_free_en) faithfully represents the free energy density
of any phase with density distribution in the family
\be
\rhosig=\rhonsig\exp\left(\sum_i \mhati\wi\right)
\label{pfamily}
\ee
where $\rhonsig$ is the density distribution of the parent. The
\moms\ $\mi$ are then related to the Lagrange multipliers
$\mhati$ by
\be
\mi = \intsig\wi\,\rhonsig\exp\left(\sum_j \mhat_j\w_j(\sig)\right)
\label{moms_lambdas}
\ee
If we regard $\shat$ as a function of the
$\mhati$, then from~\(moms_lambdas) we have
$\partial\shat/\partial\mhati = \mi$. Together with
eq.~\(mom_free_en), this implies
that the moment entropy $\smom$ has the structure of a Legendre
transform~\cite{Legendre_diffs}.
For the first derivatives of $\smom$ with respect to the
\moms, this yields
\[
\frac{\partial\smom}{\partial\mi} = -\mhati
\]
while the matrix of second derivatives is the negative inverse of the
matrix of ``second-order \moms''
$\m_{ij}$~\cite{higher_moments_not_explicit_footnote}:
\be
\frac{\partial^2\smom}{\partial\mi\partial\m_j} =
-(\mmmat^{-1})_{ij},
\quad
(\mmmat)_{ij} = \frac{\partial^2\shat}{\partial\mhati\partial\mhat_j} =
\intsig\wi\,\w_j(\sig)\,\rhonsig\exp\left(\sum_k
\mhat_k\w_k(\sig)\right) \equiv \m_{ij}.
\label{dsmom_dmi_dmj}
\ee
The chemical potentials $\mu_i$ conjugate to the \moms\ follow
as
\be
\mu_i= \frac{\partial\fmom}{\partial\mi} =
T\mhati + \frac{\partial\fexc}{\partial\mi} =  T\mhati + \muexci
\label{mom_chem_pot}
\ee
and their derivatives w.r.t.\ the $\mi$, which give the curvature of
the moment free energy, are
\be
\frac{\partial\mu_i}{\partial\m_j} =
\frac{\partial^2\fmom}{\partial\mi\partial\m_j} =
T (\mmmat^{-1})_{ij} + 
\frac{\partial^2\fexc}{\partial\mi\partial\m_j}.
\label{mom_free_en_curvature}
\ee
The pressure, finally, is given by
\be
-\pimom = \fmom - \sum_i \mu_i \mi = - T\mn + \fexc - \sum_i \muexci \mi.
\label{mom_osmotic_pressure}
\ee
Eqs.~(\ref{mom_chem_pot}-\ref{mom_osmotic_pressure}) are all
calculated via the moment free energy. The three corresponding
quantities obtained from the exact free energy~\(exact_free_en) are,
first, the chemical potentials conjugate to $\rhosig$:
\be
\musig=\frac{\delta f}{\delta \rhosig} = T \ln\rhosig +
\sum_i \muexci \wi 
\label{exact_chem_pot}
\ee
second, their derivatives w.r.t.\ $\rhosig$:
\be
\frac{\delta\musig}{\delta \rho(\sig')} = \frac{\delta^2 f}{\delta
\rhosig \delta\rho(\sig')} = \frac{T\delta(\sig-\sig')}{\rhosig} +
\sum_{i,j} \frac{\partial^2\fexc}{\partial\mi\partial\m_j}
\,\wi\,\w_j(\sig') 
\label{exact_free_en_curvature}
\ee
and, third, the resulting expression for the pressure:
\be
-\Pi = f - \intsig \musig \rhosig =  - T\mn + \fexc - \sum_i\muexci\mi.
\label{exact_osmotic_pressure}
\ee
The last of these is identical to the result~\(mom_osmotic_pressure)
derived from the moment free energy.

We note one important consequence of the form of the exact chemical
potentials~\(exact_chem_pot) for truncatable models: If two phases
$\rh\po(\sig)$ and $\rh\pt(\sig)$ have the same chemical potentials
$\musig$, then the ratio of their density distributions can be written
as a Gibbs-Boltzmann factor:
\be
\frac{\rh\po(\sig)}{\rh\pt(\sig)} = \exp\left[\sum_i 
\beta\left(\muexci\pt-\muexci\po\right)\wi\right]
\label{Gibbs_Boltzmann}
\ee
This implies that if one of the density distributions is in the
family~\(pfamily), then so is the other. The same argument obviously
applies if there are several phases with equal chemical potentials.
Conversely, we have
for the chemical potential difference between any two phases in the
family~\(pfamily)
\be
\Delta\musig = \sum_i (T\Delta\mhati+\Delta\muexci)\wi = 
\sum_i \Delta\mu_i\wi.
\label{del_musig}
\ee
The chemical potentials $\musig$ of two such phases are therefore
equal if and only if their {\em moment} chemical potentials $\mu_i$
are equal. Combining this with the fact that the
pressure~\(mom_osmotic_pressure) derived from the moment free energy
is exact (compare~\(exact_osmotic_pressure)), we conclude: Any set of
(two or more) coexisting phases calculated from the {\em moment} free
energy obeys the {\em exact} phase equilibrium conditions. That is, if
they were brought into contact, they would genuinely
coexist. (However, they will not necessarily obey the lever rule with
respect to the parent $\rhonsig$ of the given family, unless all but
one of them -- which then {\em coincides} with the parent -- are of
vanishing volume; see \(dont_sum_to_parent).)

\subsection{General criteria for spinodals and (multi-) critical points}
\label{sec:properties_general}

We now demonstrate that, for any truncatable model, the moment free
energy gives exact spinodals and (multi-) critical points. By this we
mean that, using the moment free energy, the values of external
control parameters -- such as temperature -- at which the parent phase
$\rhonsig$ becomes unstable or critical can be exactly determined.
Our argument treats spinodals and (multi-) critical points in a
unified fashion, using the fact that all of them occur when the
difference between phases with equal chemical potentials becomes
infinitesimal. Because the truncatable models that we are considering
are generally derived from mean-field approaches, we do not concern
ourselves with critical point singularities, assuming instead that the
free energy is a smooth function of all order parameters (the
densities in the system). Likewise, we do not discuss the subtle
question of how, beyond mean-field theory, free energies can actually
be defined in spinodal and unstable regions~\cite{FisZin98}.

Let us first recap the general criteria for spinodals and critical
points in multi-component systems. In order to treat the critera
derived from the exact and moment free energies simultaneously, we use
the common notation $\rhovect$ for the vector of densities specifying
the system: for the exact free energy, the components of $\rhovect$
are the values $\rhosig$; for the moment free energy, they are the reduced
set $\mi$. We write the corresponding vector of chemical potentials as
\[
\muvect(\rhovect)=\nabla f(\rhovect).
\]
This notation emphasizes that the chemical potentials are functions of
the densities. The criterion for a spinodal at the parent phase
$\rhovect\pn$ is then that there is an incipient {\em instability}
direction $\delrhovect$ along which the chemical potentials do not
change:
\be
(\delrhovect\cdot\nabla)\muvect(\rhovect\pn)=
(\delrhovect\cdot\nabla)\nabla f(\rhovect\pn)=\zerovect.
\label{spinodal_general}
\ee
As the second form of the criterion shows, an equivalent statement is
that the curvature of the free energy along the direction
$\delrhovect$ vanishes. Eq.~\(spinodal_general) can also be written as
\be
\muvect(\rhovect\pn+\epsilon\delrhovect) - \muvect(\rhovect\pn)= 
\order(\epsilon^2)
\label{spinodal_temp}
\ee
which is closely related to the critical point criterion that we
describe next.

Near a critical point, the parent $\rhovect\pn$ coexists with another
phase that is only slightly different; if, as we assume here, the free
energy function is smooth, these two phases are separated -- in
$\rhovect$-space -- by a ``hypothetical phase'' which has the same
chemical potentials but is (locally) thermodynamically unstable.
(This is geometrically obvious even in high dimensions; between any
two minima of $f(\rhovect)-\muvect\cdot\rhovect$, at given $\muvect$,
there must lie a maximum or a saddle point, which is the required
unstable ``phase".)  Now imagine connecting these three phases by a
smooth curve in density space $\rhovect(\epsilon)$. At the critical
point, all three phases collapse, and the variation of the chemical
potential around $\rhovect(\epsilon=0)=\rhovect\pn$ must therefore
obey
\[
\muvect(\rhovect(\epsilon)) - \muvect(\rhovect\pn)= 
\order(\epsilon^3)
\]
Similarly, if $n$ phases coexist, we can connect them and the $n-1$
unstable phases in between them by a curve
$\rhovect(\epsilon)$~\cite{single_instability_direction_footnote}. If
we define an $n$-critical point as one where all these phases become
simultaneously critical, we obtain the criterion
\be
\muvect(\rhovect(\epsilon)) - \muvect(\rhovect\pn)= 
\order(\epsilon^{2n-1}).
\label{n_critical_temp}
\ee
This formulation was proposed by Brannock~\cite{Brannock91}; the cases
$n=2$ and $n=3$ correspond to ordinary critical and tricritical
points, respectively. The spinodal criterion~\(spinodal_temp) is of
the same form as~\(n_critical_temp) if one chooses the curve
$\rhovect(\epsilon)=\rhovect\pn+\epsilon\delrhovect$.

In summary, we have that the phase $\rhovect\pn$ is a spinodal or
$n$-critical point if there is a curve $\rhovect(\epsilon)$ with
$\rhovect(\epsilon=0)=\rhovect\pn$ such that
\be
\Delta\muvect \equiv 
\muvect(\rhovect(\epsilon)) - \muvect(\rhovect\pn)= \order(\epsilon^l).
\label{spinodal_and_critical}
\ee
where $l=2$ for a spinodal, and $l=2n-1$ for an $n$-critical
point. Brannock~\cite{Brannock91} has shown that these criteria are
equivalent to the determinant criteria introduced by
Gibbs~\cite{pbw_gibbs}. The above forms are more useful for us because
they avoid having to define infinite-dimensional determinants (as
otherwise required to handle the {\em exact} free energy $f(\rhovect)$
of a polydisperse system, whose argument $\rhovect$ denotes an
infinite number of components, even in the truncatable case). They
also show the analogy with the standard criteria for single-species
systems (where $\rhovect$ has only a single component) more clearly.

We can now apply~\(spinodal_and_critical) to the exact free
energy~\(exact_free_en) and show that the resulting criteria are
identical to those obtained from the moment free energy. First, note
that the curve $\rhovect(\epsilon)\equiv\rh(\sig;\epsilon)$ can always
be chosen to lie within the family~\(pfamily). This follows from the
derivation of the criterion~\(spinodal_and_critical): The curve
$\rhovect(\epsilon)$ is defined as passing through $l$ phases with
equal chemical potentials, one of them being the parent
$\rhovect\pn$. As shown in~\(Gibbs_Boltzmann), all these phases are
therefore within the family~\(pfamily).
But then~\(del_musig) implies that
the condition~\(spinodal_and_critical) that $\Delta\musig$ must be
zero to $\order(\epsilon^l)$ is equivalent to the same requirement for
the differences $\Delta\mu_i$ in the {\em moment} chemical
potential. This proves that the conditions for spinodals and (multi-)
critical points derived from the exact and moment free energies are
equivalent. Intuitively, one can understand this as follows: Having
shown that the spinodal/critical point conditions can be formulated
solely in terms of density distributions $\rhosig$ within the
family~\(pfamily), it is sufficient to know the free energy of those
density distributions. This is exactly the moment free energy.

Finally, we note that the exactness of these stability and critical
point conditions holds not just for the parent $\rhonsig$, but for all
phases within the family~(\protect\ref{pfamily}). This is clear,
because any such phase could itself be chosen as the parent without
changing the family; the moment free energy would change only by
irrelevant terms linear in the \moms. In general, one will not
necessarily be interested in the properties of such ``substitute''
parents. As an exception, substitute parents that differ from the
parent only by a change of the overall density {\em are} of interest:
they lie on the dilution line of distributions from which cloud point
and shadow curves are calculated.  The dilution line is included in
the family~\(pfamily) if the overall density $\mn$ (with corresponding
weight function $\w_0(\sig)=1$) is retained in the moment free energy;
in the combinatorial derivation, this is automatically the case.

\subsection{Spinodals}
\label{sec:spinodal}

For completeness, we now give the explicit
form~\cite{IrvGor81,BeeBerKehRat86,Cuesta99} of the spinodal
criterion~\(spinodal_general) for truncatable systems; see
also~\cite{Warren99} for an equivalent derivation using the
combinatorial approach. Using~\(mom_free_en_curvature)
and abbreviating the matrix of second derivatives of the excess free
energy as $\hess$, the spinodal condition~\(spinodal_general) becomes
\be
\left|\hess + T \mmmat^{-1}\right| = 0, \qquad
(\hess + T \mmmat^{-1})\delmvect = \zerovect.
\label{mom_spinodal_aux}
\ee
As before, the (nonzero) vector $\delmvect$ with components $\delm_i$
gives the direction of the spinodal instability; the \moms\ $\mi$ and
$\m_{ij}$ that appear in $\hess$ and $\mmmat$ are to be evaluated for
the parent distribution $\rhonsig$ being studied. Note
that~\(mom_spinodal_aux) is valid for any $\rhonsig$; no specific
assumptions about the parent were made in the derivation. We can thus
simply drop the ``(0)'' superscript: for any phase with density
distribution $\rhosig$, the point where~\(mom_spinodal_aux) first
becomes zero, as external control parameters are varied, locates a
spinodal instability.

More convenient forms of~\(mom_spinodal_aux) that avoid matrix
inversions are obtained after multiplication by the second order
moment matrix $\mmmat$ (which is positive definite for linearly
independent weight functions $\wi$ and therefore has nonzero
determinant):
\be
Y = \left|\mident + \beta \mmmat \hess\right| = 0, \qquad
(\mident + \beta \mmmat\hess)\delmvect = \zerovect.
\label{mom_spinodal}
\ee
Here $\beta=1/T$ in the standard notation. From our general statements
in Sec.~\ref{sec:properties_general}, the spinodal criterion derived
from the exact free energy~\(exact_free_en) must be identical to this;
this is shown explicitly in App.~\ref{app:spinodal}.  Note that the
spinodal condition depends only on the (first-order) \moms\ $\mi$ and
the second-order \moms\ $\m_{ij}$ of the distribution $\rhosig$ (given
by \(moms_lambdas) and \(dsmom_dmi_dmj)); it is independent of any
other of its properties. This simplification, which has been pointed
out by a number of authors~\cite{IrvGor81,BeeBerKehRat86}, is
particularly useful for the case of power-law moments (defined by
weight functions $\wi=\sig^i$): If the excess free energy only depends
on the moments of order 0, 1$\ldots$ $K-1$ of the density
distribution, the spinodal condition involves only $2K-1$ moments [up
to order $2(K-1)$].

The general discussion in Sec.~\ref{sec:properties_general} as well as
the explicit calculation in App.~\ref{app:spinodal} show that the
spinodal instability direction lies within (or more precisely, is
tangential to) the family~\(pfamily) of density distributions. This
fact has a simple geometrical interpretation: By construction,
the family lies along a ``valley'' of the free energy surface (compare
Fig.~\ref{fig:proj_demo}). {\em Away from} the valley floor, any
change in $\rhosig$ increases the free energy, corresponding to a
positive curvature; the spinodal direction, for which the curvature
vanishes, must therefore be {\em along} the valley floor.

\subsection{Critical points}
\label{sec:critical}

Next, we show the explicit form of the critical point criterion for
truncatable systems. The general condition~\(spinodal_and_critical)
for an ordinary critical point ($n=2$) was shown by
Brannock~\cite{Brannock91} to be equivalent to
\be
(\delrhovect\cdot\nabla)\nabla f(\rhovect\pn)=\zerovect, \qquad
(\delrhovect\cdot\nabla)^3 f(\rhovect\pn)=0
\label{critical_simplified}
\ee
The first part of this is simply the spinodal criterion, as
expected. To evaluate the second part for the moment free
energy~\(mom_free_en), we need the third derivative of $\smom$ w.r.t.\
the \moms\ $\mi$. Writing~\(dsmom_dmi_dmj) as
\[
\sum_j \m_{ij} \, \frac{\partial^2\smom}{\partial\m_j\partial\m_k} =
 -\delta_{ik}
\]
and differentiating w.r.t.\ one of the moment densities, one finds
after a little algebra
\[
\frac{\partial^3\smom}{\partial\m_i\partial\m_j\partial\m_k} =
\sum_{lmn} (\mmmat^{-1})_{il} (\mmmat^{-1})_{jm} (\mmmat^{-1})_{kn} \,
\m_{lmn}
\]
with third-order \moms\ $\m_{lmn}$ defined in the obvious way. Thus,
evaluating~\(critical_simplified) for the moment free
energy~\(mom_free_en), we find that critical points have to obey
\be
-T \sum_{ijk} v_i v_j v_k \,\m_{ijk} + \delm_i \delm_j \delm_k \,
\frac{\partial^3\fexc}{\partial\m_i\partial\m_j\partial\m_k}
= 0,
\qquad
v_i=\sum_j (\mmmat^{-1})_{ij}\,\delm_j
\label{mom_critical}
\ee
in addition to the spinodal condition~\(mom_spinodal). As before, the
criterion has to be evaluated for the parent phase $\rhonsig$ under
consideration; but because $\rhonsig$ can be chosen arbitrarily, 
it applies to all density distributions $\rhosig$.

As expected from the general discussion in
Sec.~\ref{sec:properties_general}, the criterion~\(mom_critical) can
also be derived from the exact free energy; an alternative form
involving the spinodal determinant $Y$ is given in
App.~\ref{app:critical}. Eq.~\(mom_critical) shows that the location
of critical points depend only on the \moms\ $\mi$, $\m_{ij}$ and
$\m_{ijk}$~\cite{IrvGor81,BeeBerKehRat87}. For a system with an excess
free energy depending only on power-law moments up to order $K-1$, the
critical point condition thus involves power-law moments of the parent
only up to order $3(K-1)$.

\subsection{Onset of phase coexistence: Cloud point and shadow}

So far in this section, we have shown that the moment free energy
gives exact results for spinodals and (multi-) critical points. Now we
consider the onset of phase coexistence, where (on varying the
temperature, for example) a parent phase with density distribution
$\rhonsig$ first starts to phase separate. As explained in
Sec.~\ref{sec:intro}, this temperature together with the overall
density $\rh\pn\equiv\mn\pn$ of the parent defines a ``cloud point''; the
density of the incipient daughter phase gives the ``shadow''. If the parent
begins to coexist with $p$ phases simultaneously ($p=2$ at a triple
point, for example), there will be $p$ such shadows.

At the onset of phase coexistence, one of the coexisting phases is by
definition the parent $\rhonsig$; the lever rule does not yet play any
role because the daughters $\rhoalsig$ ($\alpha=1\ldots p$) occupy an
infinitesimal fraction of the total volume. It then follows
from~\(Gibbs_Boltzmann) that all daughters lie within the family~\(pfamily) of
density distributions.
As shown in~\(del_musig), the condition for equality of chemical potentials
$\musig$ between any two of the coexisting phases (parent and
daughters) then becomes
\[
\Delta\musig = \sum_i \Delta\mu_i\wi
\]
and is satisfied if and only if the {\em moment} chemical potentials
$\mu_i$ are equal in all phases. Likewise, the exact and moment free
energies give the same condition for equality of pressure $\Pi$ in all
phases, because they yield identical
expressions~(\ref{mom_osmotic_pressure},\ref{exact_osmotic_pressure})
for $\Pi$. In summary, we see that the conditions for the {\em onset}
of phase coexistence are identical for the exact and moment free
energies; the moment free energy therefore gives exact cloud points
and shadows.

\subsection{Phase coexistence beyond onset}
\label{sec:coex_beyond}

As stated in Sec.~\ref{sec:projection}, the moment free energy does
not give exact results beyond the onset of phase coexistence, \ie, in
the regime where the coexisting phases occupy comparable fractions of
the total system volume. As shown in
Sec.~\ref{sec:properties_general}, the calculated phases will still be
in exact thermal equilibrium; but the lever rule will now be violated
for the ``transverse'' degrees of freedom of the density
distributions. This is clear from~\(dont_sum_to_parent): In general,
no linear combination of distributions from this family can match the
parent $\rhonsig$ exactly.

A more detailed understanding of the failure of the moment free energy
beyond phase coexistence can be gained by comparing with the formal
solution of the exact phase coexistence problem.  Assume that the
parent $\rhonsig$ has separated into $p$ phases numbered by
$\alpha=1\ldots p$. The condition~\(Gibbs_Boltzmann), which follows
from equality of the chemical potentials $\musig$ in all phases,
implies that we can write their density distributions $\rhoalsig$ as
\be
\rhoalsig = \effprior \exp\left(\sum_i \mhati\pa \wi\right)
\label{exact_dens_distrib}
\ee
for {\em some} function $\effprior$. If phase $\alpha$ occupies a
fraction $\phal$ of the system volume, particle conservation
$\sum_\alpha \phal\rhoalsig = \rhonsig$ then gives
\be
\effprior = \frac{\rhonsig}{\sum_\alpha \phal\exp\left(\sum_i
\mhati\pa \wi\right)}, \qquad
\rhoalsig = \rhonsig\,\frac{\exp\left(\sum_i
\mhati\pa \wi\right)}{\sum_\beta \ph\pb\exp\left(\sum_i
\mhati\pb \wi\right)}
\label{effective_prior}
\ee
If there are $K$ \moms\ $\mi$, $i=1\ldots K$, then this exact solution
is parameterized by $(p-1)(K+1)$ independent parameters: $(p-1)K$
parameters $\mhati\pa$ (noting that the $\mhati$ of one phase can be
fixed arbitrarily), and $p-1$ parameters $\phal$ (noting that one
phase volume is fixed by the constraint $\sum_\alpha
\phal=1$). Comparing~\(exact_dens_distrib) with~\(Gibbs_Boltzmann),
one sees that the $K$ quantities
\[
T\mhati\pa+\muexci\pa \quad (i=1\ldots K)
\]
must be the same in all phases $\alpha$; the same is true for the
pressures $\Pi\pa$, and this gives the required total number of
$(p-1)(K+1)$ constraints. This formally defines the exact solution of
the phase coexistence problem for truncatable systems. Its practical
value is limited by the difficulties of finding the solution
numerically, as pointed out in Sec.~\ref{sec:intro}; see also
Sec.~\ref{sec:implementation}. There is also no interpretation of the
result in terms of a free energy depending on a small number of
densities.

Nevertheless, eq.~\(effective_prior) is useful for a comparison with
the solution provided by the moment free energy
method. From~\(exact_dens_distrib), the exact coexisting phases are
all members of a family of density distributions of the general
form~\(family); this is simply a consequence of the requirement of
equal chemical potentials~\(Gibbs_Boltzmann). This ``exact coexistence
family'' has an ``effective'' prior~\cite{exact_prior_not_known_footnote}
$\effprior\neq\rhonsig$ and is therefore different from the ``original''
family~\(pfamily). The moment free energy only gives us access to
distributions from the original family, not the exact coexistence
family, and therefore cannot yield exact solutions for phase
coexistence (beyond its onset). Fig.~\ref{fig:coex_demo} illustrates
this point explicitly in the simplified context of a bidisperse system.

It is now easy to see, however, that -- as stated in
Sec.~\ref{sec:projection} -- the exact solution can be approached to
arbitrary precision by including extra \moms\ in the moment free
energy. (This leaves the exactness of spinodals, critical points,
cloud-points and shadows unaffected, because none of our arguments
excluded a null dependence of $\fexc$ on certain of the $\mi$.)
Indeed, by adding further \moms\ one can indefinitely extend the
family~\(pfamily) of density distributions, thereby approaching with
increasing precision the actual distributions in all phases present;
this yields phase diagrams of ever-refined accuracy. 

The roles of the ``original'' \moms\ (those appearing in the excess
free energy) and the extra ones are quite different, however. To see
this, note from~\(mom_chem_pot) that equality among phases of the
moment chemical potentials implies that, for the extra moments only,
the corresponding Lagrange multipliers must themselves be {\em equal
in all coexisting phases}. (This is because there is, by construction,
no excess part to the ``extra'' chemical potentials.) We can therefore
drop the phase index on these Lagrange multipliers and write the
density distribution in phase $\alpha$ as
\be
\rhoalsig = \rhonsig \exp\left(\sum_{{\rm extra }\,i}\mhati\wi\right)
\exp\left(\sum_{{\rm original }\,i}\mhati\pa\wi\right)
\label{flexible_prior}
\ee
Comparing with~\(family) and~\(exact_dens_distrib), we see that the
extra Lagrange multipliers can be thought of as providing a ``flexible
prior'' $\prior$, allowing a better approximation to the effective
prior $\effprior$ required by particle conservation. The ``tuning'' of
the prior with these extra Lagrange multipliers also has an important
effect on the number of coexisting phases that can be found by the
moment method (see Sec.~\ref{sec:geometry}).

\setlength{\figwidth}{6.5cm}

\begin{figure}

\begin{center}
\epsfig{file=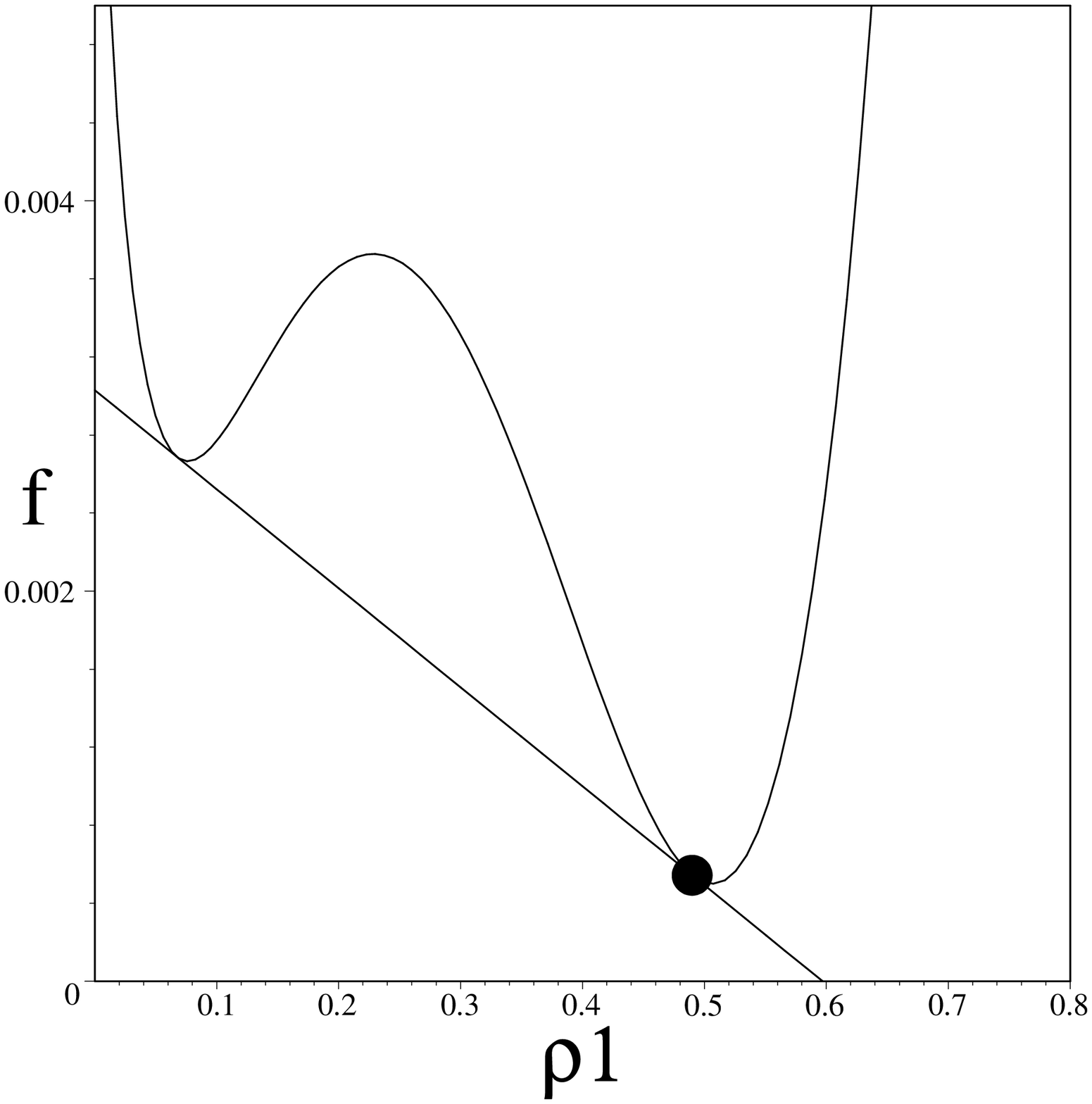,width=\figwidth}
\begin{picture}(0,0)\put(-38,152){\mbox{(a)}}\end{picture}%
\epsfig{file=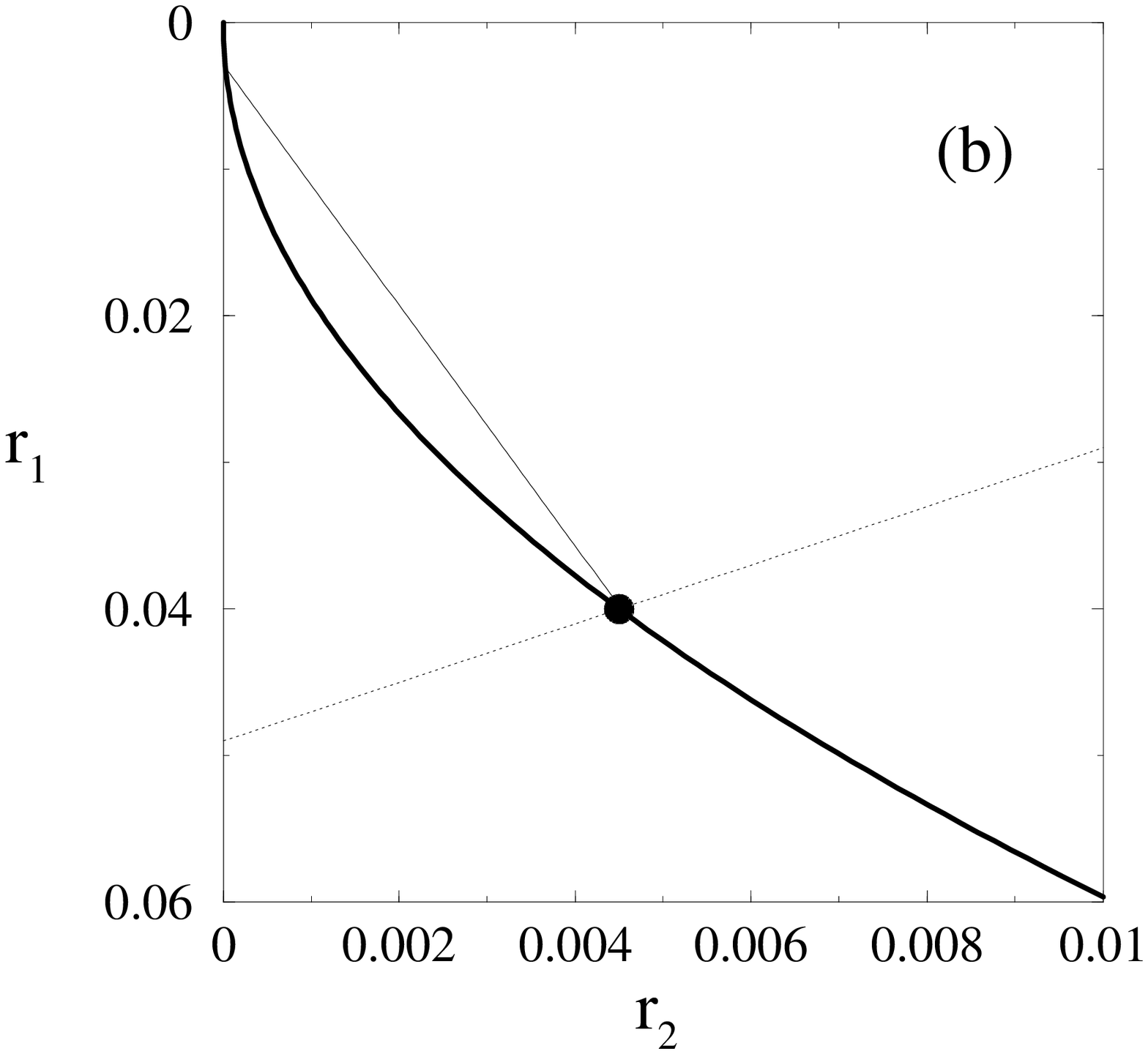,width=1.22\figwidth}

\vspace*{2mm}

\epsfig{file=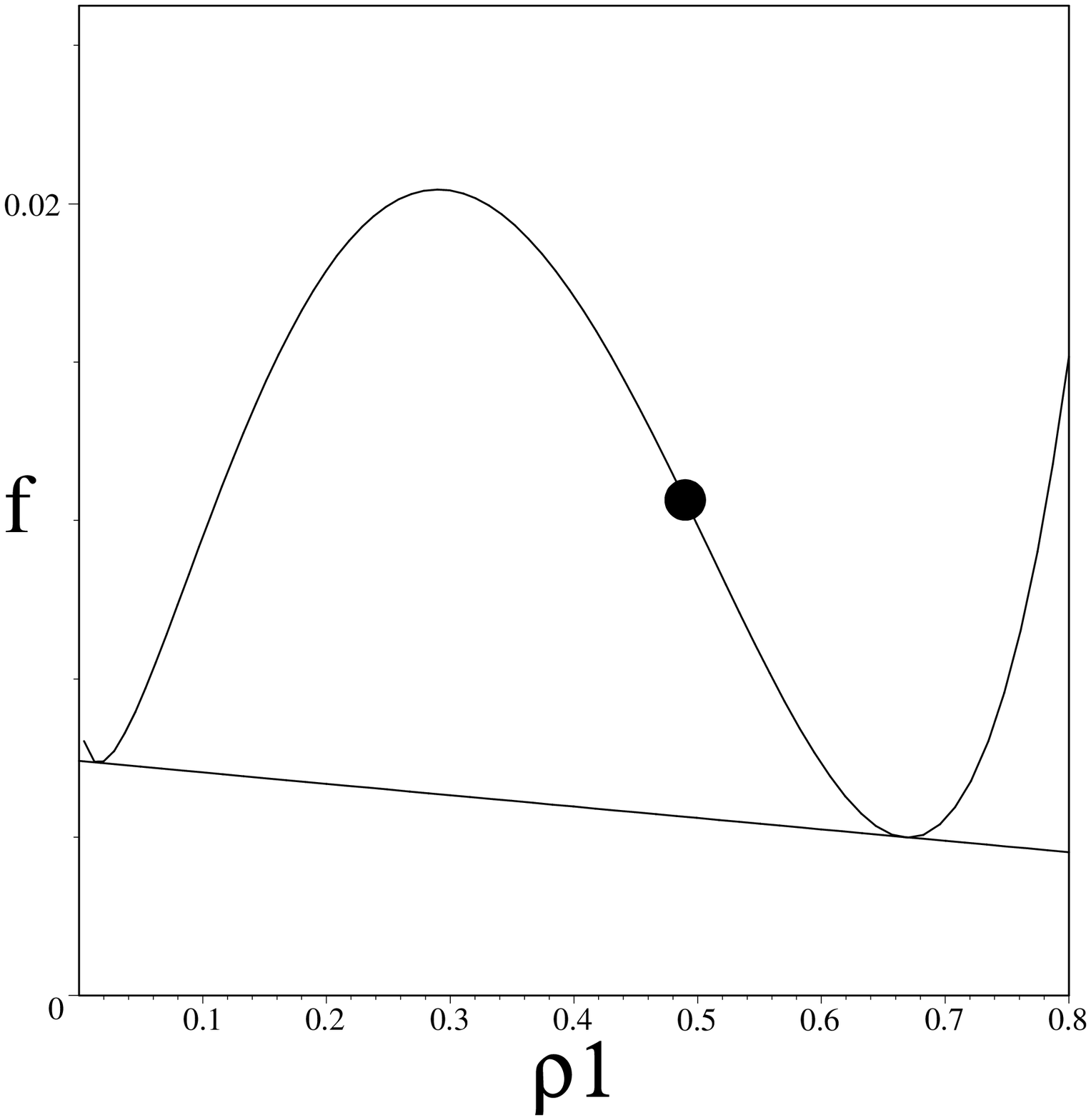,width=\figwidth}
\begin{picture}(0,0)\put(-38,152){\mbox{(c)}}\end{picture}%
\epsfig{file=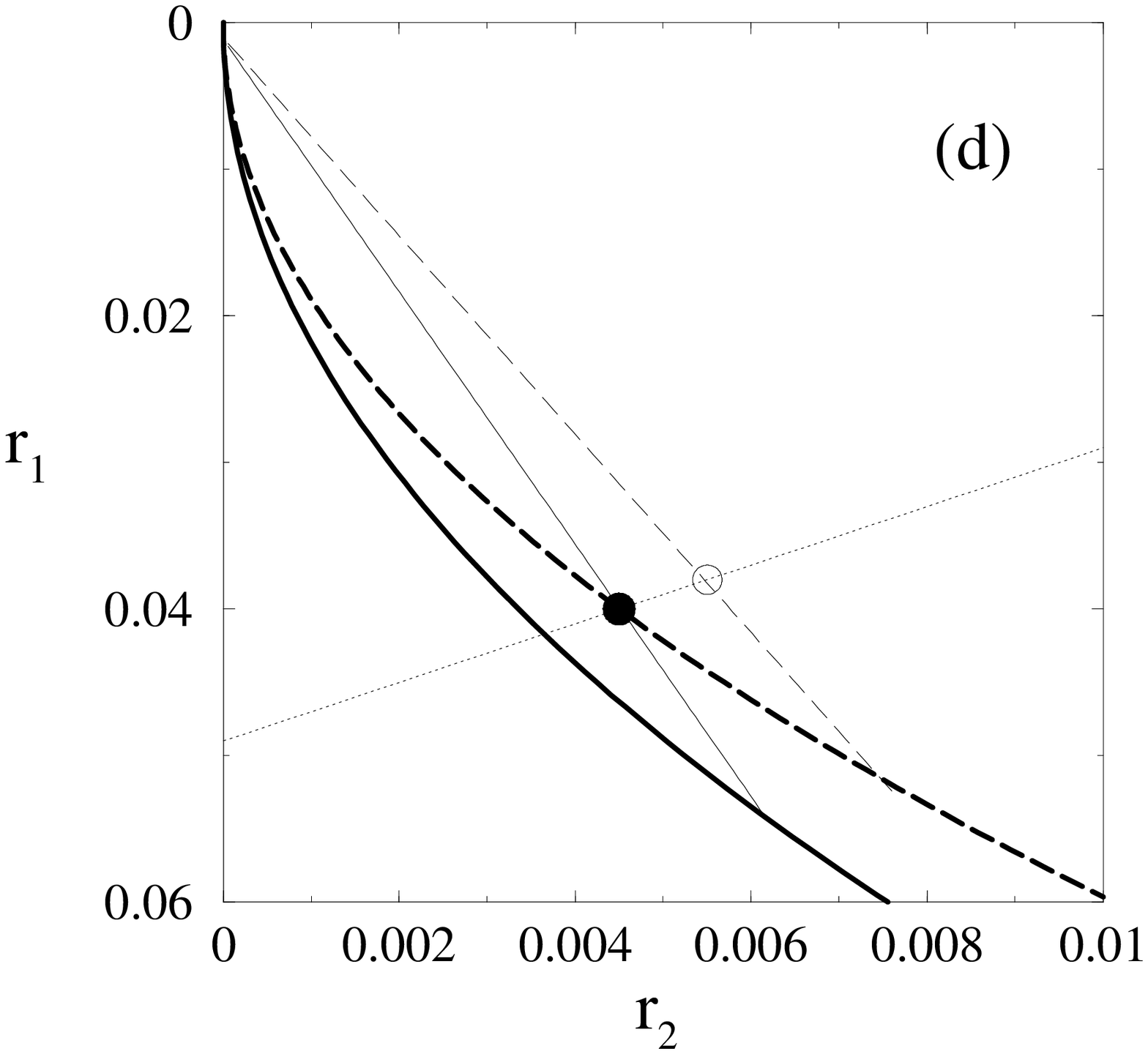,width=1.22\figwidth}
\end{center}

\caption{This figure continues the example of
Fig.~\protect\ref{fig:proj_demo}, which is a Flory-Huggins
polymer+solvent model with two chain lengths present. Now, however, we
allow for a change of the interaction parameter $\chi$ (equivalent to
changing temperature). (a) The moment free energy at the cloud point,
\ie, the value of $\chi$ for which the given parent first begins to
phase separate. (b) The tieline at the cloud point is drawn in the
($\dens_1$, $\dens_2$) plane of all density pairs (thin solid
line). It is {\em exact} and connects the parent (solid circle) with
another member of the family~(\protect\ref{pfamily}); the family is
indicated by the thick solid line. The dotted line is the
``transverse'' line of constant $\mo$ ($=\mo\pn$) passing through the
parent. (c) The moment free energy for a higher value of $\chi$. (d)
The corresponding tieline (thin dashed line) in the ($\dens_1$,
$\dens_2$) plane. As before, this connects two members of the parent's
family (now the thick dashed line), and the two phases so found are in
stable thermodynamic equilibrium with each other. Because they now
occupy comparable fractions of the system volume, however, the lever
rule is violated: the parent does not lie on the tieline. But the
lever rule violation occurs only along the transverse direction
(dotted line): the fractional phase volumes calculated from the moment
free energy result in a total density pair (indicated by the empty
circle) which has the same \mom\ $\m$ as the parent. Finally, the
exact tieline at this $\chi$ is also shown (thin solid line). As
required, it passes through the parent. But its endpoints now connect
members of a {\em different} family (thick solid line), which derives
from an ``effective parent'' (or effective prior) $(\tilde r_1,\tilde
r_2)$, which may be chosen anywhere along the thick solid line
(compare~(\protect\ref{exact_dens_distrib})). Note that if a single
extra \mom\ were added in this scenario, the resulting maximum entropy
family would in fact cover all possible density pairs ($\dens_1$,
$\dens_2$) and the corresponding two-moment free energy would
therefore give exact results in all situations. For a truly
polydisperse (rather than bidisperse) system, in which the density
pairs of this example become density {\em distributions} $\rhosig$,
each added \mom\ allows the accuracy of the calculation to be
increased.
\label{fig:coex_demo}
}
\end{figure}

\subsection{Global and local stability}
\label{sec:stability}

\begin{figure}

\begin{center}
\epsfig{file=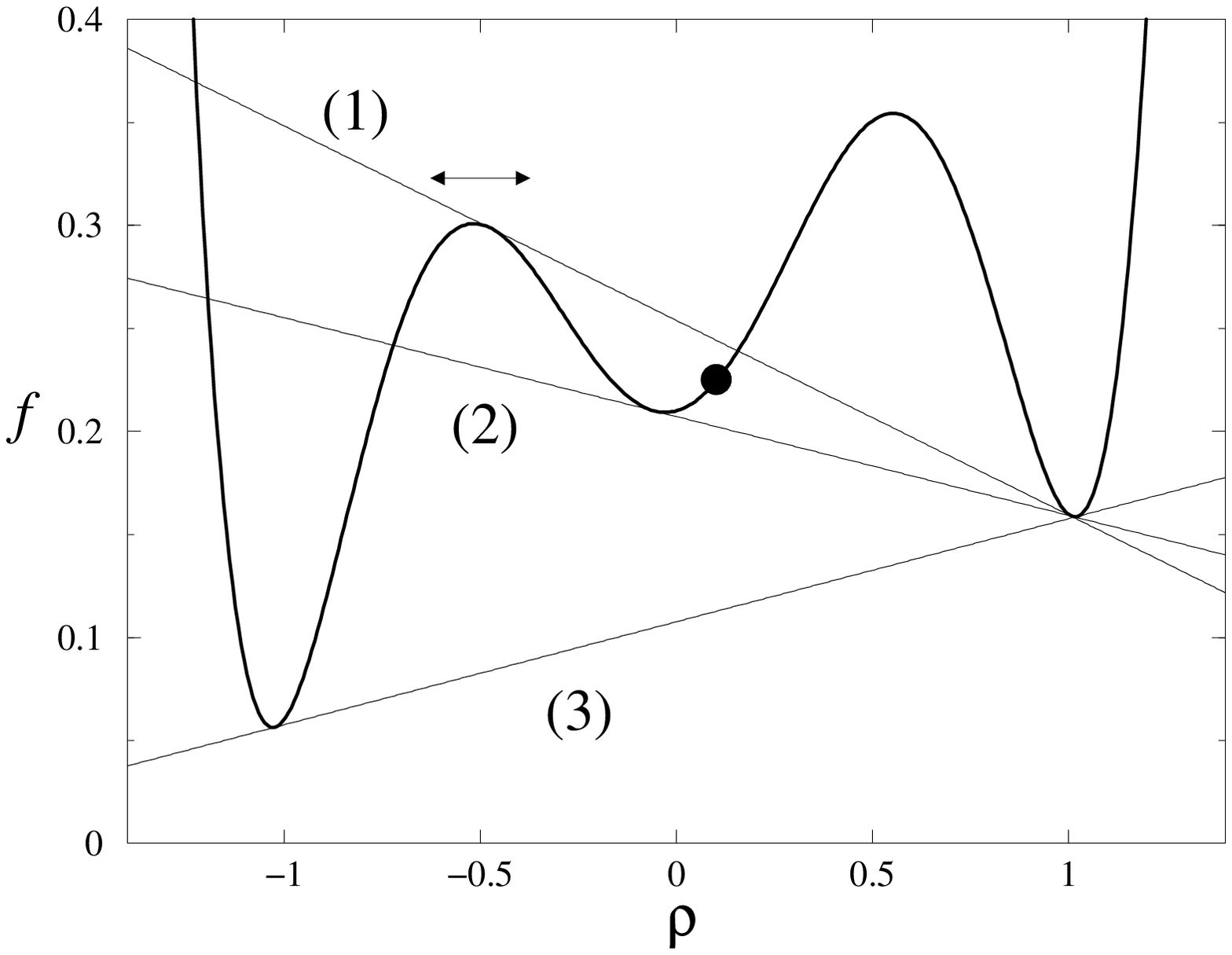, width=10cm}
\end{center}

\caption{Illustration of local and global stability of solutions of
the phase equilibrium conditions. For simplicity, we consider the case
of a monodisperse system, characterized by its free energy density $f$
(bold line) as a function of the density $\rh$. The phase split for a
parent with density $\rh\pn$ (indicated by the circle) is to be
calculated. Three double tangents to $f(\rh)$ are shown, predicting
different two-phase splits of the parent. All obey (by the double
tangent property) the conditions of equal chemical potential and
pressure in the two phases. But (1) is {\em locally} unstable: a small
fluctuation (indicated by the arrow) of the density of the more dilute
coexisting phase lowers the total free energy. (2) is locally stable,
but {\em globally} unstable: the phase split (3) gives a lower total
free energy. (3) is globally (and therefore locally) stable.
\label{fig:stability}
}
\end{figure}

Most numerical algorithms for phase coexistence calculations,
including ours, initially proceed by finding a solution to the
equilibrium conditions of equal chemical potentials and pressures in
all phases, rather than by a direct minimization of the total free
energy of the system. It is then crucial to verify whether this
solution is {\em stable}, both {\em locally} (\ie, with respect to
small fluctuations in the compositions of the phases) and {\em
globally} with respect to splitting into a larger number (or
different) phases (see Fig.~\ref{fig:stability}). Global stability is
a particularly important issue in our context because there is in
principle no limit on the number of coexisting phases in a
polydisperse system.

A useful tool for stability calculations is the ``tangent plane
distance''~\cite{Michelsen82}. Let us first define this generically
for a system with (a vector of) densities $\rhovect$, free energy
density $f(\rhovect)$ and chemical potentials
$\muvect(\rhovect)=\nabla f(\rhovect)$; this notation is the same as
in Sec.~\ref{sec:properties_general}. Assume we have found a candidate
``phase split'', that is a collection of $p$ phases $\rhovect\pa$
($\alpha=1\ldots p$) which satisfy the phase equilibrium conditions of
equal chemical potentials $\muvect=\muvect\pa$ and pressures
$\Pi=\Pi\pa$. They occupy fractions $\phal$ of the total volume, and
the overall density distribution is thus
$\rhovect\tot=\sum_\alpha\phal\rhovect\pa$.  Note that we do not yet
assume that the lever rule is satisfied, \ie, that
$\rhovect\tot=\rhovect\pn$; as explained above, this equality will
generally not hold for phase splits calculated from the moment free
energy. We now define ``global tangent plane (TP) stability'' for such
a phase split as the property that there is no other phase split (\ie,
no other tangent plane) that gives a lower total free energy for {\em
the same overall density distribution} $\rhovect\tot$. In more
intuitive language, this means that if we were to put the phases
$\rhovect\pa$ into contact with each other, the resulting system would
be thermodynamically stable; neither the composition nor the number of
phases would change over time. Note, however, that since in general
$\rhovect\tot=\sum_\alpha\phal\rhovect\pa$ is {\em not} equal to
$\rhovect\pn$, the phase split that is globally TP-stable need not
accurately reflect the number and composition of phases into which the
parent $\rhovect\pn$ would actually split under the chosen
thermodynamic conditions.

To define the tangent plane distance (TPD), note first that by virtue
of the coexistence conditions, all phases $\rhovect\pa$ lie on a
tangent plane to the free energy surface. Points $(\rhovect, f)$ on
this tangent plane obey the equation $f-\muvect\cdot\rhovect+\Pi=0$,
with $\muvect$ and $\Pi$ the chemical potentials and pressure common
to all phases. For a generic phase with density distribution
$\rhovect$ and free energy $f(\rhovect)$, the same expression will
have a nonzero value which measures how much ``below'' or ``above''
the tangent plane it lies. This defines the TPD
\be 
\tpd(\rhovect) = f(\rhovect)-\muvect\pa\cdot\rhovect+\Pi\pa.
\label{tpd}
\ee
Here we have added the superscript $\alpha$ to emphasize that the
chemical potential and osmotic pressure used in the calculation of the
TPD are those of the calculated phase equilibrium (and hence of any of
the participating phases $\al$), rather than those of the test phase
$\rhovect$. It is then clear intuitively -- and can be shown more
formally~\cite{Michelsen82} -- that the calculated phase coexistence
is globally TP-stable if the TPD is non-negative everywhere.
Geometrically, this simply means that no part of the free energy
surface must protrude beneath the tangent plane; otherwise the total
free energy of the system could be lowered by constructing a new
tangent plane that touches the protruding piece. Global TP-stability of
course encompasses {\em local} stability; the latter simply
corresponds to the requirement that the TPD be a local minimum (with
the value $\tpd=0$) at each of the phases in the candidate solution.

When verifying global TP-stability, it is obviously
sufficient to check the value of the TPD at all of its stationary
points. By differentiating~\(tpd) w.r.t.\ $\rhovect$, one sees that at
these points, the chemical potentials $\muvect(\rhovect)$ are the same
as in the calculated coexisting phases. For our truncatable
polydisperse systems, it then follows from eq.~\(Gibbs_Boltzmann) that
we only need to consider the TPD for test phases $\rhosig$ which are
in the same family~\(family) as the calculated coexisting phases. This
is a crucial point: Even though global TP-stability is a statement
about stability in the infinite-dimensional space of density
distributions $\rhosig$, it can be checked by only considering density
distributions from a $K$-dimensional family. In the (generally
hypothetical) case of an exactly calculated phase split, this family
would be~\(effective_prior), with the effective prior
$\effprior$~\cite{tpd_exact_footnote}. For a phase split calculated
from the moment free energy, it is the family~\(pfamily) with
the parent as prior, with any Lagrange multipliers for extra moments
fixed to their values in the calculated coexisting phases. The TPD of
such a test phase is (using~\(exact_free_en) for the exact free energy
and~\(exact_chem_pot) for the chemical potentials $\mu\pa(\sig)$ of
the coexisting phases)
\bea
\tpd[\rhosig] &=& f[\rhosig] - \intsig \mu\pa(\sig)\rhosig + \Pi\pa
\nonumber\\
&=& \fexc(\mi) + \intsig\rhosig\left\{ T\left[\ln\rhosig -1\right] -T
\ln\rhoalsig - \sum_i \muexci\pa \wi \right\} + \Pi\pa
\nonumber\\
&=& \fexc(\mi) + T \intsig\rhosig\left[\ln\frac{\rhosig}{\rhonsig} -1\right]
- \sum_i \left[\mhati\pa+\muexci\pa\right]\mi + \Pi\pa
\nonumber\\
&=& \fmom(\mi) - \sum_i \mu_i\pa \mi + \Pi\pa.
\nonumber
\eea
A comparison with~\(tpd) shows that this is identical to the TPD that
one would derive from the moment free energy alone. Here again, the
underlying polydisperse nature of the problem can therefore be
disregarded once the moment free energy has been obtained. In summary,
one can determine whether a phase split calculated from the moment
free energy is globally TP-stable (which means that phases of the
predicted volumes and compositions, would, if placed in contact,
indeed coexist) using only the TPD derived from the moment free
energy; the same is trivially true of the weaker requirement of local
stability~\cite{local_stab_explicit_footnote}.

\begin{figure}
\begin{center}
\epsfig{file=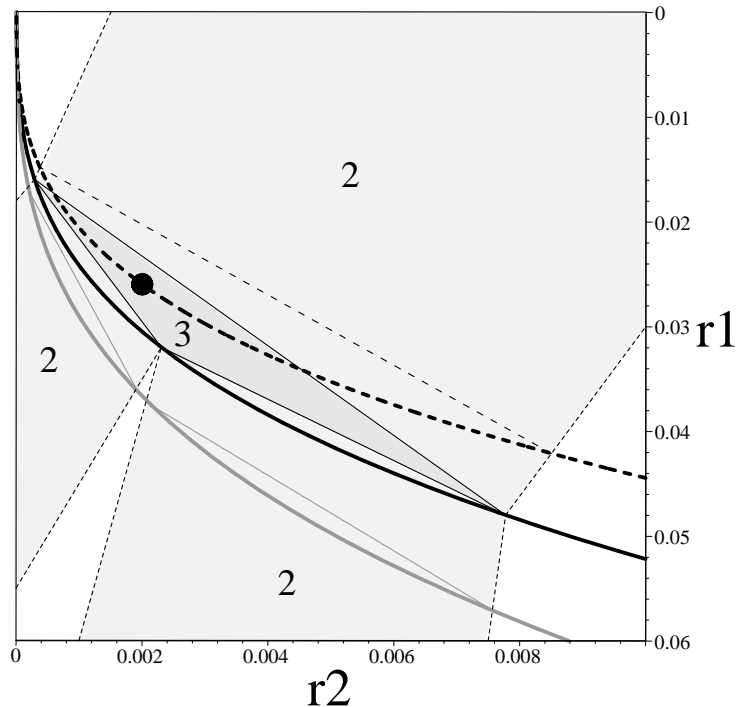, width=10cm}
\end{center}
\caption{Example of structure of the space of density
distributions. We continue the example of
Figs.~\protect\ref{fig:proj_demo} and~\protect\ref{fig:coex_demo} and
consider a bidisperse Flory-Huggins system with a density pair
($\dens_1$, $\dens_2$). As before, we take the excess free energy of
the system to depend on a single \mom, so that the
families~(\protect\ref{family}) are one-dimensional (\ie, appear as
curves in the graph). In the case sketched here, there is a
three-phase region bordered by two-phase regions delineated by dotted
lines. The bold (solid, dashed, grey) curves show three
families~(\protect\ref{family}) of density pairs; the whole space is
partitioned (or ``foliated'') into an infinite number of such
families.  All tielines must begin and end within the same family; the
thin grey tielines are shown as examples. The moment free energy
(without use of extra moments -- or in this case, with a single \mom\
retained) can only access systems from the family passing through the
parent (filled circle); it therefore predicts that the parent will
separate into the two phases within this family that are connected by
a tieline (thin dashed line). This tieline is thermodynamically
``real'': the two phases at its ends, if put into contact, would
remain in stable coexistence. But it does not give the exact phase
separation for the given parent because of the lever rule violations
allowed by the moment approach (the parent does not lie on the
tieline). Retaining one extra \mom\ in the moment free energy gives,
in this simple bidisperse case, the exact result: the parent separates
into the three phases at the corners of the three-phase triangle,
which lie on a family (bold solid curve) that does not contain the
parent.
\label{fig:foliation_three_phase}
}
\end{figure}

Recall, however, that the overall density distribution
$\rho\tot(\sig)$ for a phase split found from the moment free energy
only has the same \moms\ $\mi\pn$ as the parent $\rhonsig$, but
differs in other details (the transverse degrees of freedom).
Global TP-stability thus guarantees that such an approximate phase
split is thermodynamically stable, but does not imply that it is
identical (in either number or composition of phases) to the exact
one. Nevertheless, progressively increasing the number of extra \moms\
in the moment free energy will make $\rh\tot(\sig)$ a progressively
better approximation to $\rhonsig$, and so the exact phase split, with
the correct number of phases, must eventually be recovered to
arbitrary accuracy; see Fig.~\ref{fig:foliation_three_phase} for an
illustration.

\subsection{Geometry in density distribution space}
\label{sec:geometry}

In the above discussion of the properties of the moment free energy,
we have focused on obtaining the phase behavior of a system with a
given parent distribution $\rhonsig$. This is the point of view most
relevant for practical applications of the method, and the rest of
this Section is not essential for understanding such
applications. Nevertheless, from a theoretical angle, it is also
interesting to consider the global geometry of the space of all
density distributions $\rhosig$, without reference to a specific
parent: the exact free energy~\(exact_free_en) of the underlying
(truncatable) model induces two-phase tie lines and multi-phase
coexistence regions in this space, and one is led to ask how the
moment free energy encodes these properties.

As pointed out in Sec.~\ref{sec:projection}, the definition of the
moment free energy depends on a prior $\prior$ and represents the
properties of systems with density distributions $\rhosig$ in the
corresponding maximum entropy family~\(family).
Instead of identifying $\prior=\rhonsig$, we now allow a general prior
$\prior$. Conceptually, it then makes sense to associate the moment
free energy with the {\em family}~\(family) rather than the specific
{\em prior}. This is because any density distribution from~\(family)
can be chosen as prior, without changing the moment free energy (apart
from irrelevant linear terms in the moments $\mi$), or the identity of
the remaining family members. The construction of the moment free
energy thus partitions the space of all $\rhosig$ into (an infinite
number of) families~\(family); different families give different
moment free energies that describe the thermodynamics ``within the
family''. This procedure gives meaningful results because, as shown at
the end of Sec.~\ref{sec:properties_general}, coexisting phases are
always members of the {\em same} family.
By considering the entire ensemble of families~\(family) and their
corresponding moment free energies, one can thus in principle recover
the exact geometry of the density distribution space and the phase
coexistences within it. Note that this includes regions with more than
$K+1$ phases, even when the excess (and thus moment) free energy
depends on only $K$ \moms; see
Fig.~\ref{fig:foliation_three_phase}. Consistent with Gibbs' phase
rule, however, the families for which this occurs are exceptional:
they occupy submanifolds (of measure zero) in the space of all
families. Accordingly, to find the families on which such
``super-Gibbs'' multi-phase coexistences occur, the corresponding
prior must be very carefully tuned. Indeed, the probability of finding
such priors ``accidentally'', without either adding extra moments to
the moment free energy description or solving the exact phase
coexistence problem, is zero. Generically, one needs to retain at
least $n$ \moms\ in the moment free energy to find $n+1$ phases in
coexistence. From~\(flexible_prior), this corresponds to having $n-K$
parameters available for tuning the location of the family (or
equivalently, its prior) in density distribution space. But note that,
once the extra moments have been introduced, the required tuning need
not be done ``by hand'': it is achieved implicitly by requiring the
lever rule to be obeyed not only for the $K$ original \moms, but also
for the $n-K$ extra ones. (The solution of the phase equilibrium
conditions thus effectively proceeds in an $n$-dimensional space; but
once a solution has been found, its (global TP-)stability can still be
checked by computing the TPD only within a $K$-dimensional family of
distributions. See Sec.~\ref{sec:stability}.)

As pointed out above, phase splits calculated from the moment free
energy allow violations of the lever rule. In the global view, this
fact also has a simple geometric interpretation: the
families~\(family) are (generically) curved. In other words, for two
distributions $\rho\po(\sig)$ and $\rho\pt(\sig)$ from the same
family, the straight line $\rho(\sig)=\epsilon\rho\po(\sig) +
(1-\epsilon)\rho\pt(\sig)$ connecting them lies outside the
family. More generally, if $p$ coexisting phases $\rhoalsig$ have been
identified from the moment free energy, then the overall density
distribution $\rh\tot(\sig)=\sum_\alpha\phal\rhoalsig$ is different
from the parent $\rhonsig$; geometrically, it lies on the hyperplane
that passes through the phases $\rhoalsig$.

\section{Practical implementation of the moment method}
\label{sec:implementation}

The application of the moment free energy method to the calculation of
spinodal and critical points is straighforward using
conditions~\(mom_spinodal), and~\(mom_critical) or~\(det_critical),
respectively, and is further illustrated in Sec.~\ref{sec:examples}
below. We therefore focus in this Section on phase coexistence
calculations.

Recall that in the moment approach, each phase $\alpha$ is
parameterized by Lagrange multipliers $\mhati\pa$ for the original
moments (the ones appearing in the excess free energy of the system)
and the fraction $\phal$ of system volume that it occupies. If extra
moments are used, there is one additional Lagrange multiplier $\mhati$
for each of them; these are common to all phases. These parameters
have to be chosen such that the pressure~\(mom_osmotic_pressure) and
the moment chemical potentials $\mu_i$ given by~\(mom_chem_pot) are
equal in all phases. Furthermore, the (fractional) phase volumes
$\phal$ have to sum to one, and the lever rule has to be satisfied for
all moments (both original and extra):
\[
\sum_\al \phal \mi\pa = \mi\pn.
\]
In a system with $K$ original moments that is being studied using an
$\L$-moment free energy (\ie, with $\L-K$ extra moments), and for $p$
coexisting phases, one has $p(K+1)+\L-K=(p-1)(K+1)+\L+1$ parameters
and as many equations. Starting from a suitable initial guess, these
can, in principle, be solved by a standard algorithm such as
Newton-Raphson~\cite{PreTeuVetFla92}. Generating an initial point from
which such an algorithm will converge, however, is a nontrivial
problem, especially when more than two phases coexist.

To simplify this task, we work with a continuous control parameter
such as temperature, density $\mn$ of the parent phase, or interaction
parameter $\chi$ for polymers.  Taking the latter case as an example,
we start the calculation at a small value of $\chi$ where we are sure
to be in a single-phase region; thus the parent phase under consideration
is stable. The basic strategy is then to increment $\chi$ and detect
potential new phases as we go along. At each step, the phase
equilibrium conditions are solved by Newton-Raphson for the current
number of phases. Then we check for local stability of the solution by
calculating the Hessian of the TPD
around each of the coexisting phases and verifying that it is positive
definite. If an instability (\ie, a negative eigenvalue of the
Hessian) is found, we search for local minima starting from points
displaced either way along the instability direction. If two new local
minima are found in this way, we add them to list of phases and delete
the unstable phase; if only one new local minimum is uncovered, we add
this but retain the old phase.
Finally, we check for global (TP-)stability. As explained in
Sec.~\ref{sec:stability}, this involves scanning all ``test'' phases
from the same maximum entropy family for possible negative values of
the TPD. However, the test phase will have the same extra Lagrange
multipliers, and is therefore parameterized in terms of the $\mhati$
for the original moments only. The extra Lagrange multipliers are held
fixed, so their number is irrelevant for the computational cost of the
stability check.  Put differently, within our algorithm the {\em
precision} of the coexistence curves should depend on the {\em total}
number of moments retained, $n$, rather than the number of moments in
the excess free energy, $K$. {\em Computational effort}, on the other
hand, is dominated by the global stability check, and thus mainly
sensitive to $K$ (the dimension of the space to be searched for new
phases), rather than $n$.

This is a substantial efficiency gain, but despite it, an exhaustive search
for local TPD minima over the $K$-dimensional space of original
Lagrange multipliers is unrealistic except for $K=1$. We use instead a
Monte-Carlo-type algorithm~\cite{HesSti95} to sample the TPD at
representative points.
If any negative values of the TPD are encountered, we choose the
smallest such value, find the nearest local minimum of the TPD and add
this point to the list of phases. If any new phases have been found
during the (local and global) stability checks, we assign each of them
a default phase volume ($\ph=0.1$, say), reduce the phase volumes of
the old phases accordingly, update the number of phases and loop back
to the Newton-Raphson solution of the phase equilibrium conditions,
using the current list of phases as an initial guess. This process is
repeated until the calculated solution is found to be stable. Our
implementation of this basic scheme also contains some additional
elements (such as an adaptive choice of the stepsize for the control
parameter $\chi$, and checks for very small phase volumes) which are
useful near points where new phases appear or old ones vanish.

Comparing our own approach, which is based on the moment free energy
with extra \moms\ as just outlined, to~\(effective_prior), which is
the exact solution for truncatable systems, we see that the former
actually uses more parameters (essentially one phase-independent
Lagrange multiplier per extra moment) to represent the solution. At
first sight, this may appear counter-productive. However, the Lagrange
multipliers $\mhati\pa$ and the (fractional) phase volumes $\phal$ are
much less strongly coupled in the moment free energy approach. The
phase volumes are only determined by the lever rule, while in the
exact solution they ``feed back'' into the effective prior $\effprior$
and therefore into the equilibrium conditions of equal chemical
potentials and pressures. In the examples studied below, we have found
that the numerical advantages of this decoupling (in terms of
stability, robustness and convergence of our algorithm) can easily
outweigh the larger number of parameters that it requires.

As with any numerical algorithm, it is important to develop robust
criteria by which the convergence of the solution can be judged. This
has a novel aspect, in the moment method, since violations of the
lever rule are allowed: alongside normal numerical convergence
criteria one needs a method for deciding whether the effect of these
on the predicted phase behaviour is significant. We develop
appropriate criteria in Sec.~\ref{sec:copolymer} in the context of a
specific example. The basic ideas is that since any state of phase
coexistence predicted by the method represents the exact behaviour of
{\em some} parent, then so long as this parent is close enough to the
true one, the predicted behaviour will lie within the range of
uncertainty that arises anyway, from not knowing the true parent to
arbitrary experimental precision.

\section{Examples}
\label{sec:examples}

We now illustrate how the moment method is applied and demonstrate its
usefulness for several examples. The first two (Flory-Huggins theory
for length-polydisperse homopolymers and dense chemically polydisperse
copolymers, respectively) contain only a single \mom\ in the excess
free energy and are therefore particularly simple to analyse and
visualize. In the third example (chemically polydisperse copolymers in
a polymeric solvent), the excess free energy depends on two \moms\ and
this will gives us the opportunity to discuss the appearance of more
complex phenomena such as tricritical points.

\subsection{Homopolymers with length polydispersity}
\label{sec:length_homopolymer}

Let us start with the simple but well studied example of polydisperse
Flory-Huggins theory~\cite{pbw_flory_huggins}. One considers a system
of homopolymers with a distribution of chain lengths; the
polydisperse feature $\sig$ is simply the chain length $L$, \ie,
the number of monomers in each chain. (We treat this as a continuous
variable.) The density distribution $\rho(L)$ then gives the number
density of chains as a function of $L$. We choose the segment volume
$a^3$ as our unit of volume, making $\rho(L)$ dimensionless. The
volume fraction occupied by the polymer is then simply the first
moment of this distribution, $\phi\equiv\fhphi = \intL L\,
\rho(L)$. Within Flory-Huggins theory, the free energy density is (in
units such that $\kB T=1$)
\begin{equation}
{f}=\int dL\,\rho(L)[\loge\rho(L) -1] +
(1-\fhphi)\loge(1-\fhphi)+\chi\fhphi(1-\fhphi)
\label{FH_start}
\end{equation}
where the Flory $\chi$-parameter plays essentially the role of an
inverse temperature. Before analysing this further, we note that for a
bidisperse system with chain lengths $L_1$ and $L_2$ and number
densities $\dens_1$ and $\dens_2$, the corresponding expression would
be
\be
{f}= \dens_1 (\loge\dens_1 -1) + \dens_2 (\loge\dens_2 -1) + 
(1-\fhphi)\loge(1-\fhphi)+\chi\fhphi(1-\fhphi), \quad \fhphi =
L_1\dens_1+L_2\dens_2
\label{bidisp_FH}
\ee
This free energy, with $L_1=10$ and $L_2=20$, was used to generate the
examples shown in Figs.~\ref{fig:proj_demo} and~\ref{fig:coex_demo}.

Returning now to~\(FH_start), we note first that the excess free
energy
\be
\fexc = (1-\fhphi)\ln(1-\fhphi) + \chi \fhphi (1-\fhphi) 
\label{fexc_flory_huggins}
\ee
depends only on the \mom\ $\fhphi$. If no extra moments are used, the
moment free energy is therefore a function of a single density
variable, $\fhphi$. Let us work out its construction explicitly for
the case of a parent phase with a Schulz-distribution of lengths,
given by
\be
\rho\pn(L) = \fhrho\pn \frac{1}{\Gamma(\fhalpha)\fhbeta^\fhalpha}
\,L^{\fhalpha-1}e^{-L/\fhbeta}
\label{Schulz}
\ee
Here $\fhalpha$ is a parameter that determines how broad or peaked the
distribution is; it is conventionally denoted by $\alpha$, but we
choose a different notation here to prevent confusion with the phase
index used in the general discussion so far. The chain number density
of the parent is $\fhrho\pn$, and the {\em normalized} first moment
$\fhnmo\pn=\fhphi\pn/\fhrho\pn$ is simply the (number) average chain
length
\be
L_N \equiv \fhnmo\pn \equiv 
\frac{\fhphi\pn}{\fhrho\pn} = \fhalpha\fhbeta.
\label{LN}
\ee
The parent is thus parameterized in terms of $\fhalpha$, $\fhrho\pn$
and $\fhphi\pn$ (or $\fhbeta$). The density distributions in the
family~\(pfamily) are given by
\be
\rho(L) = \rho\pn(L) e^{\mhato L} = \fhrho\pn
\frac{1}{\Gamma(\fhalpha)\fhbeta^\fhalpha}
\,L^{\fhalpha-1}e^{-(\fhbeta^{-1}-\mhato)L}
\label{pfamily_onedim}
\ee
and the zeroth and first \moms\ are easily worked out, for members of
this family, to be
\[
\fhrho = \fhrho\pn \left(\frac{1}{1-\fhbeta\mhato}\right)^\fhalpha, \qquad 
\fhphi = \fhphi\pn \left(\frac{1}{1-\fhbeta\mhato}\right)^{\fhalpha+1}.
\]
Because the family only has a single parameter $\mhato$, these two
\moms\ are of course related to one another:
\be
\fhrho  = c \fhphi^{\fhalpha/(\fhalpha+1)}, \qquad
c = \fhrho\pn\left(\fhphi\pn\right)^{-\fhalpha/(\fhalpha+1)}.
\label{mom_rel_onedim}
\ee
Now we turn to the moment entropy, given by \(mom_free_en): $\smom =
\fhrho-\mhato\fhphi$, and to be considered as a function of
$\fhphi$. After a little algebra (and eliminating $\fhbeta$ in favor
of $\fhphi\pn$) one finds
\be
\smom = \fhrho\pn\left[
(1+\fhalpha) \left(\frac{\fhphi}{\fhphi\pn}\right)^{\fhalpha/(\fhalpha+1)} -
\fhalpha \frac{\fhphi}{\fhphi\pn} \right]
\label{FH_smom_onedim}
\ee
The last term is linear in $\fhphi$ and can be disregarded for the
calculation of phase equilibria. This then gives the following simple
result for the moment free energy
\be
\fmom = - (\fhalpha+1) c \fhphi^{\fhalpha/(\fhalpha+1)} + \fexc.
\label{FH_mom_free_en}
\ee
(In conventional polymer notation, this result would read
$\fmom=-(\alpha+1) c \phi^{\alpha/(\alpha+1)}+\fexc$.)  From this we
can now obtain the spinodal condition, for example, which identifies
the value of $\chi$ where the parent becomes unstable. The general
criterion~\(spinodal_general) simplifies in our case of a single
moment density to
\be
\left. \frac{d^2\fmom}{d\fhphi^2}\right|_{\fhphi=\fhphi\pn}=0 
\quad \Rightarrow \quad
\frac{1}{1-\fhphi\pn} - 2\chi + 
\frac{\fhalpha}{\fhalpha+1}\frac{\fhrho\pn}{
(\fhphi\pn
)^2} = 0
\label{FH_spinodal_parent}
\ee
The same condition for a phase with general density distribution
$\rh(L)$ -- rather than our specific Schulz parent $\rh\pn(L)$ --
follows from~\(mom_spinodal) as
\be
\frac{1}{1-\fhphi} - 2\chi + \frac{1}{\mt}
\label{FH_spinodal}
\ee
As expected, this becomes equivalent to~\(FH_spinodal_parent) for
parents of the Schulz form~\(Schulz), which obey
$\mt\pn=\fhrho\pn\fhalpha(\fhalpha+1)\fhbeta^2$ and eq.~\(LN). Note
that in standard polymer notation, eq.~\(FH_spinodal) would be written
as
\be
\frac{1}{1-\phi} - 2\chi + \frac{1}{L_W\phi}
\label{FH_spinodal_trad}
\ee
where $L_W=\mt/\mo$ is the weight average chain length.

The critical point condition is obtained similarly. For a single
density variable, eq.~\(critical_simplified) reduces to
\be
\left. \frac{d^3\fmom}{d\fhphi^3}\right|_{\fhphi=\fhphi\pn}=0 
\quad \Rightarrow \quad
\frac{1}{
(1-\fhphi\pn
)^2} -
\frac{\fhalpha(\fhalpha+2)}{(\fhalpha+1)^2}\frac{\fhrho\pn}{
(\fhphi\pn
)^3} = 0
\label{FH_critical_parent}
\ee
For a general distribution, the critical point
criterion~\(mom_critical) becomes instead
\[
 - \frac{\m_3}{\mt^3} + \frac{d^3\fexc}{d\m_1^3}= 0
\]
and inserting the Flory-Huggins excess free
energy~\(fexc_flory_huggins) gives
\be
\frac{1}{(1-\fhphi)^2} -\frac{\m_3}{\mt^3} = 0
\label{FH_critical}
\ee
Again, this simplifies to~\(FH_critical_parent) for Schulz parents,
for which $\m_3\pn= \fhrho\pn\fhalpha (\fhalpha+1)
(\fhalpha+2)\fhbeta^3$. The traditional statement of~\(FH_critical),
in terms of $\phi\equiv\fhphi$, $L_W$ and the $Z$-average chain length
$L_Z=\m_3/\mt$ reads~\cite{pbw_flory_huggins}
\be
\frac{1}{(1-\phi)^2} -\frac{L_Z}{L_W^2}\frac{1}{\phi^2} = 0.
\label{FH_critical_trad}
\ee

\begin{figure}
\begin{center}
\epsfig{file=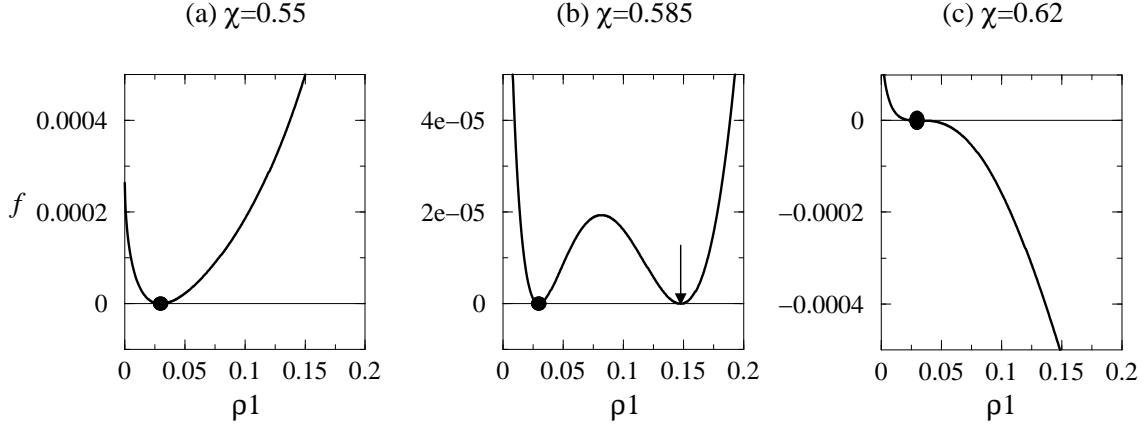,width=15cm}
\end{center}
\caption{Examples of moment free energy~(\protect\ref{FH_mom_free_en})
for Flory-Huggins theory of length-polydisperse polymers, with one
\mom, $\fhphi$, retained. The parent is of the Schulz
form~(\protect\ref{Schulz}), with $\fhphi\pn=0.03$, $L_N=100$ (hence
$\fhrho\pn=\fhphi\pn/L_N=3\cdot 10^{-4}$) and $\fhalpha=2$ (hence
$L_W=150$); the point $\fhphi=\fhphi\pn$ is marked by the filled
circles. In (a), the value of $\chi=0.55$ is sufficiently small for
the parent to be stable: The moment free energy is convex. (b) shows
the cloud point, $\chi\approx0.585$, where the parent lies on one
endpoint of a double tangent; the other endpoint gives the polymer
volume fraction $\fhphi$ in the shadow phase. Increasing $\chi$
further, the parent eventually becomes spinodally unstable
($\chi\approx 0.62$, plot (c)). Note that for better visualization,
linear terms have been added to all free energies to make the tangent
at the parent coincide with the horizontal axis.
\label{fig:FH_one_dim}
}
\end{figure}

Fig.~\ref{fig:FH_one_dim} shows the moment free
energy~\(FH_mom_free_en) for a given parent and different values of
$\chi$. The value of $\chi$ for which there is a double tangent at the
point $\fhphi=\fhphi\pn$ representing the parent gives the cloud
point; the second tangency point gives the polymer volume fraction
$\fhphi$ in the coexisting phase, hence the shadow point. Repeating
this procedure for different values of $\fhphi\pn$ (while maintaining
$L_N=\fhphi\pn/\fhrho\pn$ and $\fhalpha$ constant), one obtains the
full CPC and shadow curve for a given normalized parent length
distribution $\prob(L)=\rho\pn(L)/\fhrho\pn$.

We now consider the properties of the moment free energy with the
(chain number) density $\fhrho$ retained as an extra moment. This
provides additional geometrical insight into the properties of
polydisperse chains (while for the numerical determination of the CPC
and shadow curve, the above one-moment free energy is preferable). To
construct the two-moment free energy, we proceed as before. The
family~\(pfamily) is now
\be
\rho(L) = \rho\pn(L) e^{\mhatn+\mhato L} = \fhrho\pn
\frac{1}{\Gamma(\fhalpha)\fhbeta^\fhalpha} \,L^{\fhalpha-1} 
e^{\mhatn-(\fhbeta^{-1}-\mhato)L}
\label{FH_family_twodim}
\ee
with zeroth and first \moms
\[
\fhrho = \fhrho\pn
e^{\mhatn}\left(\frac{1}{1-\fhbeta\mhato}\right)^\fhalpha, \qquad 
\fhphi = \fhphi\pn e^{\mhatn}\left(
\frac{1}{1-\fhbeta\mhato}\right)^{\fhalpha+1}.
\]
The relations can be inverted to express the Lagrange multipliers
$\mhatn$ and $\mhato$ in terms of the \moms\ as
\begin{eqnarray*}
\mhatn &=& (\fhalpha+1)\ln\frac{\fhrho}{\fhrho\pn} -
\fhalpha\ln\frac{\fhphi}{\fhphi\pn}\\
\mhato &=& \frac{1}{\fhbeta}
\left(1-\frac{\fhrho/\fhrho\pn}{\fhphi/\fhphi\pn}\right)
\end{eqnarray*}
and one obtains the moment entropy
\be
\smom=\fhrho-\mhatn\fhrho-\mhato\fhphi =
-(\fhalpha+1)\fhrho\ln\frac{\fhrho}{\fhrho\pn} + 
\fhalpha\fhrho\ln\frac{\fhphi}{\fhphi\pn} + (\fhalpha+1)\fhrho - \fhalpha
\frac{\fhrho\pn}{\fhphi\pn}\fhphi.
\label{FH_smom_twodim}
\ee
As expected, this reduces to~\(FH_smom_onedim) for systems with
density distributions from our earlier one-parameter
family~\(pfamily_onedim), where $\fhrho$ can be expressed as a
function of $\fhphi$ according to~\(mom_rel_onedim). With both
$\fhrho$ and $\fhphi$ retained as \moms\ in the moment free energy, a
number of linear terms in~\(FH_smom_twodim) can be dropped, giving the
final result
\be
\fmom = (\fhalpha+1)\fhrho\ln\fhrho - \fhalpha\fhrho\ln\fhphi + \fexc.
\label{ps_fhfe}
\ee
Note that the dependence on the parent distribution is now only
through $\fhalpha$. This can be understood from the general discussion
in Sec.~\ref{sec:geometry}: The family~\(FH_family_twodim) of density
distributions now contains {\em all} Schulz-distributions with the
given $\fhalpha$, and the moment free energy is insensitive to which
member of this family (specified by $\fhrho\pn$ and $\fhphi\pn$) is
used as the parent.

The result~\(ps_fhfe) can of course also be obtained via the
combinatorial method, as follows. The normalized Schulz parent
distribution is given by $\parentL =
[\Gamma(\fhalpha)\fhbeta^\fhalpha]^{-1} L^{\fhalpha-1} e^{-\fhbeta
L}$. The cumulant generating function is thus $h(\lambda_1) =
-\fhalpha\loge(1-\fhbeta\lambda_1)$, giving $\fhnmo = \partial
h/\partial \lambda_1 = ab/(1-b\lambda_1)$. Dropping constants and
linear terms in the average chain length $\fhnmo$ (which do not affect
the phase behavior), the generalized entropy of mixing per particle,
$\smix=h-\lambda_1\fhnmo$ then becomes $\smix =
\fhalpha\loge\fhnmo$. Adding the ideal gas term and the excess part of
the free energy, we thus find the moment free energy density
\begin{equation}
\fmom=\fhrho\loge\fhrho-\fhalpha\fhrho\loge(\fhphi/\fhrho)
+ \fexc
\label{pbw_fhfe}
\end{equation}
in agreement with~\(ps_fhfe); it is a simple two component free
energy. In the conventional notation, eq.~\(pbw_fhfe) would read
\[
\fmom=\rho\loge\rho-\alpha\rho\loge(\phi/\rho)
+(1-\phi)\loge(1-\phi)+\chi\phi(1-\phi).
\]

The spinodal curve (SC) and critical point (CP) condition
may now be calculated from eq.~\eqref{pbw_fhfe}, using either the
methods outlined in Sec.~\ref{sec:properties_general} or the more
traditional determinant conditions~\cite{pbw_gibbs} (see also
App.~\ref{app:critical}). One finds
\begin{eqnarray}
\frac{1}{1-\fhphi}-2\chi+\frac{\fhalpha}{\fhalpha+1}
\frac{\fhrho}{\fhphi^2}=0,&&\quad{\mathrm{(SC)}}\label{spineq}\\
\frac{1}{(1-\fhphi)^2}-\frac{\fhalpha(\fhalpha+2)}{(\fhalpha+1)^2}
\frac{\fhrho}{\fhphi^3}=0.&&\quad{\mathrm{(CP)}}\label{criteq}
\end{eqnarray}
Evaluated at the parent, these conditions are identical to
eqs.~\(FH_spinodal_parent) and~\(FH_critical_parent) -- which were
derived from the one-moment free energy -- as they must be. But, as
explained at the end of Sec.~\ref{sec:properties_general}, they also
hold more generally for all systems with Schulz density distributions
with the given value of $\fhalpha$; see the remarks after
eq.~\(ps_fhfe).  Of particular interest among these systems are those
that differ from the parent only in their overall density while having
the same normalized length distribution $\prob(L)=\parentL$. They form
the ``dilution line''; their number averaged chain length satisfies
$\fhnmo=\fhnmo\pn=\int dL\,L\,\parentL$, which translates to
$\fhphi=L_N\fhrho$ (where $L_N\equiv\fhnmo\pn=\fhalpha\fhbeta$ is, as
before, the number average chain length of the parent). Inserting this
``dilution line constraint'' into the above equations, we obtain as
the spinodal and critical point conditions for systems on the dilution
line
\begin{eqnarray}
\frac{1}{1-\fhphi}-2\chi+
\frac{\fhalpha}{\fhalpha+1}
\frac{1}{L_N \fhphi}=0,&&\quad{\mathrm{(SC)}}\label{spineq2}\\
\frac{1}{(1-\fhphi)^2}-\frac{\fhalpha(\fhalpha+2)}{(\fhalpha+1)^2} 
\frac{1}{L_N\fhphi^2}=0.&&\quad{\mathrm{(CP)}}\label{criteq2}
\end{eqnarray}
Again, agreement with the general results~\(FH_spinodal_trad)
and~\(FH_critical_trad) is easily verified by noting that for Schulz
length distributions,
$L_W=\mt/\mo=(\fhalpha+1)\fhbeta=L_N(\fhalpha+1)/\fhalpha$ and
$L_Z=\m_3/\mt=(\fhalpha+2)\fhbeta=L_N(\fhalpha+2)/\fhalpha$ for the
weight and $Z$-average chain lengths, respectively.

\begin{figure}
\leavevmode
\begin{center}
\epsfig{file=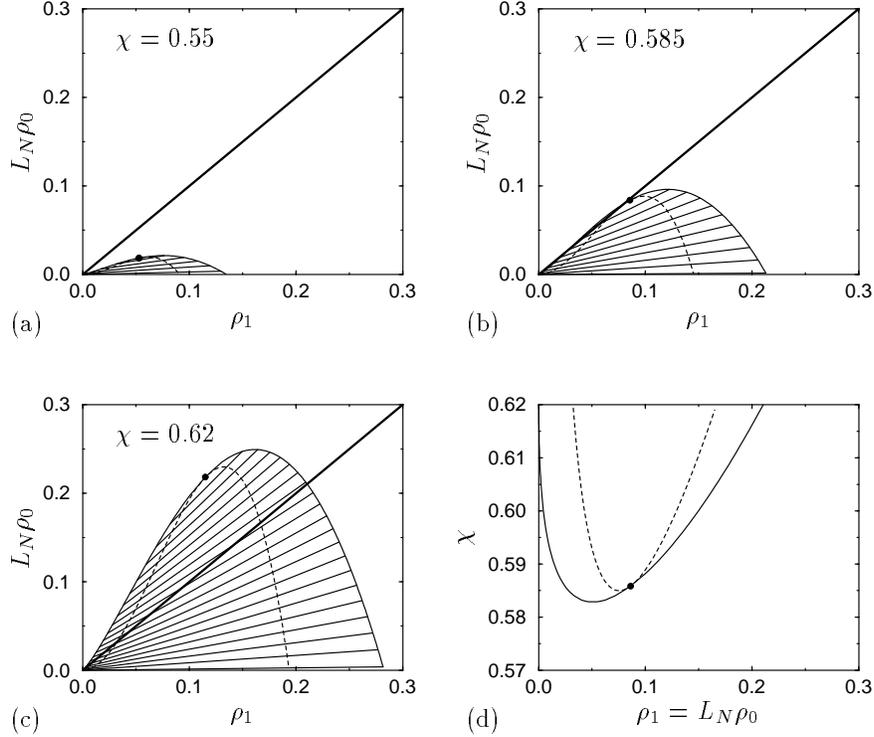}
\end{center}
\caption[?]{Conventional two component phase behavior in polydisperse
Flory-Huggins theory, shown in the $(\fhphi,\fhrho)$ plane for three
values of $\chi$. As in Fig.~\ref{fig:FH_one_dim}, the parent has
$L_N=100$ and $L_W=150$ (hence $\fhalpha=2$). Along the $y$-axis, we
plot $L_N\fhrho$ rather than $\fhrho$ so that the dilution line
$\fhphi=L_N\fhrho$, shown as the thick solid line in (a-c), is simply
along the diagonal. With $\chi$ considered as an additional variable,
the dilution line constraint defines a plane ($\fhphi=L_N\fhrho,
\chi)$. The last plot, (d), shows the cut by this plane through the
phase behavior in (a-c); the solid line is the cloud point curve, and
the dashed line is the spinodal stability condition.}
\label{pbw_fig1}
\end{figure}

The above discussion focussed on spinodals, critical points, and cloud
and shadow curves, all of which are found exactly by the moment
approach. (As usual, ``exact'' is subject to the assumed validity of
the original, truncatable free energy
expression~\(fexc_flory_huggins).) Within the moment approach, we have
also calculated the full phase behavior in the $(\fhphi, \fhrho)$
plane. This was done numerically from the two-moment free
energy~\eqref{pbw_fhfe}, facilitated by a partial Legendre
transformation from $\fhrho$ to the associated chemical potential
$\mu_0$, which can be performed analytically in this case (and
effectively brings one back to the one-moment free
energy~\(FH_mom_free_en)). Fig.~\ref{pbw_fig1}(a)--(c) shows binodal
curves, tielines, spinodal curves and critical points in the
$(\fhphi,\fhrho)$ plane, for three values of $\chi$. These plots give
an appealing geometric insight into the problem of phase separation in
this system. For instance, the slope of the tielines indicates a size
partitioning effect: the more dense phases are enriched in long
chains.

The heavy line in Fig.~\ref{pbw_fig1}(a)--(c) is the dilution line
constraint $\fhnmo=\fhnmo\pn$ or $\fhphi=L_N\fhrho$. As discussed
already, we are interested mainly in systems whose mean composition
lies on this line. (With $\chi$ as an extra variable, this line
becomes a plane ($\fhphi=L_N\fhrho, \chi$) in the space
$(\fhphi,\fhrho,\chi)$.)  Not all of the phase behavior shown in
Fig.~\ref{pbw_fig1} is then accessible. The extremities of phase
separation on the $\fhphi=L_N\fhrho$ line are points where phase
separation just starts to occur, and the locus of these points in the
$(\fhphi=L_N\fhrho, \chi)$ plane defines the cloud curve, which bounds
the phase separation region. This is shown in
Fig.~\ref{pbw_fig1}(d). This plot also includes the spinodal stability
curve and the critical point from eqs.~\eqref{spineq2}
and~\eqref{criteq2}. The spinodal curve and the cloud curve touch at
the critical point, which no longer lies at the minimum of
either. This distorted behavior is a well known feature of
polydisperse systems.  Here it is seen to be due to the way that
\emph{regular} phase behavior in $(\fhphi,\fhrho,\chi)$ space is cut
through by the dilution line constraint. Note that only two moments
are needed to understand this qualitative effect, and that
Fig.~\ref{pbw_fig1} gives direct geometrical insight into it, which
would be very hard to extract from any exact solution based
on~\(exact_dens_distrib) and~\(effective_prior).

\begin{figure}
\begin{center}
\leavevmode
\epsfig{file=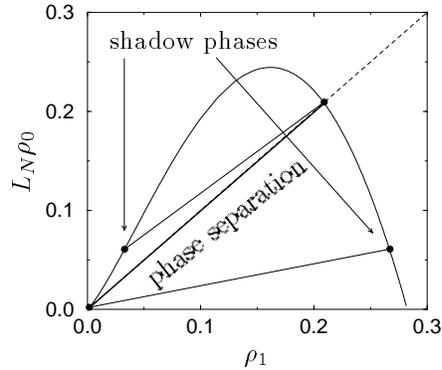}
\end{center}
\caption[?]{The phase separation region, and compositions of the
shadow phases corresponding to the two cloud points, for $\chi=0.62$
(compare Fig.~\ref{pbw_fig1}(c)).}
\label{pbw_fig1x}
\end{figure}

\begin{figure}
\begin{center}
\leavevmode
\epsfig{file=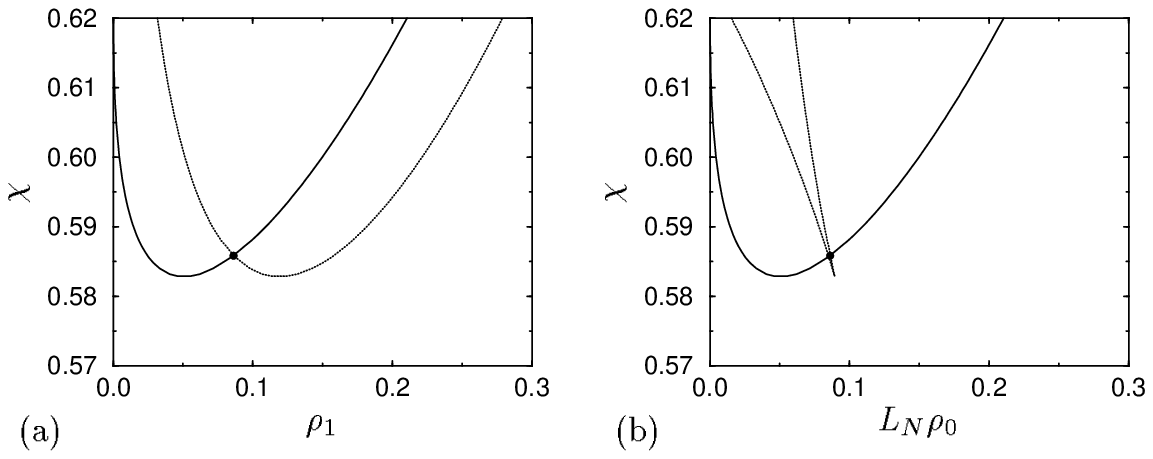}
\end{center}
\caption[?]{Phase behavior in $(\fhphi=L_N\fhrho,\chi)$ plane showing (a)
$\fhphi$-projected shadow curve, and (b) $\fhrho$-projected shadow curve.}
\label{pbw_fig2}
\end{figure}

The compositions of the phases that just start to appear as the phase
separation region is entered do not in general lie on the dilution
line $\fhphi=L_N\fhrho$. These phases are the ``shadows''; they lie at
the other ends of the cloud point tielines in
Fig.~\ref{pbw_fig1}(a)--(c). This is illustrated for $\chi=0.62$ in
Fig.~\ref{pbw_fig1x}.  The shadow phase compositions may be projected
onto the ($\fhphi=L_N\fhrho$, $\chi$) plane to give shadow curves, but
the projection is not unique. The shadow curve obtained by ignoring
the value of $\fhrho$ and projecting onto the $\fhphi$-axis, for
example, is shown in Fig.~\ref{pbw_fig2}(a). That obtained by ignoring
the value of $\fhphi$ and projecting onto the $\fhrho$-axis is shown
in Fig.~\ref{pbw_fig2}(b). The different shapes are due to the fact
that the locus of shadow phase compositions in general only intersects
the $(\fhphi=L_N\fhrho,\chi)$ plane at a single point---the critical
point.

In conclusion, we emphasize again that all the curves shown in
Fig.~\ref{pbw_fig1}(d) and Fig.~\ref{pbw_fig2} are exact, even though
they have been constructed from the phase behavior shown in
Fig.~\ref{pbw_fig1}(a)--(c) which is itself a projection of the true
phase behavior of the fully polydisperse system.

\subsection{Copolymer with chemical polydispersity}
\label{sec:copolymer}

Next, we apply the moment free energy formalism to random
AB-copolymers in the melt or in
solution~\cite{Bauer85,RatWoh91,NesOlvCri93}. For simplicity, we
assume that the copolymer chain {\em lengths} are monodisperse, so
that each chain contains an identical number $L$ of monomers. Even
with this simplification, the problem is quite challenging because, as
we shall see, a given parent can split into an arbitrarily large
number of phases as the interaction strength is increased.

For this problem we choose as our polydisperse feature, $\sig$, the
difference between the fractions of A and B-monomers on a chain. This
takes values in the range $\sig=-1\ldots 1$, with the extreme values
$\pm 1$ corresponding to pure A and B chains, respectively. We now
measure volumes in units of the chain volume (which is $L$ times
larger than the monomeric volume used in the previous example).  The
total density $\mn=\intsig\rhosig$ is then just the copolymer volume
fraction $\phi$. (Note that for the case of length polydispersity,
$\phi$ was related to the {\em first} moment of the {\em length}
distribution; here it is the {\em zeroth} moment of the distribution
of {\em chain compositions} $\sig$.)  Within Flory-Huggins theory, the
free energy density (in units of $\kB T$ per chain volume) is
\be
f = 
\intsig\rhosig[\ln\rhosig-1] + \frac{L}{\ls}(1-\mn)\ln(1-\mn) + 
L\tilde\chi\mn(1-\mn) - L\tilde\chiab\mo^2
\label{f_copol}
\ee
The interaction between A and B monomers enters through $\mo=\intsig
\sig\rhosig$. In~\eqref{f_copol}, we have specialized to the case where
the interactions of the A and B monomers with each other and with
the solvent are symmetric in A and B; otherwise, there would be an
additional term proportional to $\mn\mo$. We have also allowed the
solvent to be a polymer of length $\ls$; the case of a monomeric
solvent is recovered for $\ls=1$. In the following, we absorb factors
of $L$ into the interaction parameters by setting
\[
\chi=L\tilde\chi, \qquad \chiab=L\tilde\chiab
\]
Defining also $r=L/\ls$, the excess free energy takes the form
\be
\fexc = r(1-\mn)\ln(1-\mn) + \chi\mn(1-\mn) - \chiab\mo^2.
\label{fexc_copol}
\ee
The moment free energy~\(mom_free_en), with only $\mn$ and $\mo$
retained, is then
\be
\fmom = \mhatn\mn + \mhato\mo - \mn + 
r(1-\mn)\ln(1-\mn) + \chi\mn(1-\mn) - \chiab\mo^2.
\label{fmom_copol}
\ee
If extra \moms\ with weight functions $\wi$ are included, these simply
add a term $\mhati\mi$ each.

\subsubsection{Copolymer without solvent}
\label{sec:dense_copolymer}

We first consider the special case of a copolymer melt, for which the
overall density (\ie, copolymer volume fraction) is constrained to take everywhere its maximum
value $\mn=1$. The free energy~\(fmom_copol) then simplifies to
\be
\fmom = \mhatn + \mhato\mo - \chiab\mo^2
\label{fmom_dense_copol}
\ee
up to irrelevant constants~\cite{via_const_p}. Note that the same
expression for the excess free energy of the {\em dense} system is
obtained for a model in which the interaction energy between two
species $\sig,\sig'$ varies as $(\sig-\sig')^2$:
\be
\fexc = \frac{1}{2}\chiab\intsig d\sig'\,\rhosig \rh(\sig')
(\sig-\sig')^2 = - \chiab\mo^2 + \intsig \rhosig \sig^2
\label{fexc_dense_copol_alt}
\ee
The second term on the r.h.s.\ is linear in $\rhosig$ and can be
discarded for the purposes of phase equilibrium calculations. The
model~\(fmom_dense_copol) can therefore also be viewed as a simplified
model of chemical fractionation, with $\sig$ being related to
aromaticity, for example, or another smoothly varying property of the
different polymers. Eq.~\(fexc_dense_copol_alt) suggests that the
model should show fractionation into an ever-increasing number of
phases as $\chiab$ is increased: only the entropy of mixing prevents
each value of $\sig$ from forming a phase by itself, which would
minimize this form of $\fexc$.  It is therefore an interesting test
case for the moment free energy approach (and the method of adding
further \moms), yet simple enough for exact phase equilibrium
calculations to remain feasible~\cite{Bauer85,RatWoh91,NesOlvCri93},
allowing detailed comparisons to be made.

\begin{figure}
\vspace*{-3.3mm}
\begin{center}
\epsfig{file=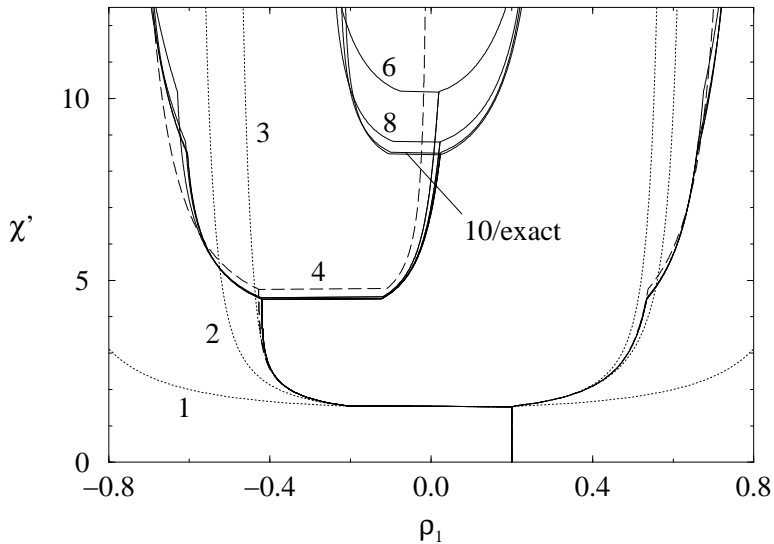,width=10cm}
\end{center}
\vspace*{-0.2cm}
\caption{Coexistence curves for a parent distribution with
$\mo\pn=0.2$. Shown are the values of $\mo$ of the coexisting phases;
horizontal lines guide the eye where new phases appear. Curves are
labeled by $n$, the number of \moms\ retained in the moment free
energy. Predictions for $n=10$ are indistinguishable from an exact
calculation (in bold).
\label{fig:dense_copol_powers}
}
\end{figure}

As a concrete example, we now consider phase separation from parent
distributions of the form $\rhonsig\propto\exp(\gamma\sig)$ (for
$-1\le\sig\le 1$, otherwise zero). The shape parameter $\gamma$ is
then a fixed function of the parental first moment density $\mo\pn =
\intsig\sig\rhonsig$. Fig~\ref{fig:dense_copol_powers} shows the exact
coexistence curve for $\mo\pn=0.2$, along with the predictions from
our moment free energy with $\L$ \moms\ ($\m_i=\intsig\sig^i\rhosig$,
$i=1\ldots \L$) retained. Comparable results are found for other
$\mo\pn$. Even for the minimal set of \moms\ ($n=1$) the point where
phase separation first occurs on increasing $\chiab$ is predicted
correctly (this is the cloud point for the given parent). As more
moments are added, the calculated coexistence curves approach the
exact one to higher and higher precision. As expected, the precision
decreases at high $\chiab$, where fractionated phases proliferate; in
this region, the number of coexisting phases predicted by the moment
method increases with $n$.  However, it is not always equal to $n+1$,
as one might expect from a naive use of Gibbs' phase rule; three-phase
coexistence, for example, is first predicted for $n=4$. In fact, for
fractionation problems such as this (but not more generally), study of
the low temperature limit (large $\chiab$) suggests that to obtain
$n+1$ phases one may have to include up to $2n$ moments.  We show
below that using localized weight functions (rather than powers of
$\sigma$) for the extra moments can reduce this number, but also gives
less accurate coexistence curves.

\begin{figure}
\begin{center}
\epsfig{file=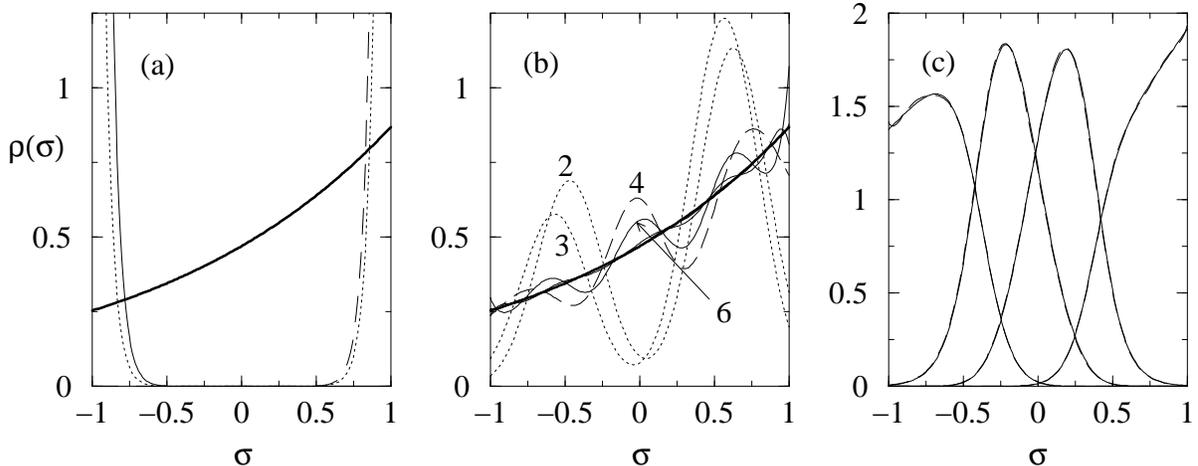, width=16cm}
\end{center}
\caption{The lever rule in phase coexistences calculated from the
moment free energy. We consider the same scenario as in
Fig.~\protect\ref{fig:dense_copol_powers} and take a horizontal cut
through that figure, varying $n$, the number of \moms\ retained in the
moment free energy, at fixed $\chiab=10$. (a) Solid and dashed lines:
The two daughter phases calculated for $n=1$ (both are exponential,
but with different signs for the exponent). Dotted line: The
corresponding total density distribution
$\rho\tot(\sig)=\sum_{\al=1,2} \phal\rhoalsig$. This deviates (rather
drastically, in this case) from the parent $\rhonsig$ (in bold); the
moment calculation with $n=1$ enforces only that the means $\mo$ of
the two distributions be equal. (Their normalizations $\mn$ are
trivially equal, because we are considering the dense limit $\mn=1$.)
(b) Total density distributions $\rh\tot(\sig)$ for increasing values
of $n$. Curves are labelled by the value of $n$; the curves for $n=8$
and $n=10$ (not labelled) are almost indistinguishable from the parent
(in bold). As $n$ increases, $\rh\tot(\sig)$ converges to the parent,
being forced to agree with it in an increasing number of its
moments. (c) Density distributions of the four daughter phases at
$\chiab=10$. For each, both the exact result and that of the moment
free energy calculation with $n=10$ are shown; they cannot be
distinguished on the scale of the plot.
\label{fig:lever_rule}
}
\end{figure}

In Fig.~\ref{fig:lever_rule}, we show how, for a given value of
$\chiab$, the lever rule is satisfied more and more accurately as the
number of \moms\ $n$ in the moment free energy is increased: the total
density distribution $\rh\tot(\sig)=\sum_\al \phal \rhoalsig$
approaches the parent $\rhonsig$. This is by design, because
the moment method forces the \moms\ $\mi=\intsig \wi \rh\tot(\sig)$
of $\rh\tot(\sig)$ to be equal to
those of the parent, for $i=1\ldots n$; as $n$ is increased, therefore,
$\rh\tot(\sig)$ approaches $\rhonsig$. 

The above results for the dense copolymer model raise two general
questions. First, how large does $n$ (the number of \moms\ retained)
have to be for the moment method to give reliable results for the
coexisting phases (and, in particular, the correct number of phases)?
And second, how should the weight functions for the extra \moms\ be
chosen? Regarding the first question, note first that the theoretical
framework developed so far only says that the exact results (and
therefore the correct number of phases) will be approached as $n$ gets
large; it does not say for which value of $n$ a reasonable
approximation is obtained. In fact it is clear that no ``universal''
(problem-independent) value of $n$ can exist. The model studied in
this section already provides a counterexample: As $\chiab$ is
increased, the (exact) number $p$ of coexisting phases increases
without limit; therefore the minimum value of $n$ ($n=p-1$) required
to obtain this correct number of phases from a moment free energy
calculation can also be arbitrarily large.

Nevertheless, we see already from Fig.~\ref{fig:dense_copol_powers}
that, for the current example, the robustness of the results to the
addition of extra \moms\ provides a reasonable qualitative check of
the accuracy of the coexistence curves. To study the convergence of
the moment free energy results to the exact ones more quantitatively,
especially in cases where the latter are not available for comparison,
we need a measure of the deviation between $\rh\tot(\sig)$ and
$\rhonsig$. To obtain a dimensionless quantity that does not scale
with the overall density of the parent, we consider the normalized
distributions $\prob\tot(\sig)=\rh\tot(\sig)/\rh\tot$ and
$\probnsig=\rhonsig/\rh\pn$ and define as our measure of deviation the
``log-error''
\be
\logerr = \intsig \probnsig
\left(\ln\frac{\probnsig}{\prob\tot(\sig)}\right)^2
\label{logerr}
\ee
This becomes zero only when $\prob\tot(\sig)=\probnsig$; otherwise it
is positive. When the deviations between the two distributions are
small, the logarithm becomes to leading order the relative deviation
$\prob\tot(\sig)/\probnsig-1$ and we can identify $\logerr^{1/2}$ as
the root-mean-square relative deviation. We work with the logarithm
rather than directly with the relative deviation because the former
gives more sensible behavior for larger deviations; in particular,
isolated points where $\prob\tot(\sig)$ is nonzero while $\probnsig$
is close to zero do not lead to divergences of
$\logerr$~\cite{not_cross_entropy}.

\begin{figure}
\begin{center}
\epsfig{file=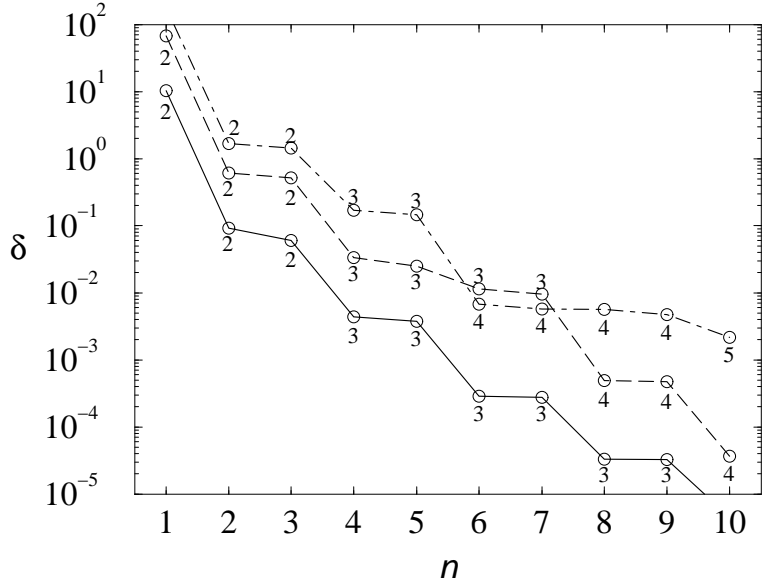, width=10cm}
\end{center}
\caption{Log-error $\logerr$ as a function of $n$ for three fixed
values of $\chiab=5, 10, 15$ (bottom to top), for the same scenario as
in Fig.~\protect\ref{fig:dense_copol_powers}. The number of phases is
indicated next to the curves. Note that in all three cases the correct
number of phases is reached for $\logerr$ of order $10^{-3}$.
\label{fig:logerr_chi5_10_15}
}
\end{figure}

Fig.~\ref{fig:logerr_chi5_10_15} shows the $n$-dependence of the
log-error $\logerr$ for three fixed values of $\chiab$, with the same
parent as in
Fig~\ref{fig:dense_copol_powers}~\cite{logerr_not_monotonic}. The
correct number of phases (3, 4, 5, respectively) is reached at values
of $\logerr\approx 4\times 10^{-3}$, $5\times 10^{-4}$, $2\times
10^{-3}$ for $\chi=5$, 10, 15, respectively. This suggests that the
log-error $\logerr$ might generally be a useful heuristic criterion
for guiding the choice of $n$, with values of $\logerr$ around
$10^{-4}$ (corresponding to an average deviation between
$\prob\tot(\sig)$ and $\probnsig$ of around 1\%) ensuring that the
correct number of phases has been detected.  Note that, although for
the current example we did know the correct number of phases (because
the simplicity of the problem made an exact calculation feasible), the
definition of $\logerr$ does {\em not} require us to know in advance
any properties of the exact solution of the coexistence problem. Also,
note that in real physical materials it is extremely unlikely for the
parent or ``feedstock'' distribution $\rhonsig$ to be known to better
than 1\% accuracy. Hence even for systems, such as polymers, where a
mean-field (and thus truncatable) free energy model is thought to be
highly accurate, deviations from the lever rule, at about the 1\%
level, will almost never be the main source of uncertainty in phase
diagram prediction.
 
We now turn to the second general question, regarding the choice of
weight functions for the extra \moms. Comparing~\(flexible_prior) with
the formally exact solution~\(effective_prior) of the coexistence
problem tells us at least in principle what is required: the
log-ratio $\ln\effprior/\rhonsig$ of the ``effective prior'' and the
parent needs to be well approximated by a linear combination of the
weight functions of the extra \moms. However, the effective prior is
unknown (otherwise we would already have the exact solution of the
phase coexistence problem), and so this criterion is of little
use~\cite{extra_moments_asymptotics}. 

\begin{figure}
\begin{center}
\epsfig{file=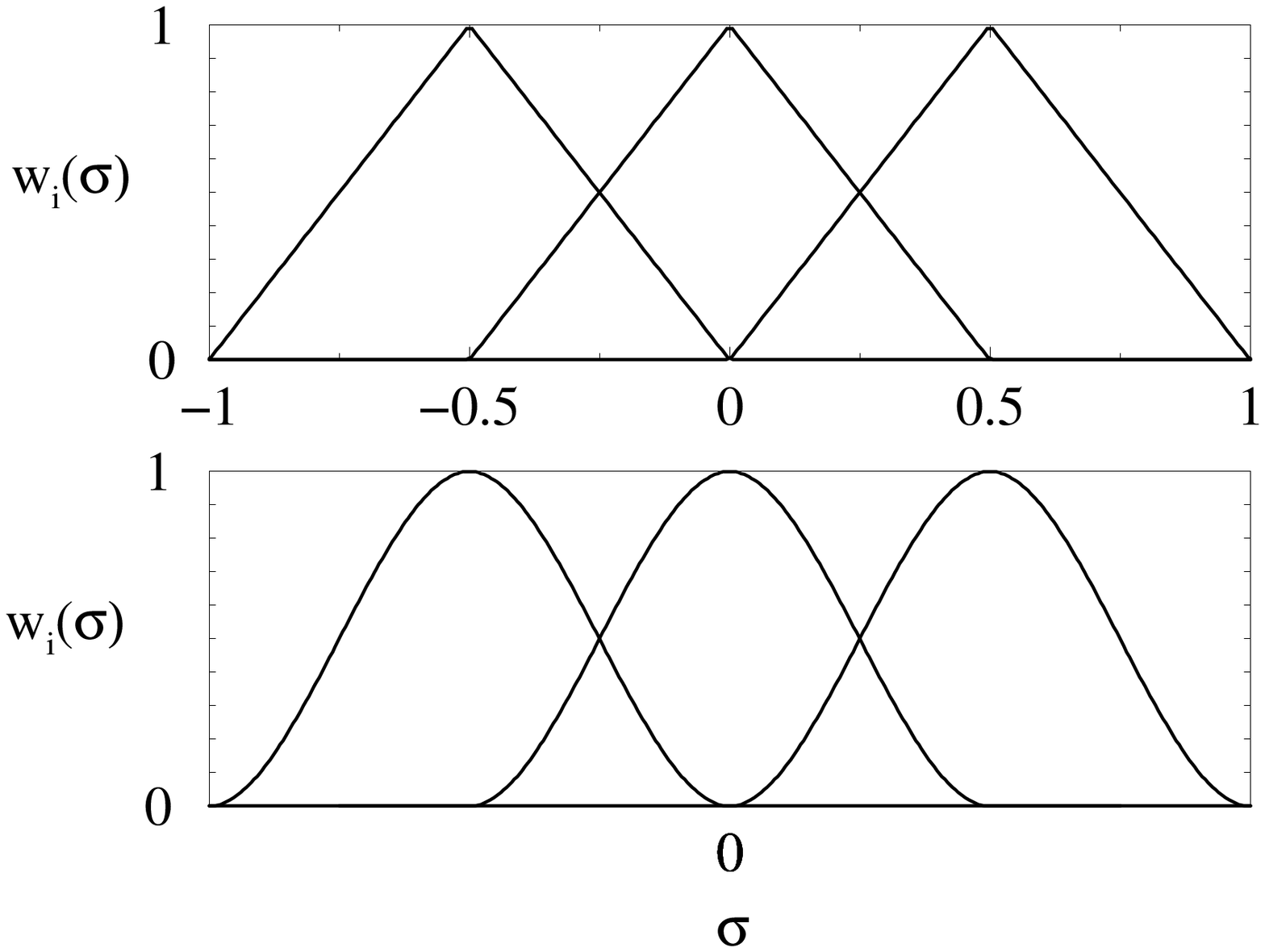,width=8cm}%
\begin{picture}(0,0)\put(-220,170){\mbox{(a)}}\end{picture}%
\hspace*{6mm}
\epsfig{file=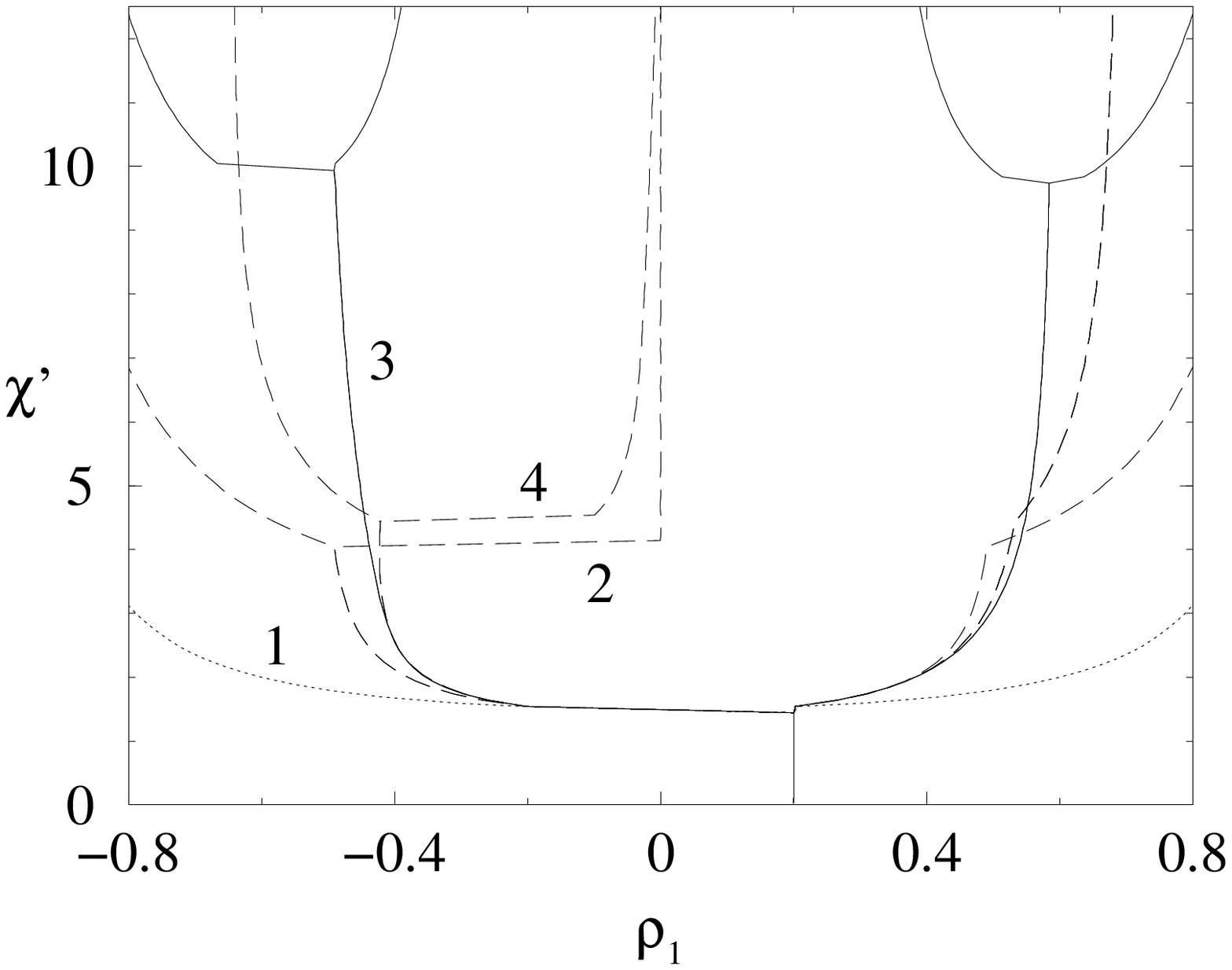,width=8cm}%
\begin{picture}(0,0)\put(-228,170){\mbox{(b)}}\end{picture}%

\vspace*{3mm}

\epsfig{file=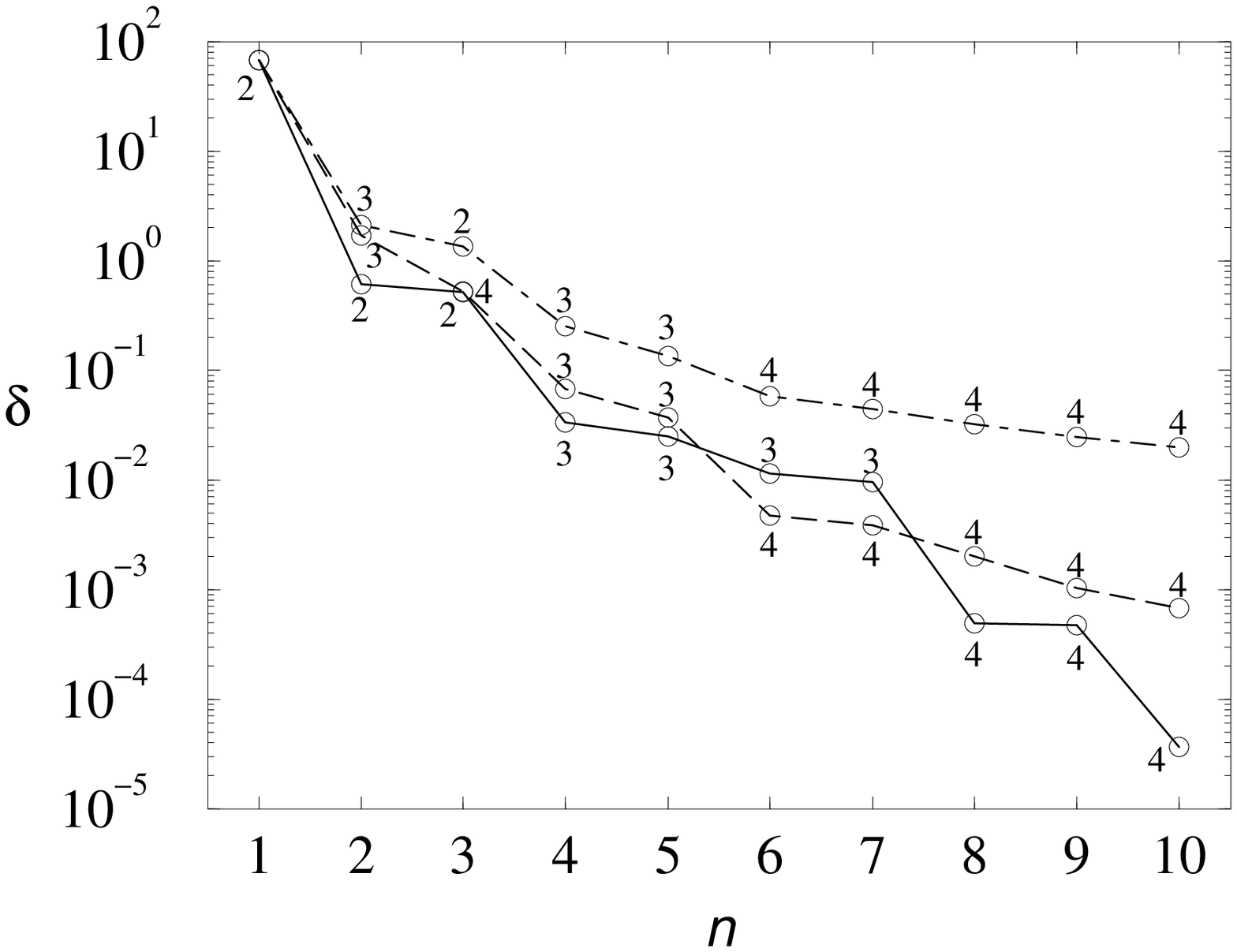,width=8cm}%
\begin{picture}(0,0)\put(-230,163){\mbox{(c)}}\end{picture}%
\end{center}
\caption{The effect of the weight functions for extra \moms\ on
coexistence calculations. (a) The ``triangle`` and ``bell'' weight
functions, for the case $n-1=3$. (b) Coexistence curves for the
triangle weight functions, for the same parent as in
Fig.~\protect\ref{fig:dense_copol_powers}. Note that multi-phase
coexistence is generally detected for smaller $n$ than in
Fig.~\protect\ref{fig:dense_copol_powers}, where power-law weight
functions were used for the extra moments. On the other hand, the
predicted number of phases now no longer varies monotonically with
$n$. (c) Dependence of log-error $\logerr$ on $n$ for fixed
$\chiab=10$ and three choices of weight functions for the extra \moms:
Power-law (solid line), triangle (dashed), bell (dot-dashed). The
number of phases is marked next to the curves.
\label{fig:triangle_bell}
}
\end{figure}

To make progress, let us try to gain some intuition from the example
at hand (dense random copolymers). So far, we have simply taken
increasing powers of $\sig$ for the extra weight functions. On the
other hand, looking at Fig.~\ref{fig:lever_rule} one might suspect
that {\em localized} weight functions might be more suitable for
capturing the fractionating of the parent into the various daughter
phases. We therefore tried the simple ``triangle'' and ``bell'' weight
functions shown in Fig.~\ref{fig:triangle_bell}(a). For a given $n$,
there are $n-1$ extra \moms\ (the first \mom, with weight function
$\w_1(\sig)=\sig$, is prescribed by the excess free energy) and we
take these to be evenly spaced across the range $\sig=-1\ldots 1$,
with overlaps as indicated in Fig~\ref{fig:triangle_bell}(a). (For the
case of triangular weight functions, the log ratio
$\ln\rhosig/\rhonsig = \sum_i\mhati\wi$ between the density
distribution of any phase and that of the parent is thus approximated
by a ``trapezoidal'' function across $n$ equal intervals spanning
$-1\ldots 1$.)  Fig.~\ref{fig:triangle_bell}(b) shows the
corresponding coexistence curves for the triangle case with
$n=1,2,3,4$. Note that multi-phase coexistence is generally detected
for smaller $n$ than for the case of power-law weight functions
(Fig.~\protect\ref{fig:dense_copol_powers}), in agreement with the
expectations outlined above. However, the predicted number of phases
does not vary monotonically with $n$: in the range of $\chiab$ shown
in Fig.~\ref{fig:triangle_bell}(b), up to four phases are predicted
for $n=3$, whereas $n=4$ gives no more than three.  This effect
illustrates a significant difference between power-law and triangle
weight functions. The former, but not the latter, make up an {\em
incremental set}: increasing $n$ by one corresponds to adding a new
weight function while leaving the existing set of weight functions
unchanged. The family~\(pfamily) of accessible density distributions
$\rhosig$ is thus enlarged without losing any of its former members.
Intuitively, one then expects that the number of phases cannot
decrease as $n$ is increased. But for nonincremental sets, such as
triangle weight functions (and bells, which give qualitatively similar
results to triangles), each increase in $n$ creates a larger family
but completely discards the previous one, so that there is then no
reason for the results to be monotonic in $n$.

We regard the fact that the number of phases can go up and down with
increasing $n$ as a disadvantage of non-incremental weight
functions. In the current example, we also find that they generally
give higher values of the log-error than power-law weight functions;
see Fig.~\ref{fig:triangle_bell}(c). Finally, incremental sets of
weight functions also have the benefit that the results for $n$ can be
used to initialize the calculation for $n+1$, thus helping to speed up
the numerics (though we have not yet exploited this property).  On the
basis of these observations, we would generally recommend the use of
incremental sets of weight functions for extra \moms. Beyond this, the
optimal choice of extra weight functions remains largely an open
problem. One avenue worth exploring in future work might be the
possibility of choosing extra weight functions {\em adaptively}. One
could imagine monitoring the log-error in a phase coexistence
calculation, and increasing the number of extra moments by one every
time it becomes larger than a certain threshold ($10^{-4}$, say). The
new weight function could then be chosen as a best fit to the current
lever-rule violation $\ln\rh\tot(\sig)/\rhonsig$ (for example as a
linear combination of weight functions from some predefined
``pool''). We are planning to investigate this method in future work.

To conclude this section on the dense random copolymer model, we
briefly discuss the spinodal criterion, and ask whether critical or
multi-critical points can exist for a general parent distribution
$\rhonsig$ (with $\mn\pn=1$). The criterion~\(spinodal_and_critical),
applied to our one-moment free energy $\fmom(\mo)$ becomes simply
\be
\fmom(\mo)=\order\left((\mo-\mo\pn)^{l+1}\right)
\ee
where $l=2$ for a spinodal and $l=2n-1$ for an $n$-critical point.
From eq.~\(fmom_dense_copol), we have that
\be
\fmom = \mhatn + \mhato\mo - \chiab\mo^2 = -\smom - \chiab\mo^2
\label{fmom_again}
\ee
The excess part $-\chiab\mo^2$ is quadratic in $\mo$ and therefore
only enters the spinodal criterion; the additional conditions for
critical points only depend on the moment entropy $\smom = - \mhatn -
\mhato\mo$. Because of the dense limit constraint $\mo=1$, $\mhatn$ is
related to $\mhato$ by
\[
- \mhatn = \ln\intsig\rhonsig e^{\mhato\sig} \equiv h(\mhato)
\]
As expected from the general discussion in Sec.~\ref{sec:equivalence},
$-\mhatn$ is thus the cumulant generating function $h(\mhato)$ of
$\rhonsig$ ($\equiv \probnsig$ in the dense limit we are considering),
introduced in eq.~\(h_def) within the combinatorial approach. The
moment entropy $\smom=h-\mhato\mo$ is its Legendre transform; its
behavior for $\mo\approx\mo\pn$ reflects that of $h$ for small
$\mhato$. We can thus use the expansion $h=\sum_{j=1}^\infty (c_j/j!)
\mhato^j$, where the $c_j$ are the cumulants of $\rhonsig$; $c_1\equiv
\mo\pn$ is its mean, $c_2$ its variance etc.. Using the properties of
the Legendre transform, it then follows that the leading term of the
moment entropy is $\smom=(\mo-\mo\pn)^2/(2c_2)$. Inserting this
into~\(fmom_again), we see immediately that the spinodal condition
becomes
\be
2\chiab c_2=2\chiab\left(\mt\pn-(\mo\pn)^2\right)=1
\label{dense_spinodal}
\ee
To find the additional conditions for critical points, consider what
happens if the cumulants $c_3$, \ldots, $c_l$ are zero, hence
$h=\mo\pn\mhato+c_2\mhato^2/2 + \order(\mhato^{l+1})$.  It can then
easily be shown that the moment entropy behaves as
$\smom=(\mo-\mo\pn)^2/(2c_2) + \order((\mo-\mo\pn)^{l+1})$;
conversely, one can show that this behavior of $\smom$ implies that
the cumulants $c_3$, \ldots, $c_l$ are zero. We thus have the simple
condition that for an $n$-critical point to be observed, the cumulants
$c_3$, $\ldots$, $c_{2n-1}$ of the parent distribution have to be
zero; the value of $\chiab$ at this critical point is given by the
spinodal condition~\(dense_spinodal). (For an ordinary critical point,
$n=2$, the condition $c_3=0$ is well known in this
context~\cite{Bauer85}.) We therefore come to the somewhat surprising
conclusion that even in a very simple polydisperse system such as this
-- with a single \mom\ occurring in the excess free energy --
multi-critical points of arbitrary order can occur, at least in
principle. In practice, the required fine-tuning of the parent will of
course make experimental study of these points difficult.

\subsubsection{Copolymer with solvent}
\label{sec:copolymer_with_solvent}

We now consider the random copolymer model in the presence of solvent,
\ie, for a copolymer volume fraction $\mo<1$. We are not aware of
previous work on this model in the literature, but will briefly
discuss below the link to models of homopolymer/copolymer
mixtures~\cite{Leibler81}. The excess free energy~\(fexc_copol) then
depends on two \moms, rather than just one as in all previous
examples. For simplicity, we restrict ourselves to the case of a
neutral solvent which does not in itself induce phase separation; this
corresponds to $\chi=0$, making the excess free energy
\be
\fexc = r(1-\mn)\ln(1-\mn) - \chiab\mo^2
\label{fexc_copol_neutral}
\ee
This is still sufficiently simple that the exact spinodal
condition~\(mom_spinodal) -- for a system with a general density
distribution $\rhosig$ -- can be worked out analytically; one finds
\be
2\chiab = \frac{r\mn + 1-\mn}{r(\mt\mn-\mo^2)+\mt(1-\mn)}
\label{copol_solvent_spinodal}
\ee
In the dense limit $\mn\to1$, the earlier result~\(dense_spinodal) is
recovered. To study critical and multi-critical points, we
use the moment free energy corresponding
to~\(fexc_copol_neutral),
\be
\fmom = \mhatn\mn + \mhato\mo - \mn + r(1-\mn)\ln(1-\mn) - \chiab\mo^2
\label{fmom_copol_neutral}
\ee
and start from the criterion~\(spinodal_and_critical). The moment
chemical potentials are given by
\begin{eqnarray}
\mu_0 &=& \mhatn -r [\ln(1-\mn)-1]
\nonumber\\
\mu_1 &=& \mhato -2\chiab\mo
\label{copol_mom_chem_pot}
\end{eqnarray}
and the curve $\rhovect(\epsilon)$ referred to
in~\(spinodal_and_critical) -- along which the critical phases merge
-- is simply parameterized in terms of $\mhatn(\epsilon)$ and
$\mhato(\epsilon)$. The restriction that the curve must pass through
the parent at $\epsilon=0$ means that $\mhatn(0)=\mhato(0)=0$, so to
get the small $\epsilon$ behavior of the chemical
potentials~\(copol_mom_chem_pot) we can expand for small $\mhatn$ and
$\mhato$. To simplify the algebra, we restrict ourselves to parent
distributions $\rhonsig$ which are symmetric about $\sig=0$. Using
this symmetry and the fact that $\epsilon$ is simply a dummy parameter
which can be reparameterized arbitrarily, we can then set
$\mhato=\epsilon$ and
$\mhatn=a_2\epsilon^2+a_4\epsilon^4+\order(\epsilon^6)$.

To have a spinodal, the moment chemical
potentials~\(copol_mom_chem_pot) must differ from those of the parent
by no more than $\order(\epsilon^2)$. We find that this gives the
condition $2\chiab=1/\mt\pn$, in agreement
with~\(copol_solvent_spinodal): symmetric parents have $\mo\pn=0$. For
a critical point, the chemical potential difference must be decreased
to $\order(\epsilon^3)$. This gives no new condition (but determines
$a_2$), implying that parents with symmetric distributions are
automatically critical at their spinodal. This result has its origin
in the symmetry $\sigma \to -\sigma$ of the excess free energy, and
can be understood by considering the $\mn$, $\mo$, $\chiab$ phase
diagram of the moment free energy~\(fmom_copol_neutral). In this phase
diagram there is a two-dimensional spinodal surface, and within this
surface lies a line of critical points. The symmetry $\sigma \to
-\sigma$ (corresponding to $\mo\to-\mo$, $\mhato\to-\mhato$) forces
this critical line to lie within the plane $\mo=0$ of symmetrical
distributions, implying that for such distributions spinodal and
critical points coincide.

\begin{figure}
\begin{center}
\epsfig{file=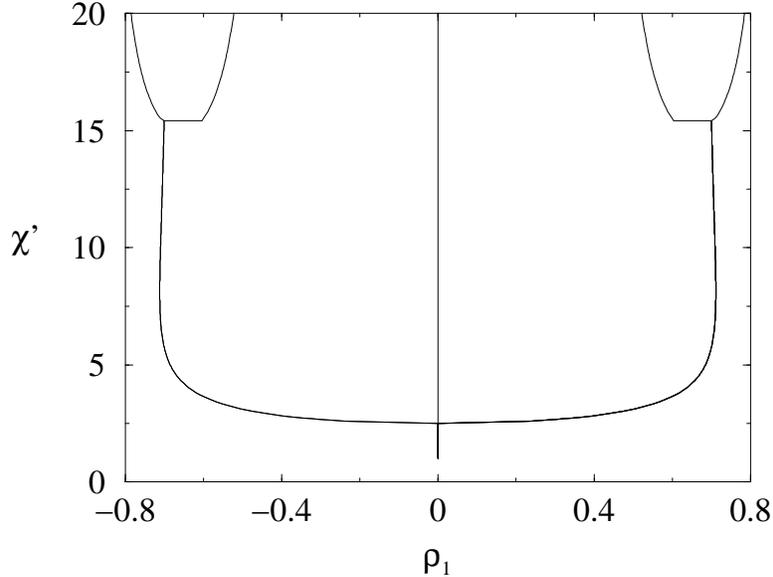, width=10cm}
\end{center}
\caption{Coexistence curve for copolymer with solvent; the parent has
a uniform density distribution $\rhonsig=\mbox{const.}=\mn\pn/2$ with
overall density (\ie, copolymer volume fraction) $\mn\pn=0.6$. The
results were obtained from the moment free energy with $n=11$ \moms\
retained (weight functions $\wi=\sig^i$, $i=0\ldots 10$). Note the
tricritical point at $\chiab=2.5$: The parent splits continuously into
three phases.
\label{fig:copolymer_tricrit_coex}
}
\end{figure}

For a tricritical point, finally, where the chemical potential
difference must be further reduced to $\order(\epsilon^3)$, we find
after some algebra the condition
\[
\m_4\pn = \frac{3r(\mt\pn)^2}{r\mn\pn + 1-\mn\pn}
\]
As an example, consider a uniform parent distribution
$\rhonsig=\mbox{const}$, which can be written as $\rhonsig=\mn\pn/2$,
using the fact that $\mn\pn=\int_{-1}^1\!d\sig\,\rhonsig$. Then
$\mt\pn=\mn\pn/3$ and $\m_4\pn=\mn\pn/5$ and so a tricritical point
occurs if the overall density (\ie, the copolymer volume fraction) is
$\mn\pn=3/(2r+3)$.  Fig.~\ref{fig:copolymer_tricrit_coex} shows the
coexistence curve calculated for this parent (with $r=1$), which
clearly shows the tricritical point at the predicted value
$\chiab=1/(2\mt\pn)=r+3/2=2.5$. Our numerical implementation manages
to locate the tricritical point and follow the three coexisting phases
without problems; we take that as a signature of its
robustness~\cite{footnote_no_symmetry}. Note that the tricritical
point that we found is closely analogous to that studied by
Leibler~\cite{Leibler81} for a symmetric blend of two homopolymers and
a symmetric random copolymer that is, nonetheless, chemically
monodisperse (in the sense that $\sigma = 0$ for all copolymers
present). In fact, in our notation, the scenario of
Ref.~\cite{Leibler81} simply corresponds to a parent density of the
form $\rhonsig \sim \delta(\sig-1)+\delta(\sig+1)$, with the copolymer
($\sigma = 0$) now playing the role of the neutral solvent.

\begin{figure}
\begin{center}
\epsfig{file=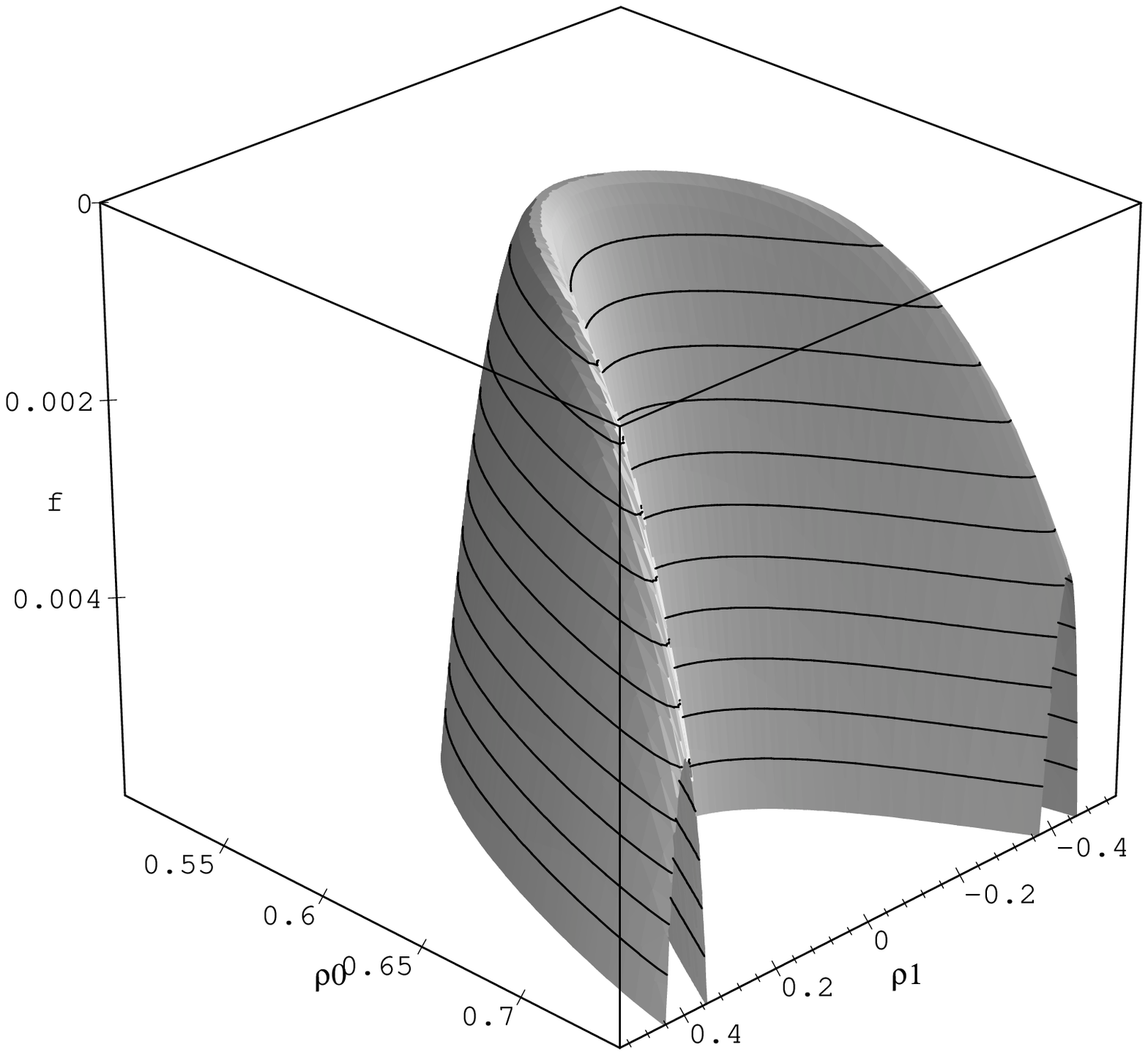, width=8.5cm}%
\begin{picture}(0,0)\put(-207,213){\mbox{(a)}}\end{picture}%
\epsfig{file=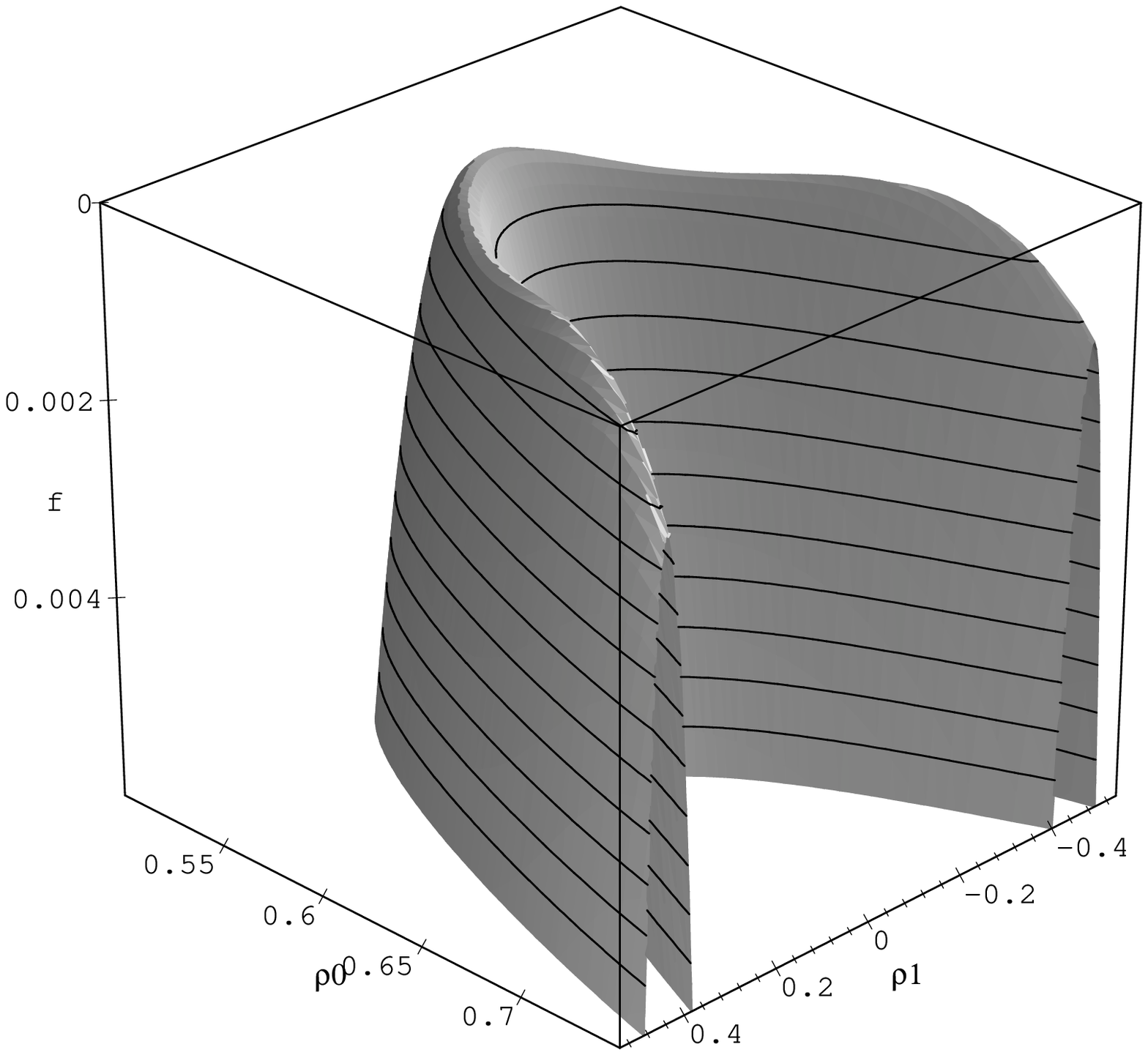, width=8.5cm}%
\begin{picture}(0,0)\put(-207,213){\mbox{(b)}}\end{picture}%
\end{center}
\caption{Moment free energy $\fmom(\mn,\mo)$ near the tricritical
point, for the same parent as in
Fig.~\protect\ref{fig:copolymer_tricrit_coex}. For ease of
visualization, the free energy surface is shown {\em upside down},
with $\fmom$ increasing in the downward direction. (a) At the
tricritical point $\chiab=2.5$, the parent is stable (note that linear
terms have been added to the free energy to make the tangent plane at
the parent horizontal). (b) As $\chiab$ is increased into the
three-phase region (here $\chiab=2.65$), the parent splits
continuously into three phases.
\label{fig:copolymer_tricrit_free_en}
}
\end{figure}

We conclude this section with another illustration of the geometrical
intuition which the moment free energy can provide. In
Fig.~\ref{fig:copolymer_tricrit_free_en}, we plot the moment free
energy $\fmom(\mn,\mo)$, eq.~\(fmom_copol_neutral), for the uniform
parent $\rhonsig$ studied above. Two values of $\chiab$ are shown: the
tricritical value ($\chiab=2.5$), and a value just in the three-phase
region ($\chiab=2.65$); the developing three minima of the free energy
can clearly be seen. Even though the underlying polydisperse system
has an infinite number of degrees of freedom $\rhosig$, the moment
method thus gives us a tool for understanding and visualizing the
occurrence of the tricritical point in terms of a simple free energy
surface with two independent variables.

\section{Conclusion}
\label{sec:conclusion}

Polydisperse systems contain particles with an attribute $\sigma$
which takes values in a continuous range (such as particle size in
colloids or chain length in polymers). They therefore have an infinite
number of conserved densities, corresponding to a density {\em
distribution} $\rhosig$. The fact that the free energy can depend on
all details of $\rhosig$ makes the analysis of the phase behavior
of such systems highly non-trivial. However, in many (especially
mean-field) models the {\em excess} free energy only depends on a
finite number of moment densities, \ie, (generalized) moments of
$\rhosig$; the only dependence on other details of $\rhosig$ is
through the ideal part of the free energy. For such models, which we
call truncatable, we showed how a {\em moment free energy} can be
constructed, which only depends on the moment densities appearing in
the excess free energy. Two possible approaches for the construction
of the moment free energy were described in
Sec.~\ref{sec:derivation}. The first, which we call the projection
method (Sec.~\ref{sec:projection}), is primarily geometrical; the
second is mainly combinatorial (Sec.~\ref{sec:combinatorial}) and
starts from first principles to derive an expression for the ideal
part of the free energy in terms of \moms. In
Sec.~\ref{sec:equivalence}, we showed that these two approaches give
essentially equivalent results.

In Sec.~\ref{sec:properties}, we explored the properties of the moment
free energy in detail. The moment free energy depends only on a finite
number of \moms, and can be used to predict phase behavior using the
conventional methods known from the thermodynamics of finite
mixtures. We showed that, for any truncatable model, this procedure
yields the same spinodals, critical and multi-critical points as the
original free energy, even though the latter depends on all details of
the density distribution $\rhosig$.  It also correctly predicts the
onset of phase coexistence, \ie, the cloud points and corresponding
shadows. Beyond the onset of phase coexistence (\ie, when one or more
coexisting phases occupy comparable volumes), the results from the
moment free energy are approximate, but can be made arbitrarily
accurate by retaining additional \moms\ in the moment free energy. We
also explained how to check the local and global stability of a phase
coexistence calculated from the moment free energy, and how the moment
free energy reflects overall phase behavior in the space of density
distributions $\rhosig$.

We discussed aspects of the numerical implementation of the moment
free energy method in Sec.~\ref{sec:implementation}, highlighting its
many advantages in terms of robustness, efficiency and stability. The
method allows violations of the lever rule, but then ensures that
these be kept small by adding $n-K$ extra moments (where $K$ is the
number of \moms\ actually present in the excess free energy of the
underlying truncatable model). The solution of the phase equilibrium
conditions thus proceeds in an $n$-dimensional space; but global
(tangent plane) stability can always be verified by minimimization (of
the tangent plane distance) in a space of dimension $K$. For $K>1$
exact minimization is not practicable, but standard search methods can
instead be employed; we chose a Monte-Carlo type algorithm for this.

We then surveyed several sample applications in
Sec.~\ref{sec:examples}. For length-polydisperse homopolymers (treated
within Flory-Huggins theory), we showed explicitly how the moment free
energy with only a single moment density (the volume fraction of
polymer) gives the correct spinodals and critical points. By retaining
the chain number density as an additional moment density, we could
rationalize the behavior of cloud point and shadow curves as cuts
through conventional two-species phase diagrams. As a second example,
the case of copolymers with chemical polydispersity was considered,
both without and with solvent. This gave us the opportunity to test
the performance of the moment free energy method for multi-phase
coexistence, and in particular to study how the accuracy of the
predictions in the coexistence region increases as more and more
\moms\ are retained.  In the presence of solvent, the same theory
predicts a tricritical point under appropriate conditions, which was
detected without problems, along with the associated three phase
region, by our numerical algorithm. A plot of the two-moment free
energy gave a simple geometrical picture of the origin of this
tricritical point.

The results for the copolymer model showed that the
log-error~\(logerr) can be used to decide when enough \moms\ have been
included to locate the correct number of phases and their
compositions. For general purposes we suggest a log-error tolerance of
order $10^{-4}$; this corresponds to limiting the rms lever-rule
violations at the predicted phase coexistence (found by global
minimization of the moment free energy) to 1\% or so of the parent
distribution. Given that in practice the parent is itself not known to
such high precision, these violations lie well within an acceptable
range of error. Indeed, every state of coexistence actually predicted
by our algorithm represents the {\em exact} phase equilibrium of {\em
some or other} parent, indistinguishable from the real one to within
experimental accuracy. Here, as elsewhere in this paper, ``exact''
means exactly as predicted by a truncatable model free energy; such
models are, of course, themselves approximate, which is another reason
not to expend resources reducing the log-error below the level
suggested above.

In summary, with exact results for cloud point and shadow curves,
critical points and spinodals, as well as refinably accurate
coexistence curves and multi-phase regions, the moment free energy
method allows rapid and accurate computation of the phase behavior of
many polydisperse systems. Moreover, by establishing the link to a
projected free energy $\fmom(\mi)$ which is a function of a finite set
of conserved \moms\ $\mi$, it restores to the problem much of the
geometrical interpretation and insight (as well as the computational
methodology) associated with the phase behavior of finite mixtures.
This contrasts with many procedures previously in use for truncatable
polydisperse systems
\cite{Bauer85,RatWoh91,NesOlvCri93,Solc93,SolKon95,GuaKinMor82}. Some
previous approximations to the problem have used (generalized) moments
as thermodynamic coordinates; see
\eg~\cite{IrvGor81,HenVan92,CotPra85,Bartlett97}.  The moment free
energy method provides a rational basis for these methods and, so long
as the generalized moments are correctly chosen, guarantees that many
properties of interest are found exactly. Note also that the
commonplace method of ``binning'' the $\sigma$-distribution into
discrete ``pseudo-components'' can be seen as a special case of the
moment method in which each weight function is zero outside the
corresponding bin; but since these moments do not, in general,
coincide with those contained in the excess free energy the advantages
described above are then lost.

We finish with some remarks about extensions to the moment method,
within and outside the sphere of phase diagram prediction. First,
methods of this type may extend to nontruncatable models, for which
the excess free energy {\em cannot} be written directly in terms of a
finite number of moments as in~\(free_en_decomp). For example, many
mean-field theories correspond to a variational minimization of the
free energy: $F\le \langle E\rangle_0 - TS_0$, where the subscript $0$
refers to a trial Hamiltonian~\cite{Feynman72}. In such a case, one
might choose to {\em first} make a physically motivated decision about
which (and how many) \moms\ $\mi$ to keep, and then include among the
variational parameters the ``transverse" degrees of freedom of the
density distribution $\rhosig$.  We have not pursued this approach,
though it may form a promising basis for future
developments. Secondly, we note that physical insight based on moment
free energies may increasingly play a part in understanding kinetic
problems. For example, one of us~\cite{warren99b} has argued that in
many systems the zeroth moment (mean particle density) can relax more
rapidly (by collective diffusion) than can the higher moments (by
interdiffusion of different species), and that this may lead to novel
kinetics in certain areas of the phase diagram. And finally, we point
to recent work~\cite{PagCatAck99} in which the moment method has been
extended, via density functional theory, to allow study of
inhomogeneous states; this opens the way to studying the effect of
polydispersity on interfacial tensions and other interfacial
thermodynamic properties.

{\em Acknowledgements:} We thank P.\ Bladon, N.\ Clarke, R.\ M.\ L.\
Evans, T.\ McLeish, P.\ Olmsted, I. Pagonabarraga, W.\ C.\ K.\ Poon
and R.\ Sear for helpful discussions. PS is grateful to the Royal
Society for financial support. This work was also funded in part under
EPSRC GR/M29696 and GR/L81185.

\appendix

\section{Moment (Gibbs) free energy for fixed pressure}
\label{app:const_p}

In this appendix, we explain how the construction of the moment free
energy is extended to scenarios where the pressure (rather than the
overall system volume) is held constant. One then describes the system
by the particle number distribution $N(\sig)=V\rhosig$; this is
normalized so that its integral gives the total number of particles,
$N=\intsig N(\sig)$. The relevant thermodynamic potential is the Gibbs
free energy $G=\min_V F+\Pi V$, which we now construct. For a
truncatable system, whose free energy {\em density} obeys
eqs.~(\ref{free_en_decomp_gen},\ref{puremoment_excess}) by definition,
the Helmholtz free energy can be written as
\[
F = T \intsig N(\sig) \left[\ln\frac{N(\sig)}{V}-1\right] + \tilde{F}
\]
where the excess free energy $\tilde{F}(N_i,V)=V\fexc(\mi)$ depends on
volume $V$ and on the moments $N_i=V\mi=\intsig N(\sig)\wi$ of the
particle number distribution. If $\tilde{F}=0$, a Legendre transform
gives the Gibbs free energy of an ideal mixture:
\[
G_{\rm id} = \min_V F_{\rm id} + \Pi V = T \intsig N(\sig)
\ln\frac{\Pi N(\sig)}{NT} = NT \intsig \nsig \ln\nsig + NT\ln\frac{\Pi}{T}
\]
Here $\nsig=N(\sig)/N=\rhosig/\rho$ is the normalized particle
distribution. In the general case, we split off the ideal part
\[
G = G_{\rm id} + \tilde{G} = \min_V F + \Pi V 
\]
and obtain for the excess Gibbs free energy
\[
\tilde{G}=\min_{V} NT\ln\frac{NT}{\Pi V} + \tilde{F}(N_i,V) + \Pi V
\]
This is a function of $N$ and the $N_i$ (as well as the fixed control
parameters $\Pi, T$, which we suppress in our notation). We can
therefore write the excess Gibbs free energy per particle,
$\tilde{G}(N_i,N)/N=\gexc(\nmi)$, as a function of the moments
$\nmi=N_i/N$ ($i=1\ldots K$) of the normalized particle distribution
$\nsig$. Altogether, the Gibbs free energy per particle of a
truncatable system becomes
\be
g[\nsig]=G/N=T \intsig \nsig \ln\nsig + T\ln\frac{\Pi}{T} + \gexc(\nmi)
\label{g_decomp}
\ee
This has again a truncatable structure; the excess part $\gexc$
``inherits'' its moment structure from $\fexc$. Note that because of
the normalization of the $\nmi$, $\gexc$ does not depend on the
density $\rh\equiv\m_0$. It is therefore normally a function of one
less variable than $\fexc$, unless $\fexc$ is already independent of
$\rh$ (as is the case in Flory-Huggins theory,
eq.~\(fexc_flory_huggins), for example). The chemical potentials
$\musig$ follow from~\(g_decomp) as
\bea
\musig &=& \frac{\delta}{\delta N(\sig)} Ng[N(\sig)/N]
\nonumber\\
&=& \frac{\delta g}{\delta \nsig} + g - \int\!d\sig'\, \prob(\sig')
\frac{\delta g}{\delta \prob(\sig')}
\nonumber\\
&=& T\ln\nsig + \sum_i \wi \frac{\partial\gexc}{\partial\nmi} + \left[
T \ln\frac{\Pi}{T} + \gexc - \sum_i \nmi
\frac{\partial\gexc}{\partial\nmi}\right]
\label{g_chem_pot}
\eea
Multiplying by $\nsig$ and integrating over $\sig$, one has
$g=\intsig\musig\nsig$ as it should be (compare eq.~\(g_decomp)).
Comparing~\(g_chem_pot) with~\(exact_chem_pot), one deduces that
\[
\frac{\partial\gexc}{\partial\nmi} = \muexci
\]
and that the density $\rh$ of a phase with moments $\nmi$ is given by
\[
T \ln\left(\frac{T\rh}{\Pi}\right) = \gexc - \sum_i \nmi \muexci.
\]

To construct the moment Gibbs free energy, one can now proceed as in
the constant volume case: in the ideal contribution, we replace
$\ln\nsig\to\ln[\nsig/\nnsig]$, with a normalized parent distribution
$\nnsig$, and then minimize $g$ at fixed values of the $\nmi$. The
minimum occurs for distributions $\nsig$ from the family
\be
\nsig = \nnsig \exp\left(\mhatn + \sum_i \mhati \wi\right)
%
%
\label{g_pfamily}
\ee
The extra Lagrange multiplier $\mhatn$ comes from the normalization
condition $\nmn=\intsig\nsig=1$. Inserting into~\(g_decomp), one
obtains the moment Gibbs free energy
\be
\gmom(\nmi)=T\left(\mhatn + \sum_i \mhati \nmi\right) + T\ln\frac{\Pi}{T} +
\gexc(\nmi)
\label{g_mom}
\ee
where the $\nmi$ are given by
\[
\nmi = \intsig \wi \nnsig \exp\left(\mhatn + \sum_i \mhati \wi\right) 
= \frac{\intsig \wi \nnsig \exp\left(\sum_i \mhati \wi\right)}
{\intsig \nnsig \exp\left(\sum_i \mhati \wi\right)}
\]
The derivatives of $\gmom$ can again be identified as moment chemical
potentials and obey
\be
\mu_i \equiv\frac{\partial\gmom}{\partial\nmi}
= T\mhati + \frac{\partial\gexc}{\partial\nmi} = T\mhati + \muexci.
\label{g_mom_chem_pot}
\ee
The constraint $\nmn=1$ does not affect this result because the
partial derivative is taken with the values of all moments other than
$\nmi$ fixed anyway. For the same reason, the second derivatives of
the ideal part of $\gmom$ are (up to a factor $T$) given by the
inverse of the matrix of second-order normalized moments $\nm_{ij}$,
by analogy with~\(mom_free_en_curvature). (Note that the first row and
column of the second-order moment matrix, corresponding to $i=0$ or
$j=0$, need to be retained; they can only be discarded {\em after} the
inverse has been taken.)

From~\(g_chem_pot), one deduces immediately that if one of a number of
coexisting phases is in the family~\(g_pfamily), then so are all
others. Equality of the exact chemical potentials $\musig$ in all
phases is then equivalent to equality of the quantities
\be
\muexci+T\mhati = \mu_i\quad (i=1\ldots K), \qquad
T\mhatn + \gexc - \sum_i \nmi \muexci.
\label{g_eq_cond}
\ee
But from~(\ref{g_mom},\ref{g_mom_chem_pot}), these are simply the
moment chemical potentials $\mu_i=\partial\gmom/\partial\nmi$ and (up
to an unimportant additive constant) the associated Legendre transform
$\gmom-\sum_i \nmi\mu_i$. Therefore, phase equilibria can be
constructed by applying the usual tangent plane construction to the
moment Gibbs free energy, in analogy with the constant volume
case. Note, however, that this tangent plane construction now takes
place in the space of the normalized moments $\nmi$, rather than that
of the unnormalized \moms\ $\mi$. As pointed out above, this
generically reduces the dimension by one. For the trivial case of a
monodisperse system, $\fexc$ is a function of $\rh$ only and $\gexc$
does not depend on any density variables at all, so that the tangent
plane condition becomes void as expected.

The exactness statements in Sec.~\ref{sec:properties} can also be
directly translated to the constant pressure case. The arguments above
imply directly that the onset of phase coexistence is found exactly
from the moment Gibbs free energy: all phases are in the
family~\(g_pfamily), because one of them (the parent) is, and the
requirement of equal chemical potentials is satisfied.  Spinodals and
(multi-) critical points are also found exactly. Arguing as in
Sec.~\ref{sec:properties_general}, and using the vector notation
of~\(spinodal_and_critical), the criterion for such points is found as
\be
\Delta\muvect \equiv 
\muvect(\nvect(\epsilon)) - \muvect(\nvect\pn)= \order(\epsilon^l)
\label{g_spinodal_and_critical}
\ee
where $l=2$ for a spinodal and $l=2n-1$ for an $n$-critical
point. Again, the curve $\nvect(\epsilon)$ only has to pass through
phases with equal chemical potentials and can therefore be chosen to
lie within the family~\(g_pfamily). As discussed above
(eq.~\(g_eq_cond)), the chemical potential difference on the left hand
side of~\(g_spinodal_and_critical) is then zero to the required order
in $\epsilon$ if the same holds for the $\Delta\mu_i$ and $\Delta h$,
where $h=\gmom-\sum_i\nmi\mu_i$. It may look as if the constraint on $h$
gives one more condition here than in the constant volume case; but in
fact $\Delta\mu_i=\order(\epsilon^l)$ for all $i$ already implies that
$\Delta\gmom=\order(\epsilon^{l+1})$ and hence $\Delta
h=\order(\epsilon^l)$. The exact condition~\(g_spinodal_and_critical)
for spinodals and critical points is therefore again equivalent to the
same condition obtained from the moment Gibbs free energy.

\section{Moment entropy of mixing and large deviation theory}
\label{app:LDT}

In this appendix, we discuss some interesting properties of the
Legendre transform result~\eqref{pbw_legendre} for the moment entropy of
mixing, in particular its relation to large deviation theory (LDT).

We first note that~\eqref{pbw_legendre}, which we derived by taking
the limit $x\to 0$ of the result~\eqref{sN_exact} for general $x$, can
also be obtained by a more direct route. In the limit $x\to0$, the
sizes of particles in the smaller system become independent random
variables drawn from $\parent$; the second phase can be viewed as a
{\em reservoir} to which the small phase is connected.  One writes the
moment generating function for $\myP(m)$ in the small phase as a
product of $xN$ independent moment generating functions of $\parent$,
and evaluates the integral over $\myP(m)$ by a 
saddle point method~\cite{pbw_schrodinger}.

One can also view~\eqref{pbw_legendre} as a generalisation of
Cram\'er's large deviation theory (LDT)~\cite{pbw_cramer}.  Recall
that this theory treats the ``wings'' of a distribution like $\myP(m)$
correctly, to which the central limit theorem (CLT) does not
apply~\cite{pbw_clt}. However, the CLT approximation in this problem
has quite a beautiful interpretation, and is worth describing
separately. It can be derived by assuming that $\parent$ is itself a
Gaussian in eq.~\eqref{pbw_legendre}, or just written down directly:
\[
\myP(m)\sim \exp\left[-N\frac{(m-\meanm)^2}{2\,\varm}\right]
\]
where $\varm$ is the variance of the parent distribution.  We see that
this gives a term in the free energy of the form
\[
\Delta f=-\kT\rho\sN=\rho\kT\frac{(m-\meanm)^2}{2\,\varm}.
\]
This result states that there is an {\em entropic spring} term in the
free energy that penalizes deviations of the mean size $m$ from the
mean size in the parent distribution.  The spring constant is
inversely proportional to the variance in the parent distribution,
thus if the parent is narrower, it is harder to move away from the
parental mean size~\cite{pbw_polymer_analogy}.  However, as indicated
above, the CLT is not sufficiently accurate for our purposes.  For
instance it gives a finite weight to $m<0$ even if $w(\sigma)$ is
strictly positive.  That said, the CLT may be attractive for narrow
size distributions, and one might make a connection with the recent
results of Evans {\em et al}~\cite{pbw_evans} (even though the latter
do not seem to rely on a specific shape of the distribution).  We have
not fully explored this avenue.

\section{Spinodal criterion from exact free energy}
\label{app:spinodal}

In this appendix, we apply the spinodal criterion~\(spinodal_general)
to the exact free energy~\(exact_free_en) and show that it can be
expressed in a form identical to~\(mom_spinodal).  This result has
been given by a number of
authors~\cite{IrvGor81,BeeBerKehRat86,Cuesta99}, but we include it
here for the sake of completeness.

Choosing for $\rhovect$ the vector $\rhosig$ (whose ``components'' are the
values of $\rhosig$ for all $\sigma$), the spinodal
criterion~\(spinodal_general) applied to~\(exact_free_en) becomes
\be
\int\!d\sig'\, \frac{\delta^2 f}{\delta \rhosig \delta\rho(\sig')} \,
\delta\rho(\sig') = \sum_{i,j} 
\frac{\partial^2\fexc}{\partial\mi\partial\m_j}\,\wi\,\delta\m_j
+ T \, \frac{\delta\rhosig}{\rhosig} = 0
\label{zero_eigenval_cond}
\ee
The change of $\ln\rhosig$ along the instability direction (which is
$\delta\rhosig/\rhosig$) is therefore a linear combination of the
weight functions $\wi$. This means that the instability direction is
within the family~\(pfamily), consistent with our general discussion
in Sec.~\ref{sec:properties_general}.  One can now
rewrite~\(zero_eigenval_cond) in the form
\be
\delta\rhosig = - \beta \sum_{i,j} \wi\,\rhosig\,
\frac{\partial^2\fexc}{\partial\mi\partial\m_j}\,\delta\m_j
\label{exact_instab_dir}
\ee
and take the moment with the $k$-th weight function $\w_k(\sig)$ to
get
\[
\delta\m_k + 
\beta \sum_{i,j} \m_{ki}
\frac{\partial^2\fexc}{\partial\mi\partial\m_j}\,\delta\m_j 
= 0.
\]
As promised, this is identical to the spinodal
condition~\(mom_spinodal) derived from the moment free energy, with
the matrix multiplications written out explicitly.

\section{Determinant form of critical point criterion}
\label{app:critical}

In this appendix, we give the form of the critical point
criterion~\(mom_critical) that uses the spinodal determinant $Y$
from~\(mom_spinodal)~\cite{pbw_gibbs}. At a critical point, the
instability direction connects two neighboring points on the
spinodal. The first order variation of the spinodal determinant $Y$
along this direction must therefore vanish. This gives
\be
\delta Y = \sum_i \frac{\partial Y}{\partial \mi}\, \delta\mi + 
\sum_{i\leq j} \frac{\partial Y}{\partial \m_{ij}}\, \delta\m_{ij} = 0.
\label{crit_point_cond}
\ee
Here the $\delta\m_{ij}$ are the changes in the second order \moms\
along the instability direction. We can express these as a function of
the $\delm_i$ via the change in the Lagrange multipliers $\mhati$:
From the definition~\(moms_lambdas), a change in the $\mhati$ changes
the first and second order \moms\ according to
\be
\frac{\partial\mi}{\partial\mhat_j}=\m_{ij}=\mmmat_{ij}, \quad
\frac{\partial\m_{ij}}{\partial\mhat_k}=\m_{ijk}.
\label{m_lambda_jacobian}
\ee
Hence changes in second and first order \moms\ are related by
\[
\delta\m_{kl} = \sum_{i} \m_{kli}\, \delta\mhati =
\sum_{ij} \m_{kli}\, (\mmmat^{-1})_{ij}\, \delta\m_j
\]
Inserting this into~\(crit_point_cond) gives the determinant form of
the critical point condition
\be
\sum_i \frac{\partial Y}{\partial \mi}\, \delta\mi + 
\sum_{ij}\sum_{k\leq l} \frac{\partial Y}{\partial \m_{kl}}\, 
 \m_{kli}\, (\mmmat^{-1})_{ij}\, \delta\m_j = 0.
\label{det_critical}
\ee
Of course, if in an application of the moment free energy method one
succeeds in calculating $Y$ explicitly as a function of the $\mi$
alone (rather than as a function of the $\mi$ and the $\m_{ij}$), then
one can determine critical points simply from the condition
\[
\delta Y = \sum_i \frac{d Y}{d \mi}\, \delm_i = 0.
\]
We have written a total derivative sign here to indicate that the
dependence of $Y$ on the $\m_{ij}$ is accounted for implicitly.


\begin{thebibliography}{10}
\bibitem{discrete_sigma_footnote}
Some components of $\sig$ can also be discrete. For example, in a blend of two
  homopolymers A and B, both of which are length-polydisperse, one would choose
  $\sig=(L,a)$, where $L$ is the chain length and $a=1,2$ labels polymer A and
  B, respectively. All integrations $\intsig$ are then to be read as the
  combined sum and integral $\int dL \sum_{a=1,2}$.
\bibitem{dew_bubble}
The parts of the CPC where the incipient phase has higher or lower density than
  the original phase are sometimes referred to as dew and bubble curve,
  respectively.
\bibitem{osmotic_footnote}
In systems containing solvent whose degrees of freedom have been eliminated
  (integrated out) from the free energy, this pressure is the osmotic pressure.
\bibitem{moment_equivalence_class_footnote}
Formally, one can partition the space of all density distributions $\rhosig$
  into equivalence classes, defined by having the same values for the \moms\
  $\mi$. The moment subspace is then the space of these equivalence classes;
  the transverse subspace corresponds to variations of $\rhosig$ within a given
  equivalence class.
\bibitem{conditioning_footnote}
In the standard notation of probability theory, the quantity $\myP(m)$ should
  really be written as $\myP(m|x)$ -- the distribution of $m$, given the fixed
  value of $x$. Eq.~(\protect\ref{Pm_norm}) is then recognized as Bayes'
  theorem.
\bibitem{exact_form_footnote}
As expected, this is of the form required for an exact solution of the phase
  coexistence problem; compare~\(effective_prior).
\bibitem{Legendre_diffs}
The derivation of the moment entropy using the \patrick\ method already
  suggests this property. The only difference is that we now treat the {\em
  \moms} $\mi$ as conjugate to the Lagrange multipliers $\mhati$, whereas in
  the \patrick\ derivation -- which works with the normalized particle
  distribution $\nsig$ -- this role is played by the {\em normalized} moments
  $\nmi$. Compare eq.~(\protect\ref{pbw_legendre}).
\bibitem{higher_moments_not_explicit_footnote}
Note that if an analytic expression for $\smom$ in terms of the $\mi$ can be
  obtained, then derivatives such as $\partial^2\smom/\partial\mi\partial\m_j$
  can of course be evaluated directly in terms of the $\mi$. The second order
  moments $\m_{ij}$ then never appear explicitly in the calculation.
\bibitem{single_instability_direction_footnote}
To simplify matters, we make the common assumption that there is only a single
  instability direction. This implies that the $n$ coexisting phases have to
  meet along a line (curve). For the more general case, where two or more
  instability directions appear simultaneously, see~\protect\cite{Brannock91}.
\bibitem{exact_prior_not_known_footnote}
Of course, this effective `prior' is not known in advance (so the name is a
  slight misnomer); it can only be determined by an exact solution of the phase
  coexistence problem. As a consequence, it also changes as control parameters
  such as temperature are varied.
\bibitem{local_stab_explicit_footnote}
More explicitly, the exact free energy and TPD are convex w.r.t.\ the
  transverse degrees of freedom of $\rhosig$; local stability therefore only
  needs to be checked for the moment degrees of freedom, where again the moment
  free energy is sufficient.
\bibitem{via_const_p}
This result can also be obtained by considering a system {\em with} solvent at
  fixed osmotic pressure $\Pi$ (see App.~\protect\ref{app:const_p}), and then
  taking the limit $\Pi\to\infty$. Note that the normalized moment
  $\nmo=\mo/\mn$ and the (unnormalized) moment density $\mo$ are identical for
  this model (and in our chosen units), because the dense limit corresponds to
  $\mn=1$.
\bibitem{not_cross_entropy}
One might also have chosen the cross-entropy as an error measure; this
  corresponds to leaving out the square in~(\protect\ref{logerr}), thus
  allowing cancellations between positive and negative values of the logarithm.
  In our numerical experiments, this tended to give misleadingly small values
  of $\logerr$.
\bibitem{logerr_not_monotonic}
One may expect intuitively that, starting from $\logerr=0$ at the cloud point,
  $\logerr$ should increase as one moves deeper into the coexistence region by
  increasing $\chiab$. We find that this is so, but only as an overall trend:
  when new phases are detected as $\chiab$ is increased, $\logerr$ may
  temporarily decrease before increasing again. This explains the crossing of
  the curves for $\chiab=10$ and $\chiab=15$ in
  Fig.~\protect\ref{fig:logerr_chi5_10_15}.
\bibitem{extra_moments_asymptotics}
One exception might be the case where the polydispersity variable $\sig$ can
  assume values from an infinite range (as in the length polydisperse polymer
  problem treated in Sec.~\protect\ref{sec:length_homopolymer}, where there is
  no upper limit on $\sig\equiv L$). The form of the formally exact
  solution~(\protect\ref{effective_prior}) can then give information about the
  asymptotic behavior of the extra weight functions. We leave this issue for
  future work.
\bibitem{footnote_no_symmetry}
Note that our algorithm does not make use of the symmetry $\sigma\to-\sigma$
  which could in principle be used to reduce the complexity of the three-phase
  coexistence problem to that of ordinary two-phase coexistence. The scenario
  therefore provides a genuine test of the multi-phase capabilities of the
  algorithm.
\bibitem{pbw_clt}
The CLT only claims that $\myP(m)$ converges to a Gaussian if one
  simultaneously rescales $m-\meanm$ by $N^{1/2}$, and is thus strictly limited
  to the region $|m-\meanm|\alt N^{-1/2}$. LDT, on the other hand, addresses
  the asymptotic behaviour of the Cram\'er `rate function' $\loge\myP$ for all
  $m$.
\end{thebibliography}

\end{document}